\documentclass[aps,floats,floatfix,amssymb,amsmath,prl,nofootinbib,
twocolumn,superscriptaddress,groupedaddress]{revtex4-1}

\usepackage{tabularx}
\usepackage{capt-of}
\usepackage{diagbox}
\usepackage{calc}
\usepackage{psfrag}
\usepackage{graphicx}
\usepackage{color}
\usepackage[dvipsnames]{xcolor}
\usepackage{colortbl}
\usepackage[unicode=true,pdfusetitle, hypertexnames=false,
 bookmarks=true,bookmarksnumbered=false,bookmarksopen=false,
 breaklinks=false,pdfborder={0 0 0},backref=false,colorlinks=true]
 {hyperref}

\definecolor{plava_boja}{rgb}{0.0, 0.0, 0.6}
\usepackage{hyperref}
\hypersetup{
    colorlinks=true,
    linkcolor=plava_boja,
    filecolor=plava_boja,      
     urlcolor=plava_boja,
     citecolor=plava_boja
}

\usepackage{geometry}
\usepackage{physics}
\usepackage{soul}
\pdfpageattr {/Group << /S /Transparency /I true /CS /DeviceRGB>>}

\newcommand{\im}[1]{\mathrm{Im}\, #1}
\newcommand{\re}[1]{\mathrm{Re}\, #1}

\usepackage[makeroom]{cancel}


\bibpunct{[}{]}{,}{n}{}{}
\usepackage[normalem]{ulem}

\makeatletter
\let\ps@titlepage\ps@plain
\makeatother

\begin{document}

\title{Spectral Functions of the Holstein Polaron: Exact and Approximate Solutions}
\author{Petar Mitri\'c}
\affiliation{Institute of Physics Belgrade,
University of Belgrade, Pregrevica 118, 11080 Belgrade, Serbia}
\author{Veljko Jankovi\'c}
\affiliation{Institute of Physics Belgrade,
University of Belgrade, Pregrevica 118, 11080 Belgrade, Serbia}
\author{Nenad Vukmirovi\'c}
\affiliation{Institute of Physics Belgrade,
University of Belgrade, Pregrevica 118, 11080 Belgrade, Serbia}
\author{Darko Tanaskovi\'c}
\affiliation{Institute of Physics Belgrade,
University of Belgrade, Pregrevica 118, 11080 Belgrade, Serbia}

\begin{abstract}
It is generally accepted that the dynamical mean field theory gives a good solution of the Holstein model, but only in dimensions greater than two. Here, we show that this theory, which becomes exact in the weak coupling and in the atomic limit, provides an excellent, numerically cheap, approximate solution for the spectral function of the Holstein model in the whole range of parameters, even in one dimension. To establish this, we make a detailed comparison with the spectral functions that we obtain using the newly developed  momentum-space numerically exact hierarchical equations of motion method, which yields electronic correlation functions directly in real time. We crosscheck these conclusions with our path integral quantum Monte Carlo and exact diagonalization results, as well as with the available numerically exact results from the literature.
\end{abstract}
\maketitle

\begin{figure}[!t]
 \includegraphics[width=\columnwidth,trim=0cm 0cm 0cm 0cm]{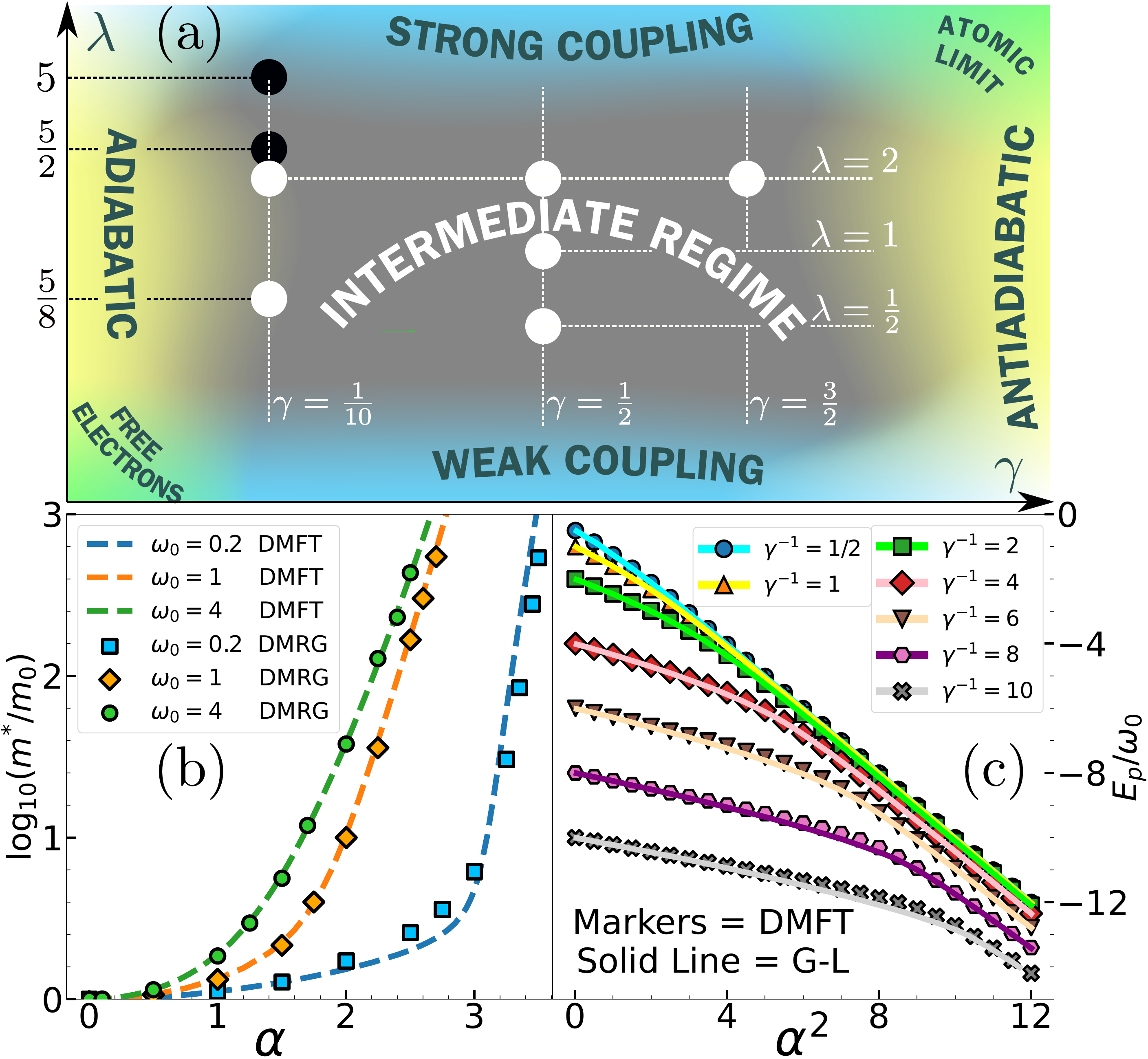}
 \caption{
 (a) Schematic plot of different regimes in the $(\gamma, \lambda)$ parameter space. The white (black) circles correspond to parameters for which both HEOM and QMC (just QMC) calculations were performed. The DMFT results are obtained in practically whole space of parameters.
 (b) Comparison of the DMFT and DMRG (taken from Refs.~\cite{JeckelmannWhite_1998,Romero_1998}) renormalized electron mass at $T=0$. 
 (c) Comparison of the ground state energy from the DMFT and the global-local variational approach (taken from Ref.~\cite{Romero_1998}) at $T=0$.
}
\label{fig:scheme} 
\end{figure}

\hspace{8pt}
The Holstein model is the simplest model that describes an electron that
propagates through the crystal and interacts with localized optical
phonons \cite{HolsteinI_1959}. On the example of this model, numerous many-body methods were developed and tested \cite{Alexandrov_book2007}. The Holstein molecular crystal model is also very important in order to understand the role of polarons (quasiparticles formed by an electron dressed by lattice vibrations) in real materials \cite{Franchini_NatRevMater2021}.
This is still a very active field of research fueled by new directions in
theoretical studies \cite{Vidmar_2011,Kloss_2019,Jeckelmann_2017,Prodanovic_2019,Stolpp_2020,
Murakami_2019,Jansen_2019,Fetherolf_2020, 2014_Mishchenko} and advances in experimental techniques \cite{Kang_NatRevMater2018}. 
\newline
\hspace*{8pt} 
The Holstein model can be solved analytically only in the limits of weak and strong electron-phonon coupling \cite{Mahanbook,LangFirsov,Alexandrov_book2010}. Reliable numerical results for the ground state energy and quasiparticle effective mass were obtained in the late 1990s using the density matrix renormalization group (DMRG) \cite{JeckelmannWhite_1998,ZhangWhite_1999} and path integral quantum Monte Carlo (QMC) methods \cite{Kornilovitch_1998}, and also within
variational approaches \cite{Romero_1998,Romero_1999,Bonca_1999}. At the time, numerically exact spectral functions for one-dimensional (1D) systems were obtained only within
the DMRG method \cite{JeckelmannWhite_1998,ZhangWhite_1999}. The main drawback of the QMC method is that it gives
correlation functions in imaginary time and obtaining spectral
functions and dynamical response functions is often impossible since the
analytical continuation to the real frequency is a numerically ill-defined
procedure. Interestingly, at finite temperature the spectral functions
were obtained only very recently using finite-$T$ Lanczos (FTLM) \cite{Bonca_Lanczos_2019} and finite-$T$ DMRG \cite{JansenBonca_2020} methods. All these methods have their strengths and
weaknesses depending on the parameter regime and temperature. As
usually happens in a strongly interacting many-body problem, a complete
physical picture emerges only by taking into account the solutions obtained with different methods.
\newline
\hspace*{8pt} 
The hierarchical equations of motion (HEOM) method is a numerically exact technique that has recently gained popularity in the chemical physics community \cite{Tanimura_2020,Xu_HEOM_2007,Jin_Yan_JChemPhys_2008,DongHou_PRB2014}. It has been used to explore the dynamics of an electron (or exciton) linearly coupled to a Gaussian bosonic bath. Within HEOM, we calculate the correlation functions directly on the real time (real frequency) axis \cite{ZhenHua_PRL2012}. Nevertheless, the applications of the HEOM method to the Holstein model \cite{ChenTanimura_2015,SongShi_2015I,SongShi_2015II,Dunn_2019,YanShi_2020} have been, so far, scarce because of the numerical instabilities stemming from the discreteness of the phonon bath on a finite lattice.
\newline
\hspace*{8pt} 
Along with numerically exact methods, a number of approximate techniques
have been developed and applied to the Holstein model \cite{Hohenadler_2003,Filippis_2005,Berciu_PRL2006,Berciu_PRB2006}. The dynamical mean field theory (DMFT) is a simple nonperturbative technique that has emerged as a method of choice for the studies of the Mott physics within the Hubbard model \cite{DMFT_RMP1996,martin_reining_ceperley_2016}. It can also be applied to the Holstein model giving numerically cheap results directly on the real frequency axis \cite{Ciuchi_1997}. This method fully takes into account local quantum
fluctuations and it becomes exact in the limit of infinite coordination
number when the correlations become completely local. It was soon
recognized \cite{Fratini_2001,Fratini_2003} that the DMFT gives qualitatively correct spectral functions
and conductivity for the Holstein model in three dimensions. In low-dimensional systems  the solution is approximate as it neglects the nonlocal
correlations and one might expect that the DMFT solution would not be accurate, particularly in one dimension. Surprisingly, to our knowledge, only the DMFT solution for the Bethe lattice
was used in comparisons with the numerically exact results for the
ground state properties in one dimension \cite{Romero_1998,Ku_Bonca_2002}. The  quantitative
agreement was rather poor suggesting that the DMFT
cannot provide a realistic description of the low-dimensional Holstein
model due to the importance of nonlocal correlations \cite{Romero_1998,Ku_Bonca_2002,Alexandrov_book2010}.
\newline
\hspace*{8pt} 
In this Letter, we present a comprehensive solution of the 1D Holstein model:
(i) We solve the DMFT equations in all parameter regimes. At zero temperature we find a remarkable agreement of the DMFT ground state energy and effective mass with the available results from the literature in one, two, and three dimensions.
(ii) For intermediate electron-phonon coupling, we obtain numerically exact spectral functions using the recently developed momentum-space HEOM approach \cite{2022_Jankovic}. For strong coupling we calculate the spectral functions using exact diagonalization (ED). We find a very good agreement with DMFT results and therefore demonstrate that the DMFT is rather accurate, in sharp contrast to current belief in the literature.
(iii) 
We crosscheck the results with our QMC calculations in imaginary time.
Overall, we demonstrate that the DMFT emerges as a unique method that gives close to exact spectral functions in the whole parameter space of the Holstein model, both at zero and at finite temperature.
\newline
\hspace*{8pt} 
\emph{Model and methods.---}
We study the 1D Holstein model given by the Hamiltonian
\begin{align} \label{eq:Holstein_hamiltonian}
H =& -t_0 \sum_i \left( c_i^\dagger c_{i+1} + \mathrm{H.c.} \right) \nonumber\\
   & -g \sum_i n_i \left( a_i^\dagger + a_i \right) + \omega_0 \sum_i a_i^\dagger a_i .
\end{align}
Here, $c_i^\dagger$($a_i^\dagger$) are the electron (phonon) creation  operators, $t_0$ is the hopping parameter and, $n_i = c_i^\dagger c_i$.  
We consider dispersionless optical phonons of frequency $\omega_0$, and 
$g$ denotes the electron-phonon coupling parameter. $t_0$, $\hbar$, $k_B$, and lattice constant are set to $1$. We consider the dynamics of a single electron in the band. It is common to define several dimensionless parameters: adiabatic parameter $\gamma = \omega_0 / 2t_0$, electron-phonon coupling  $\lambda = g^2/2t_0 \omega_0$, and $\alpha = g /\omega_0$. These parameters correspond to different physical regimes of the Holstein model shown schematically in Fig.~\ref{fig:scheme}(a). 
\newline
\hspace*{8pt} 
In order to obtain reliable solutions in the whole parameter space, we use two approximate methods and three methods that are numerically exact. In the
Holstein model, the DMFT reduces to solving
the polaron impurity problem in the conduction electron band
supplemented by the self-consistency condition~\cite{Ciuchi_1997}. The impurity problem can be solved in terms of the continued fraction expansion, giving the local Green's function on the real frequency axis (see Ref.~\cite{Ciuchi_1997} and Supplemental Material (SM) \cite{SuppMat}, Sec.~\ref{DMFT_1D}, for details). A crucial advantage of the DMFT for the Holstein model is that it becomes exact in both the weak coupling and in the atomic limit, and that it can be easily applied in the whole parameter space both at zero and at finite temperature. 
The DMFT equations can be solved on a personal computer in just a few seconds to a few minutes depending on the parameters.
On general grounds, the DMFT is expected to work particularly well at high temperatures when the correlations become more local due to the thermal fluctuations \cite{Vucicevic_2019,Vranic_2020}. We will compare the DMFT with the well-known self-consistent Migdal approximation (SCMA) \cite{Migdal_1958}, which becomes exact only in the weak coupling limit; see Sec.~\ref{Migdal_App} of SM \cite{SuppMat}. 
\newline
\hspace*{8pt} 
We have recently developed the momentum-space HEOM method \cite{2022_Jankovic} that overcomes the numerical instabilities 
originating from the discrete bosonic bath. Within this method we calculate the time-dependent greater Green's function $G^>(k,t)$, which presents the root of the hierarchy of the auxiliary Green's functions. The hierarchy is, in principle, infinite, and one actually solves the model by truncating the hierarchy at certain  depth $D$.  
The HEOM are propagated independently for each allowed value of $k$ up to long times ($\omega_0t_\mathrm{max}\sim 500$). 
The propagation takes 5 to 10 hours on 16 cores per momentum $k$. The discrete Fourier transform is then used to obtain spectral functions without introducing any artificial broadening.
Numerical error in the HEOM solution can originate from the finite-size effects since the method is applied on the lattice with $N$ sites, and also from the finite depth $D$. 
We always use $N$ and $D$, as given in SM \cite{SuppMat}, which correctly represent the thermodynamic limit.
Generally, for larger $g$ we need smaller $N$ and larger $D$. This is why the ED method with a small number of sites could be a better option in the strong coupling regime. The ED method can be used more efficiently after the initial Hamiltonian is  transformed by applying the Lang-Firsov transformation; see SM \cite{SuppMat}, Sec.~\ref{ED_method}.
\newline
\hspace*{8pt} 
In the QMC method we calculate the correlation function 
 $C_{k}(\tau)= 
\langle c_{k}(\tau) c_{k}^\dagger \rangle_{T,0}$  in imaginary time.  The thermal expectation value is performed over the states with zero electrons and $c_k(\tau) = e^{\tau H} c_k e^{-\tau H}$. We use the path integral representation, the discretization of imaginary time, and analytical calculation of integrals over the phonon coordinates. We then evaluate a multidimensional sum over the electronic coordinates by a Monte Carlo method. This method is a natural extension of early works where such approach was applied just to thermodynamic quantities \cite{Raedt_1982,Raedt_1983,Raedt_1984}. Details of the method are presented in Ref.~\cite{2022_Jankovic}. 
%
\newline
\hspace*{8pt} 
\emph{Results at zero temperature.---} In Fig.~\ref{fig:scheme}(b), we show the DMFT results for the electron effective mass at the bottom of the band, $m^*/m_0 = 1-d \mathrm{Re} \Sigma(\omega)/d\omega|_{E_p}$ (where $\Sigma(\omega)$ is the self-energy), over a  broad range of parameters covering practically the whole parameter space in the $(\gamma,\lambda)$ plane. We see that the mass renormalization is in striking agreement with the DMRG result \cite{JeckelmannWhite_1998,Romero_1998} that presents the best available result from the literature. Small discrepancies are visible only for stronger interaction with small $\omega_0$. A similar level of agreement can be seen in the comparison of the ground state (polaron) energy $E_p$ in Fig.~\ref{fig:scheme}(c). Here, the results obtained with variational global-local method \cite{Romero_1998,Romero_1999} are taken as a reference. While the agreement in the weak coupling and in the atomic limit could be anticipated since the DMFT becomes exact in these limits, we find the quantitative agreement in the crossover regime between these two limits rather surprising, having in mind that the DMFT completely neglects nonlocal correlations. It is also interesting that this was not observed earlier. The only difference from the standard reference of Ciuchi {\it et al.}~\cite{Ciuchi_1997} is that we applied the DMFT to the 1D case, as opposed to the Bethe lattice. This is, however, a key difference. Otherwise the DMFT provides only a qualitative description of the Holstein model \cite{Romero_1998,Ku_Bonca_2002,Barisic2006,Alexandrov_book2010,Franchini_NatRevMater2021}. From the technical side, the only difference as compared to the case of the Bethe lattice is in the self-consistency equation. For obtaining a numerically stable and precise solution, it was crucial to use an analytical expression for the self-consistency relation (see Sec.~\ref{practical_dmft} in SM \cite{SuppMat}). We have also calculated the effective mass for two- and three-dimensional lattices (see Sec.~\ref{Supp:Effective_mass} in SM \cite{SuppMat}) and the agreement with the QMC calculation from Ref.~\cite{Kornilovitch_1998} is excellent. This was now expected since the importance of nonlocal correlations decreases in higher dimensions. A comparison with the Bethe lattice effective mass is illustrated in SM \cite{SuppMat}, Sec.~\ref{Supp:bethe_lattice_comp}.
\begin{figure}[!t]
  \includegraphics[width=\columnwidth,trim=0cm 0cm 0cm 0cm]{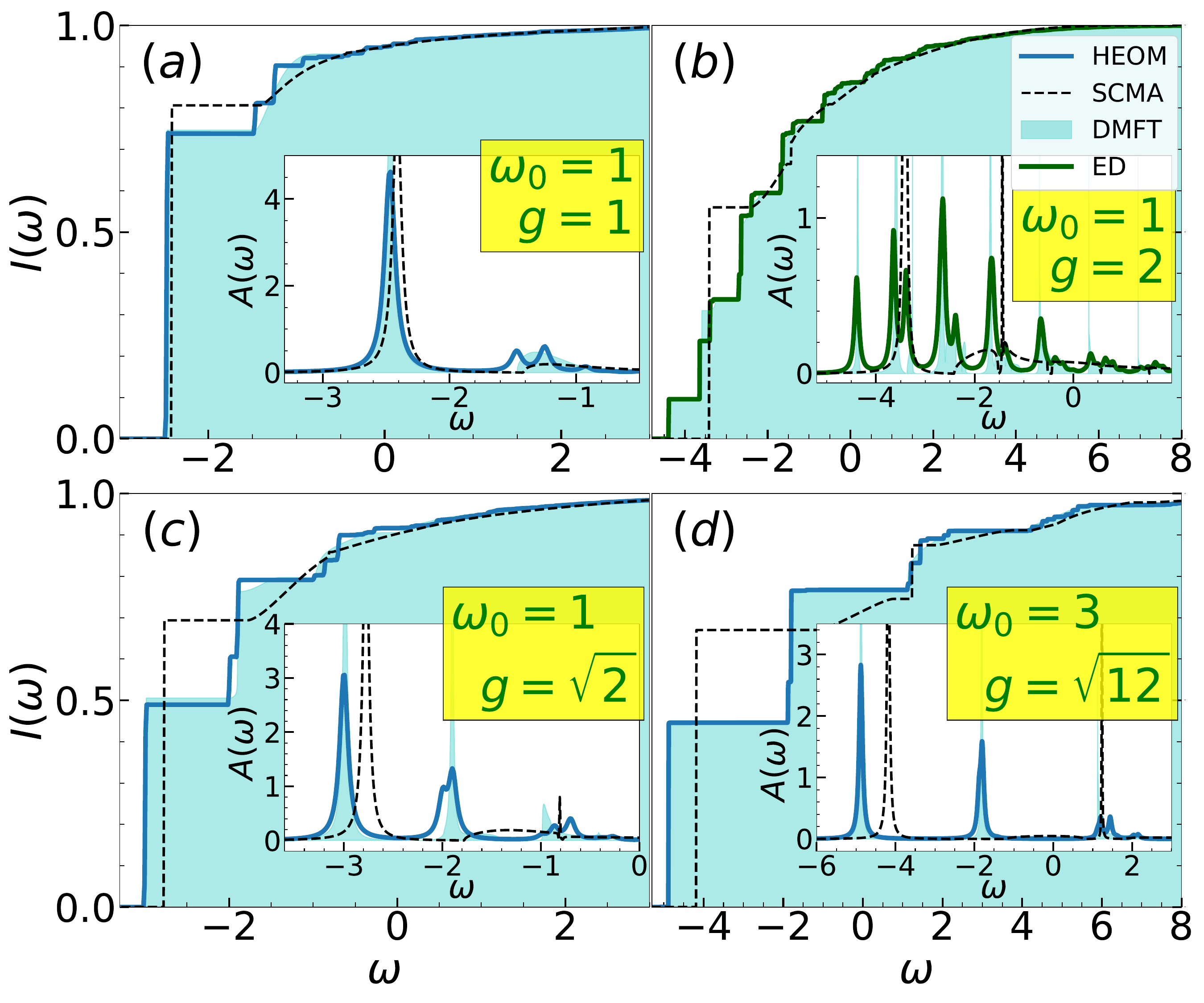} 
 \caption{(a)--(d) Integrated HEOM, DMFT, SCMA and ED spectral weight, $I(\omega) = \int_{-\infty}^{\omega} d\nu A_k(\nu)$, for $k=0$ and $T=0$. The insets show comparisons of the spectral functions. $I(\omega)$ is obtained without broadening, whereas $A(\omega)$ is broadened by Lorentzians of half-width $\eta = 0.05$. 
 }
 \label{fig:zeroT}
\end{figure}
\newline
\hspace*{8pt} 
The next step is to check if the agreement with the numerically exact solution extends also to spectral functions. Typical results at $k=0$ are illustrated in Fig.~\ref{fig:zeroT}. We note that at $T=0$ the DMFT quasiparticle peak is a delta function (broadened in Fig.~\ref{fig:zeroT}), while satellite peaks are incoherent having intrinsic nonzero width.
In HEOM, the peak broadening due to the finite lattice size $N$ and finite propagation time $t_\mathrm{max}$ is generally much smaller than the Lorentzian broadening used in the insets of Figs.~\ref{fig:zeroT}(a)--(d). The weights of the DMFT and HEOM quasiparticle peaks correspond to the $m_0/m^*$ ratio. 
The satellite peaks are also very well captured by the DMFT solution  in all parameter regimes.
For $g = 1$ we can see two small peaks in the first satellite structure of the HEOM solution. We find very similar peaks also in the DMFT solution when applied on a lattice of the same size, which is here equal to $10$ (see SM \cite{SuppMat}, Sec.~\ref{Sec:finitesize}).
Hence, we conclude that these peaks are an artefact of the finite
lattice size.
%
In the strong coupling regime $\omega_0=1, g=2$, the DMFT is compared with ED since the thermodynamic limit is practically reached for $N=4$; see SM \cite{SuppMat}, Sec.~\ref{Sec:finitesize}. 
%
Here, we notice a pronounced excited quasiparticle peak \cite{Bonca_1999,Bonca_Lanczos_2019} whose energy is below $E_p + \omega_0$. This peak, which consists of a polaron and a bound phonon, is also very well resolved within the DMFT solution. 
%
For parameters in Fig.~\ref{fig:zeroT}(d) the lattice sites are nearly decoupled, approaching the atomic limit ($t_0 \ll g, \omega_0$), when the DMFT becomes exact (see Sec.~\ref{Atomic_App} in SM \cite{SuppMat}).
For a comparison, we show also the SCMA spectral functions.  As the interaction increases, the SCMA solution misses the position and the weight of the quasiparticle peak and the satellite peaks are not properly resolved. Further comparisons of zero temperature spectral functions are shown in Sec.~\ref{T=0SUPP} of SM \cite{SuppMat}.\newline
\begin{figure}[!t]
  \includegraphics[width=\columnwidth,trim=0cm 0cm 0cm 0cm]{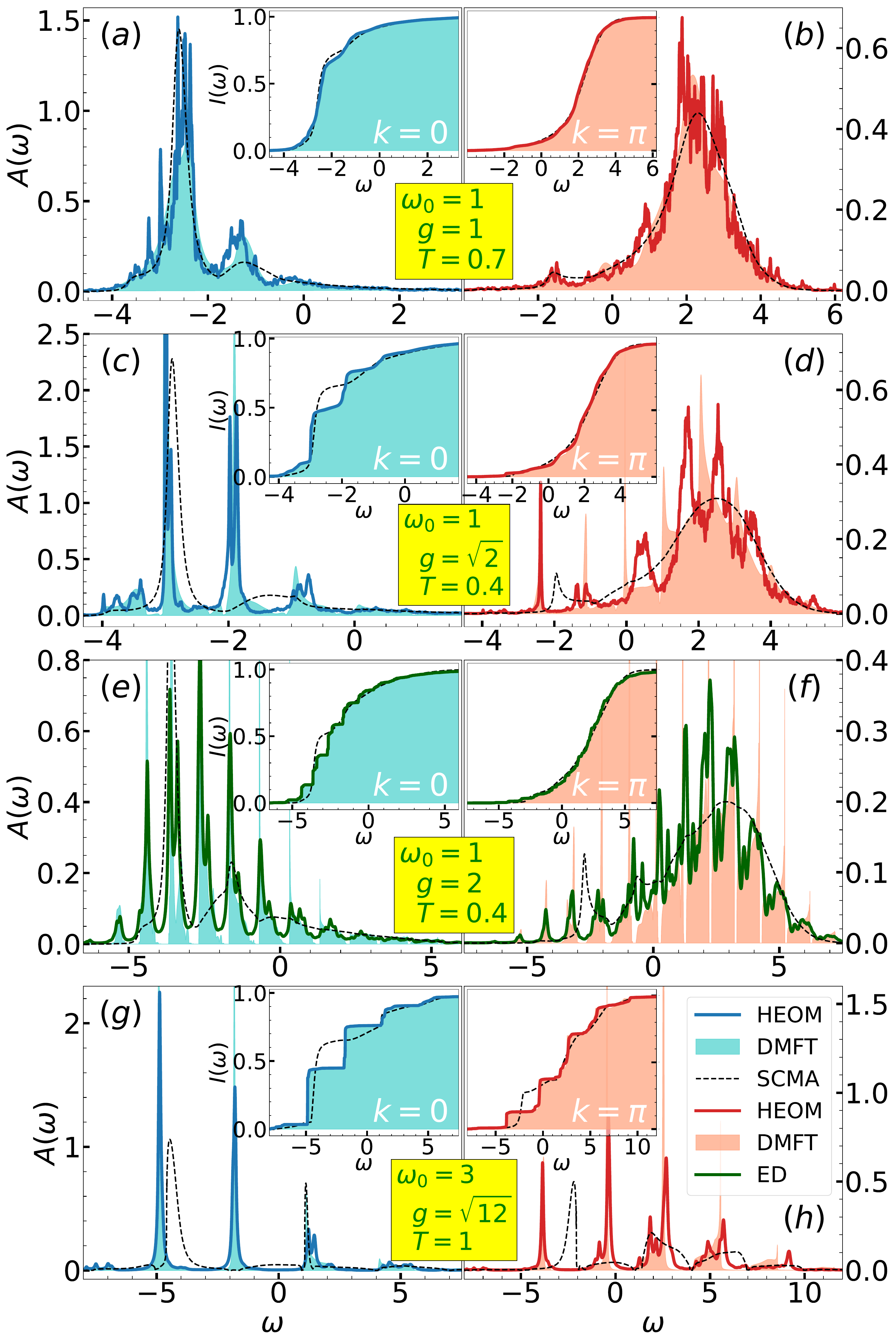} 
 \caption{(a)--(h) Spectral functions at $T>0$ for $k=0$ and ${k=\pi}$.  In panels (e)--(f) only the ED results are broadened by Lorentzians of half-width $\eta=0.05$, while all the curves are broadened in (g)--(h) with the same $\eta$. All insets are shown without broadening.
 }
 \label{fig:finiteT}
\end{figure}
%
%
\hspace*{8pt} 
\emph{Results at finite temperature.---} Reliable finite-$T$ results for the spectral functions of the Holstein model have been obtained only very recently using the FTLM \cite{Bonca_Lanczos_2019} and finite-$T$ DMRG methods \cite{JansenBonca_2020}.  Here, we  calculate the spectral functions using HEOM or ED and compare them extensively with the DMFT. The results are crosschecked using the QMC results in imaginary time. 
\newline
\hspace*{8pt} 
Typical results for the spectral functions are shown in Fig.~\ref{fig:finiteT}, while additional results for other momenta and other parameters are shown in Sec.~\ref{TfiniteSUPP} of SM \cite{SuppMat}. We see that for $T>0$ the satellite peaks appear also below the quasiparticle peak. The agreement between the DMFT and HEOM (ED) spectral functions is very good.  The agreement remains excellent even for $g=2$ where the electrons are strongly renormalized $m^* / m_0 \approx 10$, which is far away from both the atomic and weak coupling limits, where the DMFT is exact. A part of the difference between the DMFT and HEOM (ED) results can be ascribed to the small finite-size effects in the HEOM and ED solutions, as detailed in SM \cite{SuppMat}, Sec.~\ref{Sec:finitesize}. In accordance with the presented results, it is not surprising that the self-energies are nearly $k$ independent, as shown in SM \cite{SuppMat}, Sec.~\ref{self_energies}.
It is also instructive to examine the difference between the SCMA and DMFT(HEOM) solutions. For moderate interaction [Figs.~\ref{fig:finiteT}(a)~and~\ref{fig:finiteT}(b)], the weight of the SCMA quasiparticle peak is nearly equal to the DMFT(HEOM) quasiparticle weight, and the overall agreement of spectral functions is rather good. This is not the case for stronger electron-phonon coupling [Figs.~\ref{fig:finiteT}(c)--(h)] where the SCMA poorly approximates the true spectrum.
\begin{figure}[!t]
\includegraphics[width=\columnwidth,trim=0cm 0cm 0cm 0cm]{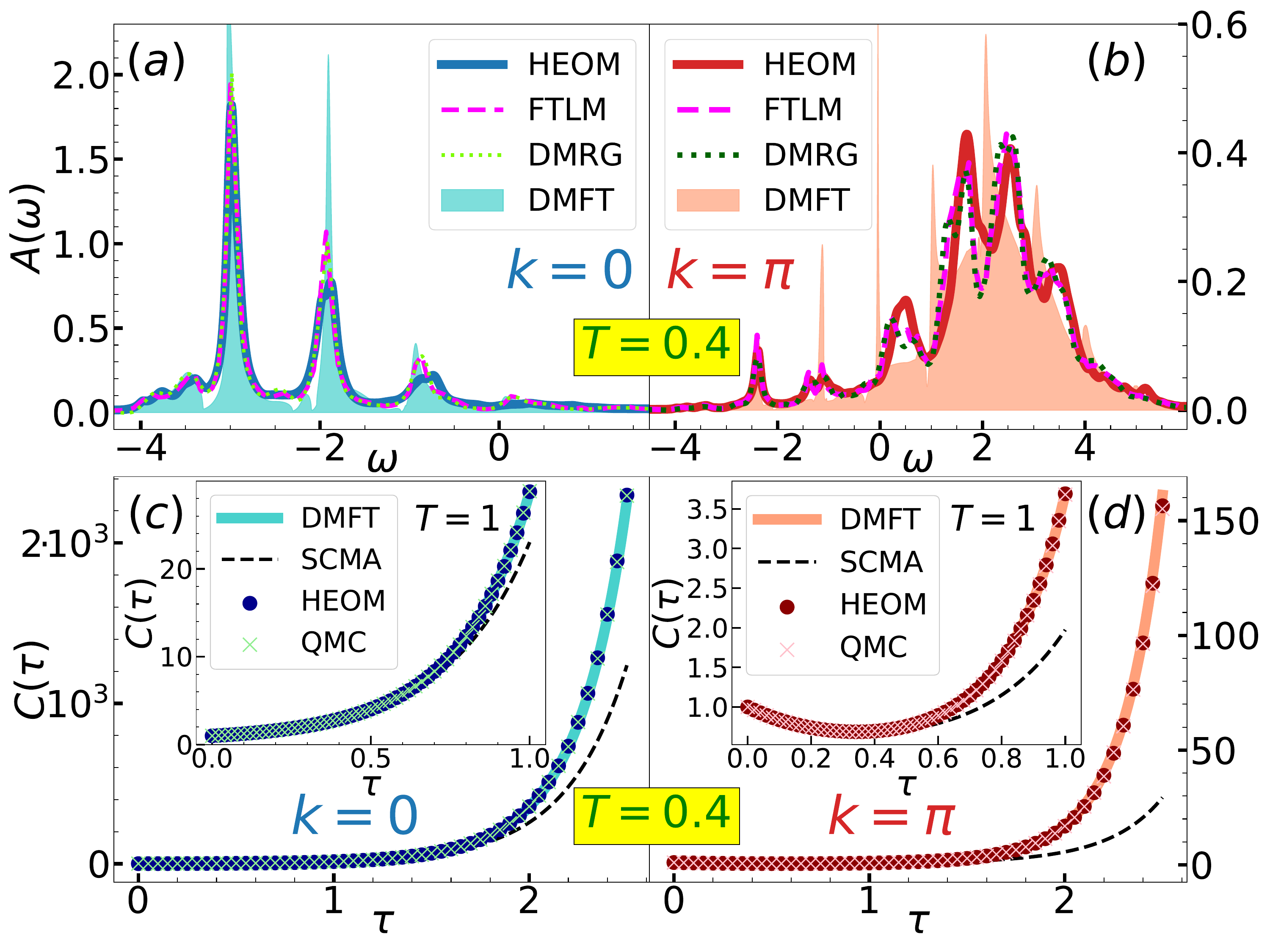} 
 \caption{(a)--(b) Comparison of DMFT, HEOM, and finite-$T$ DMRG and FTLM (taken from Ref.~\cite{JansenBonca_2020}) spectral functions at $T=0.4$. All the lines are here broadened by Lorentzians of half-width $\eta=0.05$. (c)--(d)  DMFT, QMC, HEOM, and SCMA imaginary time correlation functions at $T=0.4$ ($T=1$ in the insets). Here, $g=\sqrt{2}$, $\omega_0=1$.
 }
 \label{fig:corF}
\end{figure}
\newline
\hspace*{8pt} 
We observe that for $g=\sqrt 2$ and $k=\pi$ the DMFT and HEOM satellite peaks are somewhat shifted with respect to one another, see Figs.~\ref{fig:finiteT}(c)~and~\ref{fig:finiteT}(d). This is the most challenging regime for the DMFT, representing a crossover ($\lambda=1$) between the small and large polaron. Nevertheless, the agreement remains very good near the quasiparticle peak for $k=0$, which will be the most important for transport in weakly doped systems. 
In order to gain further confidence into the details of the HEOM spectral functions for $g=\sqrt{2}$, we compare them with the available results obtained within the finite-$T$ DMRG and Lanczos methods. We find an excellent agreement, as shown in Figs.~\ref{fig:corF}(a)~and~\ref{fig:corF}(b).
\newline
\hspace*{8pt} 
The DMFT and HEOM results are crosschecked with the path integral QMC calculations. 
The quantity that we obtain in QMC is the single electron correlation function in imaginary time, which can be expressed through the spectral function as $C_{k}(\tau) = \int_{-\infty}^{\infty} d\omega \, e^{-\omega \tau} A_k(\omega )$. 
Typical results are illustrated in Figs.~\ref{fig:corF}(c)~and~\ref{fig:corF}(d), while extensive comparisons are presented in Sec.~\ref{CorrF} of SM \cite{SuppMat}. At $T=0.4$ we can see a small difference in $C_{\pi}(\tau)$ between the DMFT and QMC(HEOM) results. At $T=1$, both for $k=0$ and $k=\pi$, the difference in $C_{k}(\tau)$ is minuscule, well below the QMC error bar, which is smaller than the symbol size. 
This confirms that nonlocal correlations are weak.
Similarly, as for the spectral functions, the SCMA correlation functions show clear deviation from other solutions.
We, however, note that great care is needed when drawing conclusions from the imaginary axis data since a
very small difference in the imaginary axis correlation functions can   correspond to substantial differences in spectral functions. 
\newline
\hspace*{8pt} 
\emph{Conclusions.---} 
In summary, we have presented a comprehensive solution of the 1D Holstein polaron covering all parameter regimes. We showed that the DMFT is a remarkably good approximation in the whole parameter space. This approximation is simple, numerically efficient, and can also be easily applied  in two and three dimensions.
We successfully used momentum-space HEOM  and ED methods
for comparisons with the DMFT spectral functions both at zero and at finite temperature.  
The comparisons showed an excellent agreement between the spectral functions in most of the parameter space. For parameters that are most challenging for the DMFT, a very good agreement was found around $k=0$ and a reasonably good agreement was obtained at larger values of $k$.
All of the results are crosschecked with the imaginary axis QMC calculations and with the available results from the literature. 
Both the DMFT and HEOM methods are implemented directly in real frequency, without artificial broadening of the spectral functions. This will be crucial in order to calculate dynamical quantities and determine a potential role of the vertex corrections to conductivity by avoiding possible pitfalls of the analytical continuation, which we leave as a challenge for future work.
%
%
\begin{acknowledgments}
\hspace*{8pt} 
D.$\,$T.~acknowledges useful discussions with V.~Dobrosavljevi\'c.
 We thank J.~Bon\v ca for sharing with us the data from Ref.~\cite{Bonca_Lanczos_2019}.
The authors acknowledge funding provided by the Institute of Physics Belgrade, through the grant by the Ministry of Education, Science, and Technological Development of the Republic of Serbia.
Numerical simulations were performed on the PARADOX supercomputing facility at the Scientific Computing Laboratory, National Center of Excellence for the Study of Complex Systems, Institute of Physics Belgrade. 
\end{acknowledgments}

\bibliographystyle{apsrev4-1}
%


\clearpage
\pagebreak
\newpage

%


\onecolumngrid
\begin{center}
  \textbf{\large Supplemental Material: Spectral functions of the Holstein polaron: exact and approximate solutions}\\[.2cm]
  Petar Mitri\'c,$^{1}$ Veljko Jankovi\'c,$^{1}$ Nenad Vukmirovi\'c,$^{1}$ and Darko Tanaskovi\'c$^1$\\[.1cm]
  {\itshape ${}^1$Institute of Physics Belgrade,
University of Belgrade, Pregrevica 118, 11080 Belgrade, Serbia}
  \\[1cm]
\end{center}
\twocolumngrid

\setcounter{equation}{0}
\setcounter{figure}{0}
\setcounter{table}{0}
\makeatletter
\renewcommand{\theequation}{S\arabic{equation}}
\renewcommand{\thefigure}{S\arabic{figure}}
\renewcommand{\bibnumfmt}[1]{[S#1]}
\renewcommand{\citenumfont}[1]{S#1}
\renewcommand{\thetable}{S\arabic{table}}

Here we present numerical results that complement the main text and we also show some technical details of the calculations. The Supplemental Material is organized as follows. The DMFT for the Holstein polaron is briefly reviewed in Sec.~\ref{DMFT_1D}. Numerical implementation of the DMFT self-consistency loop is presented in detail and it is used to calculate the mass renormalization in one, two and three dimensions and for the Bethe lattice as well. In Sec.~\ref{Migdal_App} the self-consistent Migdal approximation is briefly reviewed and used as a benchmark for the DMFT in the weak-coupling limit. 
Sec.~\ref{ED_method} presents the ED method. In Sec.~\ref{Sec:finitesize} we investigate how the results depend on the chain length $N$ and on hierarchy depth $D$.
Sec.~\ref{Atomic_App} examines the DMFT solution close to the atomic limit.  Additional DMFT, SCMA, ED and HEOM results for the spectral functions at $T=0$ and $T>0$ for various parameter values and for different momenta $k$ are shown in Secs.~\ref{T=0SUPP} and \ref{TfiniteSUPP}, respectively. 
The $k$-dependence of the self-energies is shown in Sec.~\ref{self_energies}.
A detailed comparison of the DMFT, HEOM and QMC correlation functions is presented in Sec.~\ref{CorrF}. 
Sec.~\ref{Integral_A} presents a numerical procedure that was used for the calculation of the integrated spectral weight. In Sec.~\ref{def_spectral} we show that the different definitions of spectral functions used by various methods are all in agreement.
 

\section{DMFT for the Holstein polaron}
\label{DMFT_1D}

The DMFT solution for the Holstein polaron on the infinitely-connected Bethe lattice was presented by Ciuchi {\it et al.}~in 1997 \cite{SMCiuchi_1997}. Interestingly, to our knowledge, this method has not been so far implemented on a finite-dimensional lattice. Details of the implementation in 1d and in arbitrary number of dimensions are the main content of this Section.


\subsection{Physical content of the DMFT approximation}
\label{physical_content_dmft}

The DMFT was developed in the early 1990's in the context of the Hubbard model \cite{SMDMFT_RMP1996} and has since significantly contributed to our understanding of the systems with strong electronic correlations \cite{SMmartin_reining_ceperley_2016}. The DMFT is a non-perturbative method that fully takes into account local quantum fluctuations. It becomes exact in the limit of infinite coordination number \cite{SMDMFT_RMP1996}, while it can be considered as an approximation in finite number of dimensions that keeps only local correlations by assuming that the self-energy $\Sigma(\omega)$ is ${\bf k}$-independent. 


\begin{figure}[!t]
 \includegraphics[width=3.0in,trim=0cm 0cm 0cm 0cm]{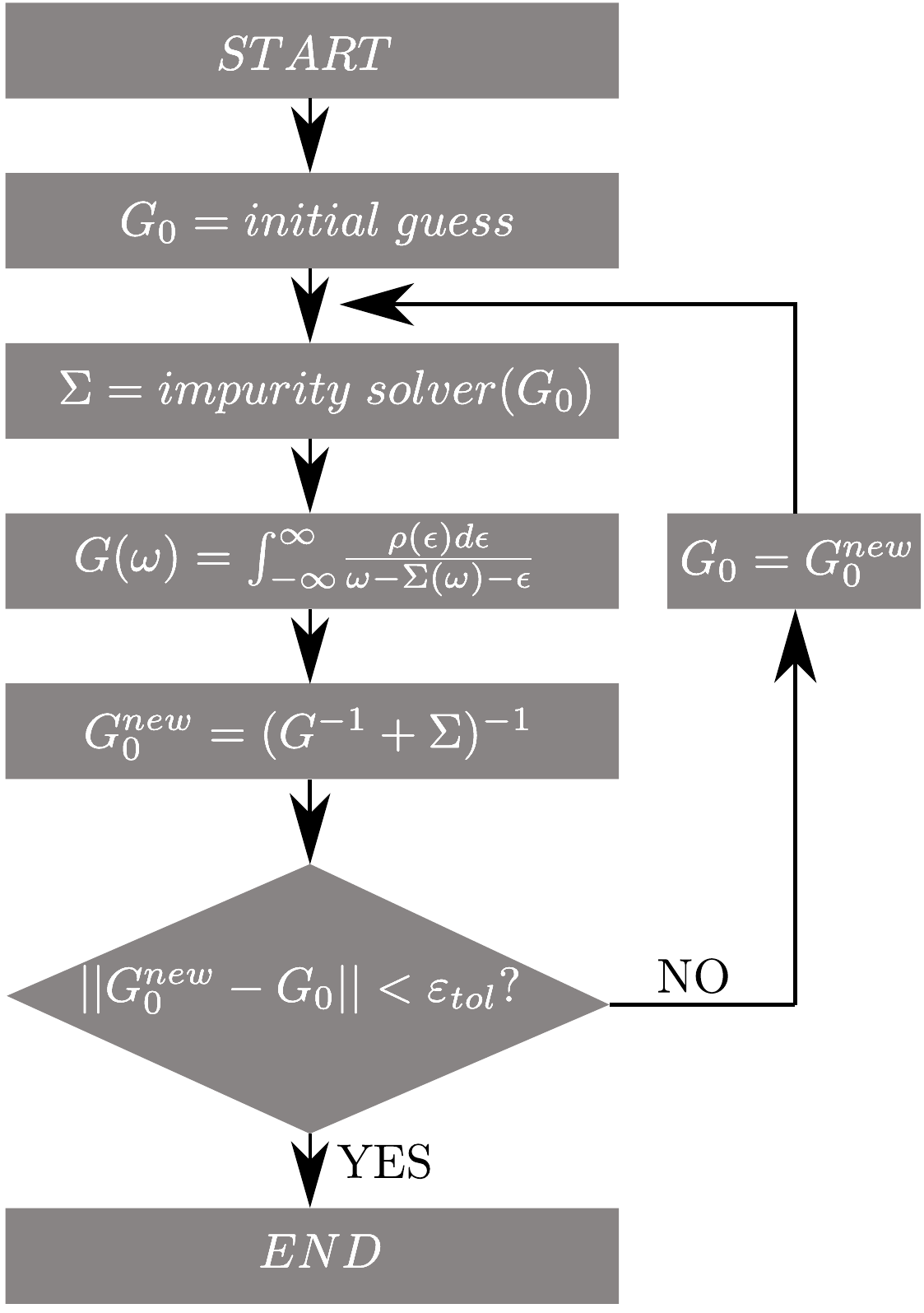}

 \caption{
DMFT self-consistency loop.
}
\label{Supp:fig:DMFT_algorithm} 
\end{figure}

In practice, the DMFT reduces to solving the (Anderson) impurity problem in a frequency dependent Weiss field $G_0(\omega)$ that needs to be determined self-consistently. The bare propagator (Weiss field) $G_0(\omega)$ is responsible for the electron fluctuations between the impurity and the reservoir (conduction bath). {On-site} correlation is taken into account through the self-energy. The connection with the lattice problem is established by the requirement that the impurity self-energy $\Sigma_{\mathrm{imp}}(\omega)$ is equal to the lattice self-energy $\Sigma_{ii}(\omega)$ (while the nonlocal components $\Sigma_{ij}(\omega)$ are equal to zero within DMFT) and that the impurity Green's function $G_{\mathrm{imp}}(\omega)$ is equal to the local lattice Green's function $G_{ii}(\omega) = \frac{1}{N} \sum_{\bf k} G_{\bf k}(\omega)$. The DMFT equations are solved iteratively as shown schematically in Fig.~\ref{Supp:fig:DMFT_algorithm}. For a given bare propagator $G_0$ an {\it impurity solver} is used to obtain the self-energy, and then the self-consistency is imposed by the Dyson equation. The subscripts for the impurity and the local lattice Green's function are omitted since these two quantities coincide when the self-consistency is reached.

\begin{figure}[t]
 \includegraphics[width=\columnwidth,trim=0cm 0cm 0cm 0cm]{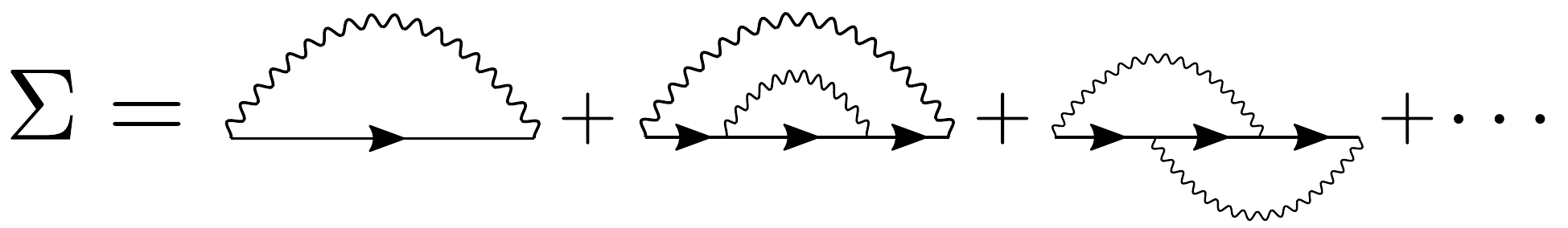}

 \caption{
First few DMFT Feynman diagrams of the self-energy in 
the expansion over $G_0$.
}
\label{Supp:fig:digrams} 
\end{figure}

The DMFT solution for the Holstein polaron follows the general concepts introduced for the Hubbard model with an important simplification which comes from the fact that we consider the dynamics of just a single electron. We briefly review some key aspects and for details we refer the reader to Ref.~\cite{SMCiuchi_1997}.

The self-energy for the polaron impurity, which is coupled to the reservoir by the bare propagator $G_0(\omega)$, can be simply expressed in a form of the continued-fraction expansion (CFE), which is in a sharp contrast with the Hubbard model where the numerical solution of the Anderson impurity model is the most difficult step.  Here, the self-energy at $T=0$ is simply given by
\begin{equation}\label{CFE}
 \Sigma(\omega) = \frac{g^2}{G_0^{-1}(\omega-\omega_0)- \cfrac{2g^2}{G_0^{-1}(\omega-2\omega_0)- \frac{3g^2}{G_0^{-1}(\omega-3\omega_0)-\dots}}}
\end{equation}
(For a derivation and generalization to $T>0$ see Ref.~\cite{SMCiuchi_1997}.)
This expansion has an infinite number of terms and in practice it needs to be truncated. In order to understand which condition needs to be fulfilled for a truncation, we will look at the diagrammatic expansion of the self-energy. 

For a single electron (i.e.~in the zero density limit) the Feynman diagrams of the self-energy consist of a single electron line accompanied by the lines that describe the emission and the absorption of phonons. There are no bubble diagrams and hence there is no renormalization of the phonon propagator. As an illustration,
a diagrammatic expansion over $G_0(\omega)$ up to the order $g^4$ is shown in Fig.~\ref{Supp:fig:digrams}. These diagrams are included if we keep the terms up to the second stage in the CFE. 

There are two important implications from this diagrammatic expansion. First, if we keep in the expansion terms up to the order $g^{2N}$ then only the phonon states $|n\rangle$ with $n\leq N$ appear as intermediate states. Therefore, since the importance of the multiphonon effects can be estimated by the parameter $\alpha^2 = g^2/\omega_0^2$ \cite{SMMahanbook}, we need to keep $N \gg \alpha^2$ terms in the CFE. Second, we see that the vertex corrections (involving the phonons on the same site in the real-space representation \cite{SMBarisic2006}) are included in the DMFT solution. This should be contrasted with the self-consistent Migdal approximation (SCMA) which completely neglects the vertex corrections in the self-energy. However, we note that one should be careful in making a direct comparison to the SCMA, since the DMFT diagrams are expanded using $G_0$, unlike the SCMA.

\vspace*{-.3cm}

\subsection{Numerical implementation of the DMFT loop}
\label{practical_dmft}
\vspace*{-.1cm}

We will now discuss step by step the self-consistency loop shown in Fig.~\ref{Supp:fig:DMFT_algorithm}. The DMFT loop starts by guessing the solution for the free propagator $G_0(\omega)$. Better guesses lead to fewer number of iterations, so depending on the parameter regime we take $G_0(\omega)$ to be either the Green's function in the Migdal approximation~\eqref{Supp:eq:Migadal_app} or the Green's function in the atomic limit~\eqref{Supp:eq:atomic}, since both of these expressions are analytically known. They correspond to the cases of very weak coupling and vanishing hopping, respectively. Next, the self-energy $\Sigma(\omega)$ is calculated using the impurity solver~(\ref{CFE})
and its generalization to finite temperatures \cite{SMCiuchi_1997}.
In practice these are implemented using the recursion relations, which at finite temperature read as:
\begin{subequations} \label{Supp:eq:im_solver}
\begin{align}
\Sigma(\omega) &= G_0^{-1}(\omega) - G^{-1}(\omega), \label{Supp:num_dyson_eq}\\
G(\omega) &= \sum_{n=0}^{\infty} \frac{(1-e^{- \omega_0/T})  e^{- n \omega_0/T}}{G_0^{-1}(\omega) - A_n^{(0)}(\omega) - B_n^{(0)}(\omega)},  \label{Supp:eq:trace}\\
A_n^{(p)} (\omega) &= \frac{(n-p)g^2}{G_0^{-1}(\omega+(p+1)\omega_0) - A_n^{(p+1)}(\omega)} \label{Supp:eq:As},\\
B_n^{(p)} (\omega) &= \frac{(n+p+1)g^2}{G_0^{-1}(\omega - (p+1)\omega_0) - B_n^{(p+1)}(\omega)} \label{Supp:eq:Bs},\\
A_n^{(n)}(\omega) & = 0,\quad B_n^{(\infty)}(\omega)=0. \label{Supp:eq:As_in_con}
\end{align} 
\end{subequations}
Quantities $A_n^{(p)}$ and $B_n^{(p)}$ are determined recursively, starting from~\eqref{Supp:eq:As_in_con} and going back to~\eqref{Supp:eq:Bs} and~\eqref{Supp:eq:As}. Then, $G(\omega)$ is calculated using~\eqref{Supp:eq:trace}, which enables us to use Dyson Eq.~\eqref{Supp:num_dyson_eq} to obtain $\Sigma(\omega)$.
For $T=0$ the equations simplify and the self-energy can be written as $\Sigma(\omega) = B_0^{(0)}(\omega)$, which coincides with Eq.~\eqref{CFE}. The physical interpretation of the quantities in Eq.~\eqref{Supp:eq:im_solver} is the following: $G(\omega)$ is the interacting Green's function of the impurity. The quantity $A_n^{(0)}(\omega)$  is just a finite fraction that takes into account the emission of phonons. Similarly, $B_n^{(0)}(\omega)$ is an infinite continued fraction, which takes into account the absorption of phonons. The infinite fraction $B_n^{(0)}(\omega)$ can be calculated accurately even if we truncate it $B_n^{(N)}(\omega)=0$, taking $N$ to be a number much larger  than $\alpha^2$. The infinite series~\eqref{Supp:eq:trace} can also be truncated by using the number of terms $n_{max} \gg T/\omega_0$ \cite{SMCiuchi_1997}. 

Next step in the DMFT loop is calculating the local Green's function of the lattice using the self-energy $\Sigma(\omega)$ from the impurity solver. It is calculated as
\begin{equation} \label{Supp:eq:Local_Green}
G(\omega) = \int_{-\infty}^\infty \frac{\rho(\epsilon)d\epsilon}{\omega - \Sigma(\omega)-\epsilon},
\end{equation}
where $\rho(\epsilon)$ is the noninteracting density of states. This integral is convergent since we are integrating below the complex pole $\epsilon=\omega - \Sigma(\omega)$, as a consequence of the causality $\im \Sigma(\omega)<0$. However, numerical instabilities can arise due to the fact that the complex pole can be arbitrarily close to the real axis. Hence, the numerical integration of Eq.~\eqref{Supp:eq:Local_Green} requires additional care. In Sec.~\ref{DMFT_2_3D} we present a numerical procedure which solves this problem. However, in the 1d case these  numerical instabilities are completely avoided since Eq.~\eqref{Supp:eq:Local_Green} admits an analytical solution, as shown in Sec.~\ref{calc_LGF_1d}.

Following the DMFT algorithm from Fig.~\ref{Supp:fig:DMFT_algorithm}, we now calculate the next iteration of the free propagator using the Dyson equation
\begin{equation}
G_0^{\mathrm{new}}(\omega)= [G^{-1}(\omega)+\Sigma(\omega)]^{-1}.
\end{equation}
We check if $|G_0^{new}(\omega) - G_0(\omega)| < \varepsilon_{\mathrm{tol}}$ (for each $\omega$), where $\varepsilon_{\mathrm{tol}}$ is the tolerance parameter that we typically set to $\varepsilon_{tol}\sim 10^{-4}$ or smaller. If this condition is satisfied, the DMFT loop terminates and $\Sigma$, $G_0$ and $G$ are found. Otherwise,  $G_0^{\mathrm{new}}$ is used in the impurity solver and the procedure is repeated until convergence is reached. 

After the DMFT loop has been completed, we can use
the calculated self-energy $\Sigma(\omega)$ to find the retarded
Green’s function of our original problem
\begin{equation}
G_{\bf k}(\omega) = \frac{1}{\omega - \Sigma(\omega) - \varepsilon_{\bf k}}.
\end{equation}
The spectral function is then simply given by
\begin{equation}
A_{\bf k}(\omega) = -\frac{1}{\pi} \mathrm{Im}G_{\bf k}(\omega).
\end{equation}

\vspace*{-0.7cm}
\subsubsection{Self-consistency equation for the local Green's function in one dimension}
\label{calc_LGF_1d}
\vspace*{-0.3cm}
Let us now show how the local Green's function~\eqref{Supp:eq:Local_Green} can be analytically evaluated in a 1d system with nearest neighbor hopping $t_0$. The noninteracting density of states reads as
\begin{equation}
\rho(\epsilon) = \frac{\theta(4t_0^2-\epsilon^2)}{\pi\sqrt{4t_0^2-\epsilon^2}},
\end{equation}
where $\theta$ is the Heaviside step function. Equation~\eqref{Supp:eq:Local_Green} can be rewritten using the substitution $\epsilon = 2t_0 \sin x$
\begin{equation}
G(\omega) = \frac{1}{4t_0  \pi} \int_{-\pi}^{\pi} \frac{dx}{B - \sin x},
\end{equation}
where we introduced 
\begin{equation}
B = (\omega - \Sigma(\omega)) / 2t_0.
\end{equation}
Additional substitution $z = e^{ix}$ leads us to
\begin{equation} \label{Supp:eq:Contour_integral}
G(\omega) = -\frac{1}{2t_0  \pi} \oint\limits_{C} \frac{dz}{(z- z_+)(z - z_-)},
\end{equation}
where this represents the counterclockwise complex integral over the unit circle $C$ and $z_{\pm} = iB \pm \sqrt{1-B^2}$. In order to apply the method of residues, we first need to find out if $z_{\pm}$ are inside the complex unit circle $|z| = 1$. Causality implies that $\im{\Sigma(\omega)} < 0$ which means that $\im{B} > 0 $. In this case one can show that $|z_+| < 1$ and $|z_-| > 1$, which means that only the pole at $z_+$ gives a non-vanishing contribution to the Eq.~\eqref{Supp:eq:Contour_integral}
\begin{equation} \label{Supp:eq:Local_Green_Solution}
G(\omega) = \frac{-i}{2t_0 \sqrt{1-B^2}} = \frac{1}{2t_0  B \sqrt{1 - \frac{1}{B^2}}}.
\end{equation}
In Eq.~\eqref{Supp:eq:Local_Green_Solution} we wrote the solution in two ways. They are completely equivalent in our case when $\im{B} > 0$, but can otherwise give different results. Since $B$ can be arbitrarily close to the real axis, it is important to ensure additional numerical stability by requiring that the expression for $G(\omega)$  satisfies that the $\im{B} = 0$ solution coincides with the solution in the limit $\im{B} \to 0$. This is not satisfied by the expressions in Eq.~\eqref{Supp:eq:Local_Green_Solution}, but it can be achieved by combining their imaginary and real parts 
\begin{equation} \label{Supp:eq:Local_Green_Solution_2}
G(\omega) = \re{\frac{1}{2t_0 a B \sqrt{1 - \frac{1}{B^2}}}} + i\; \im{\frac{-i}{2t_0a \sqrt{1-B^2}}}.
\end{equation}

\subsubsection{Self-consistency equation for the local Green's function in arbitrary number of dimensions}
\label{DMFT_2_3D}
Here we present a numerical procedure for the calculation of the local Green's function \eqref{Supp:eq:Local_Green} for arbitrary density of states $\rho(\epsilon)$, that completely eliminates the potential numerical singularity at $\epsilon = \omega - \Sigma(\omega)$. This is particularly important since the techniques presented in Sec.~\ref{calc_LGF_1d} fail when the dispersion relation even slightly changes. It is also relevant in the higher-dimensional systems where the density of states is not necessarily analytically know.

Let us suppose that the self-energy and the density of states are known only on a finite, equidistant grid $\omega_0, \omega_1 ... \omega_{N-1}$, where $\Delta \omega = \omega_{i+1}-\omega_i$. Further, suppose that the density of states is vanishing outside some closed interval $[D_1, D_2]$ and that the grid is wide enough so that there are at least a couple of points outside that closed interval: $\rho(\omega_0)=...=\rho(\omega_3)=0$ and $\rho(\omega_{N-1})=...=\rho(\omega_{N-4})=0$. These are quite general assumptions that are always satisfied in the systems we are examining. The local Green's function can now be rewritten as
\begin{equation} \label{Supp:Eq:ref:LGF_1}
G(\omega) = \sum_{i=0}^{N-2} \int_{\omega_i}^{\omega_{i+1}} d\epsilon
\frac{\rho(\epsilon)}{\omega-\Sigma(\omega)-\epsilon}.
\end{equation}

At each sub-interval $[\omega_i, \omega_{i+1}]$ the density of states is only known at the endpoints, so it is natural to approximate it using a linear function
\begin{equation}
\rho(\epsilon) = a_i + b_i (\epsilon - \omega_i),
\end{equation}
where $a_i = \rho(\omega_i)$, $b_i = (\rho(\omega_{i+1})-\rho(\omega_i))/\Delta \omega$.
Introducing a shorthand notation $\xi = \omega - \Sigma(\omega)$,  we evaluate Eq.~\eqref{Supp:Eq:ref:LGF_1} analytically
\begin{align}
G(\omega) &= \sum_{i=0}^{N-2} b_i (\omega_i - \omega_{i+1}) \nonumber \\
          &+ \sum_{i=0}^{N-2}  a_i  
          	\left[ \ln(\xi-\omega_i) - \ln(\xi-\omega_{i+1}) \right] \nonumber\\
          &+ \sum_{i=0}^{N-2}b_i (\xi-\omega_i)
          \left[ \ln(\xi-\omega_i) - \ln(\xi-\omega_{i+1}) \right]. \label{Supp:eq:LGF_2}
\end{align}
The first line is just a telescoping series that is vanishing 
\begin{equation}
\sum_{i=0}^{N-2} b_i (\omega_i - \omega_{i+1})  = \rho(\omega_0)-\rho(\omega_{N-1})=0.
\end{equation}
The last two lines in Eq.~\eqref{Supp:eq:LGF_2} can be transformed by shifting the indices $i+1 \to i$, taking into account that a few boundary terms are vanishing and using the identity $a_i - a_{i-1} = (\omega_i - \omega_{i-1})b_{i-1}$
\begin{align} \label{Supp:eq:LGF_final}
G(\omega) =& \sum_{i=0}^{N-2} \frac{\rho(\omega_{i+1})-2\rho(\omega_{i})+\rho(\omega_{i-1})}{\Delta \omega}  \nonumber \\
& \times (\omega - \omega_i - \Sigma(\omega)) \ln\left(\omega - \omega_i - \Sigma(\omega)\right) .
\end{align}
This expression now has no numerical instabilities. This is most easily seen from the fact that it has the form $x\ln x$ which is well defined even in the limit $x\to 0$, where it vanishes. Of course, the results were obtained by using the linear interpolation of the density of states. This is completely justified if $\rho(\epsilon)$ is smooth or has finitely many cusps. However, the presence of van Hove singularities in $\rho(\epsilon)$ may require some special analytical treatment around them.

\subsection{Effective mass in 1d, 2d and 3d}
\label{Supp:Effective_mass}

\begin{figure}[!t]
\includegraphics[width=\columnwidth,trim=0cm 0cm 0cm 0cm]{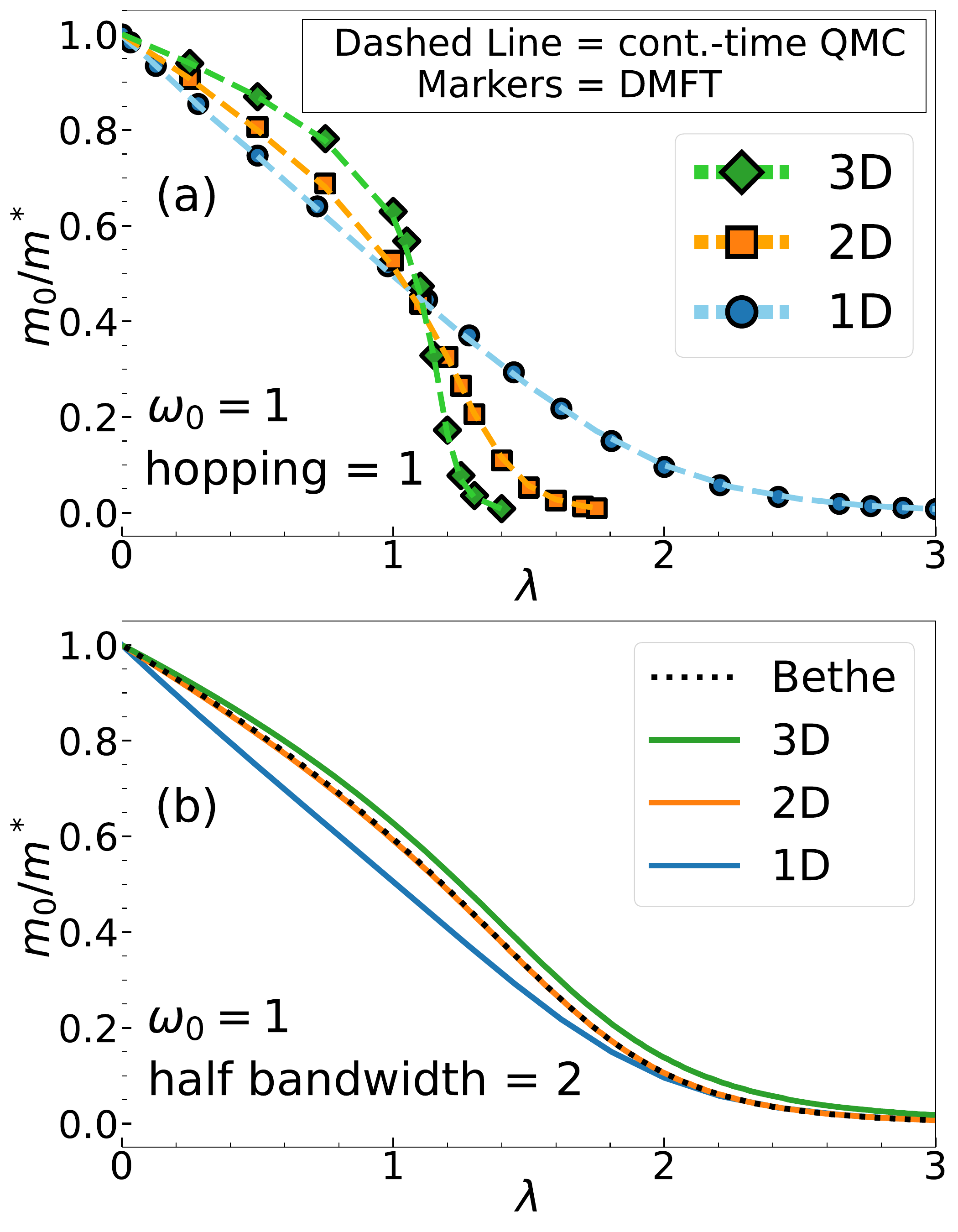}
 \caption{ (a)
Continuous-time QMC (taken from Ref.~\onlinecite{SMKornilovitch_1998}) vs. DMFT mass renormalization in 1d, 2d and 3d, with $\omega_0=1$. (b) Comparison of the DMFT mass renormalization on different lattices.
}
\label{Supp:fig:mass_higher_D} 
\end{figure}

The DMFT mass renormalization is calculated in one, two and three dimensions. These are then compared to the continuous-time path-integral quantum Monte Carlo (QMC) results from Ref. \onlinecite{SMKornilovitch_1998}. In that paper it was noted that the numerical accuracy of the QMC method is $0.1\%-0.3\%$. The results are presented in Fig. \ref{Supp:fig:mass_higher_D}(a).

We note that the definition of $\lambda$ and $\gamma$ is slightly different than the one we gave in the main text. Here
\begin{equation}
\lambda = \frac{g^2}{\omega_0 W/2};\quad \gamma = \frac{\omega_0}{W/2},
\end{equation}
where $W/2$ is the half bandwidth. This coincides with our previous definition in 1d, but gives an extra normalization in higher dimensions.

\subsection{Comparisons with the Bethe lattice results}
\label{Supp:bethe_lattice_comp}

\begin{figure}[!t]
\includegraphics[width=\columnwidth,trim=0cm 0cm 0cm 0cm]{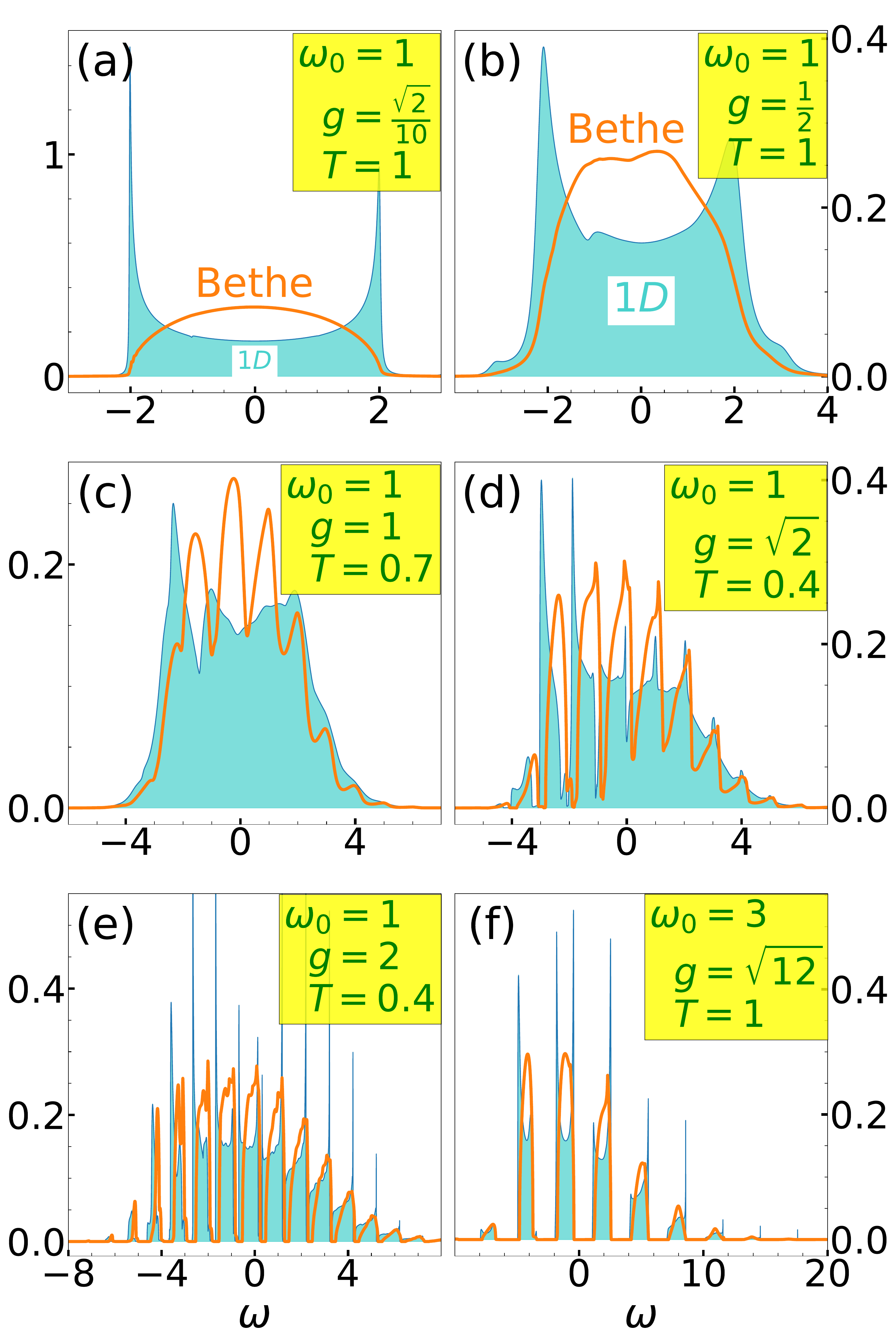}
 \caption{1d vs Bethe DMFT local spectral functions.
  \vspace*{-0.18cm} 
}
\label{Supp:fig:dmft_Bethe_vs_1D} 
\end{figure}
\begin{figure}[!t]
\includegraphics[width=\columnwidth,trim=0cm 0cm 0cm 0cm]{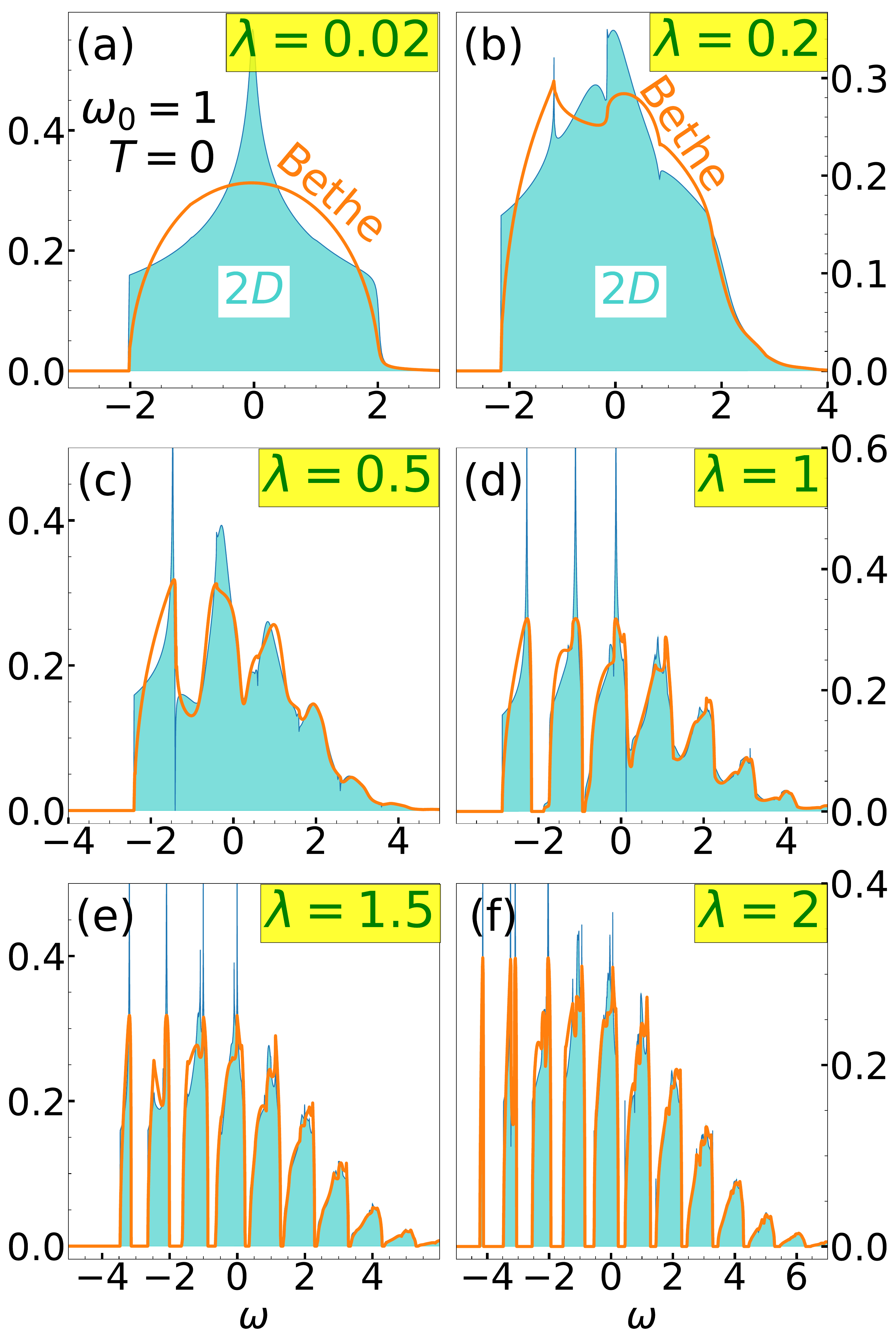}
 \caption{2d vs Bethe DMFT local spectral functions. 
 \vspace*{-0.18cm}
}
\label{Supp:fig:dmft_Bethe_vs_2D} 
\end{figure}

In the main text we emphasized that the misconception about the validity of the DMFT in 1d appeared since only the DMFT results on the Bethe lattice were used in comparisons with other methods \cite{SMRomero_1998,SMKu_Bonca_2002}. In this section we illustrate why such comparison is inappropriate.

The main difference in practical implementation, compared to 1d, can be ascribed to the self-consistency condition for the Bethe lattice (corresponding to the semi-elliptic density of states) which can be formulated using a simple algebraic equation \cite{SMCiuchi_1997}
%
%
\begin{equation} \label{Supp:eq:bethe_sc}
G_0(\omega) = \left(\omega - \frac{(W/2)^2}{4} G(\omega) \right)^{-1}.
\end{equation}
%
In Fig.~\ref{Supp:fig:mass_higher_D}(b) we compare the DMFT mass renormalization on different lattices using the same half-bandwidth. There is a clear discrepancy between the 1d and the Bethe lattice results, in accordance with the already mentioned earlier works.

The Bethe lattice lacks a dispersion relation since it has no translational symmetry. Therefore in Fig.~\ref{Supp:fig:dmft_Bethe_vs_1D} we compare only the local spectral functions $A(\omega)=-\frac{1}{\pi}\mathrm{Im}G(\omega) = -\frac{1}{\pi}\mathrm{Im} \frac{1}{N} \sum_k G_k(\omega) $ of the Bethe and 1d lattice. For small couplings, the spectral functions resemble the noninteracting density of state and we find a large discrepancy, as shown in panels (a) and (b). In contrast, close to the atomic limit in Fig.~\ref{Supp:fig:dmft_Bethe_vs_1D}(f) spectral functions become more alike. We note that the regimes at panels (c)-(f) are the same as in Fig.~\ref{fig:finiteT} from the main text.



It is rather surprising that there is a striking agreement between the effective mass for 2d and the Bethe lattice as shown in Fig.~\ref{Supp:fig:mass_higher_D}(b), even though the noninteracting density of states are different, Fig.~\ref{Supp:fig:dmft_Bethe_vs_2D}(a). Interestingly, we can see from Fig.~\ref{Supp:fig:dmft_Bethe_vs_2D} that the local spectral functions become very similar already for moderate interactions.

\newpage
\clearpage

\section{Weak-coupling limit}
\label{Migdal_App}

In this section we introduce the self-consistent Migdal approximation (SCMA) and use it as a benchmark for  the DMFT in the weak-coupling limit, where  SCMA is exact.  More importantly, we can examine a deviation of SCMA from DMFT for stronger couplings, which is shown in the main text and in the following sections of the  SM.

\subsection{Migdal approximation}
\label{Migdal_App_Reg}

\begin{figure}[!h]
 \includegraphics[width=3.2in,trim=0cm 0cm 0cm 0cm]{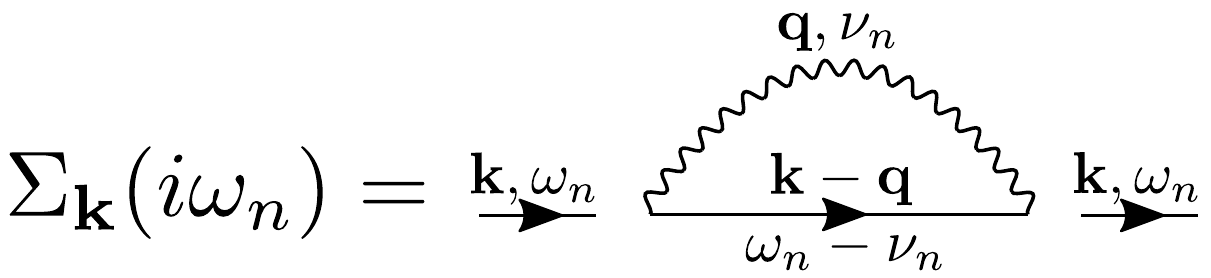}
 \caption{
Feynman diagrams of the self-energy in the Migdal approximation
}
\label{Supp:fig:Migdal_approx} 
\end{figure}

The Migdal approximation \cite{SMMigdal_1958},  as shown in Fig.~\ref{Supp:fig:Migdal_approx}, is defined by taking into account only the lowest order Feynman diagram in the perturbation expansion of the self-energy. 

Due to its simplicity it can be evaluated analytically
\begin{equation}\label{Supp:eq:Migadal_app}
\Sigma_{k}(\omega) = g^2 (b + 1) S(\omega-\omega_0) + g^2 b \, S(\omega+\omega_0),
\end{equation}
where $b\equiv b(\omega_0)=(e^{\omega_0/T}-1)^{-1}$ and 
\begin{equation}
S(\omega) = (\omega^2 - 4t_0^2)^{-1/2} \quad \text{for}\;\omega>0,\nonumber 
\end{equation}
while the solution for $\omega<0$ can be obtained by noting that $\mathrm{Im}S(\omega)$ and $\mathrm{Re}S(\omega)$ are symmetric and antisymmetric functions, respectively.
However, this solution is accurate only for very small coupling $g$. For larger coupling a much better solution is obtained within the self-consistent Migdal approximation.

\vspace*{-0.1cm}

\subsection{Self-consistent Migdal approximation}
\label{Self_Cons_Migdal_App}
\vspace*{-0.1cm}
\begin{figure}[!h]
 \includegraphics[width=3.2in,trim=0cm 0cm 0cm 0cm]{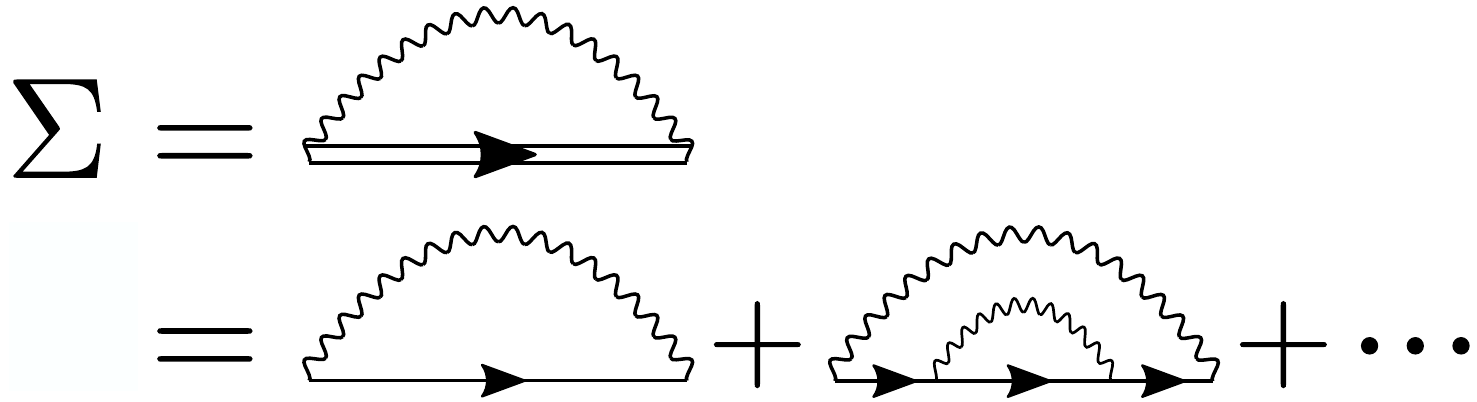}
 \caption{
Feynman diagrams in the SCMA approximation.
}
\label{Supp:fig:SCMA_approx} 
\end{figure}

\begin{figure}[!h]
\centering$
  \begin{array}{c}
   \includegraphics[width=3.05in,trim=0cm 0cm 0cm 0cm]{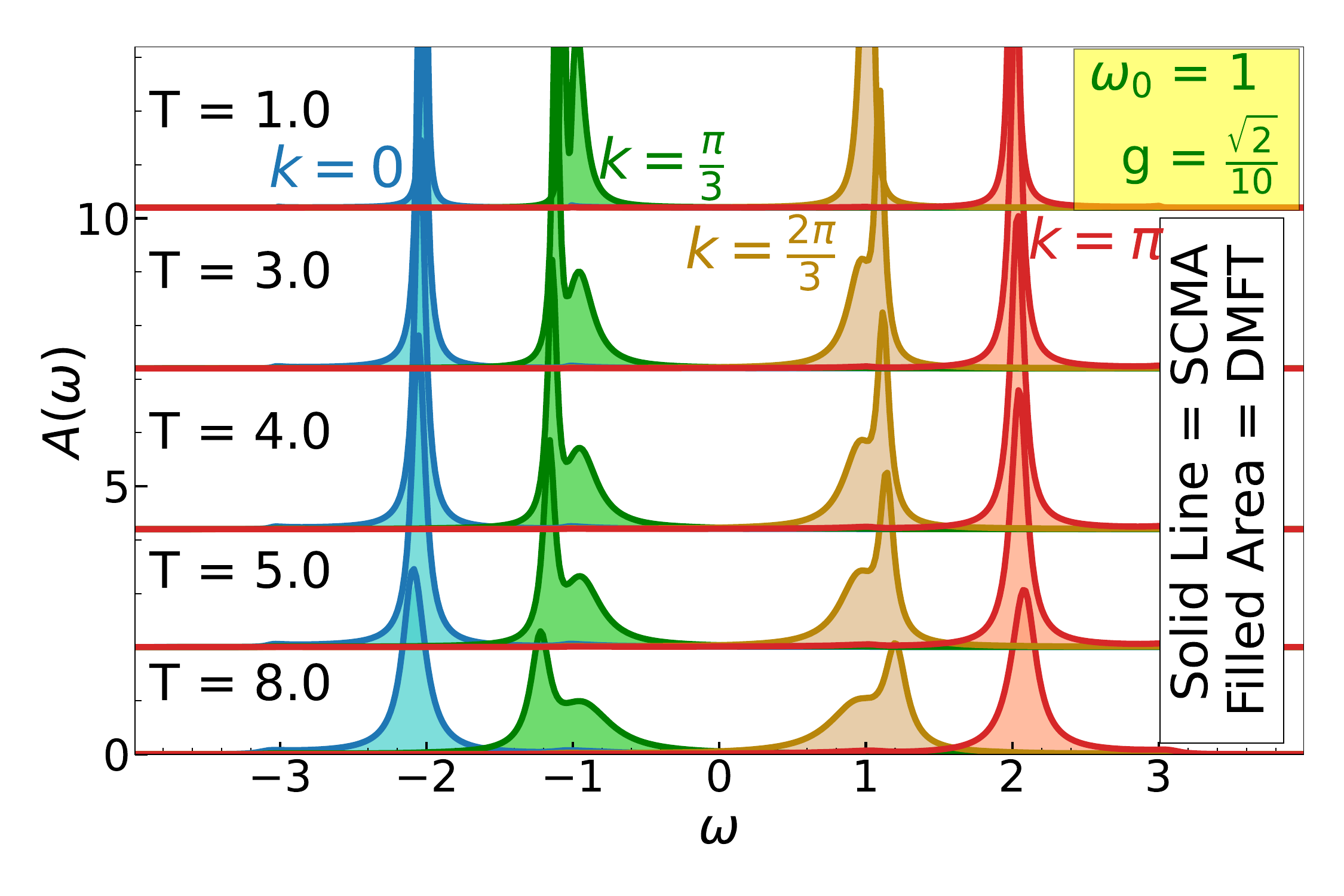} \\
   \includegraphics[width=3.05in,trim=0cm 0cm 0cm 0cm]{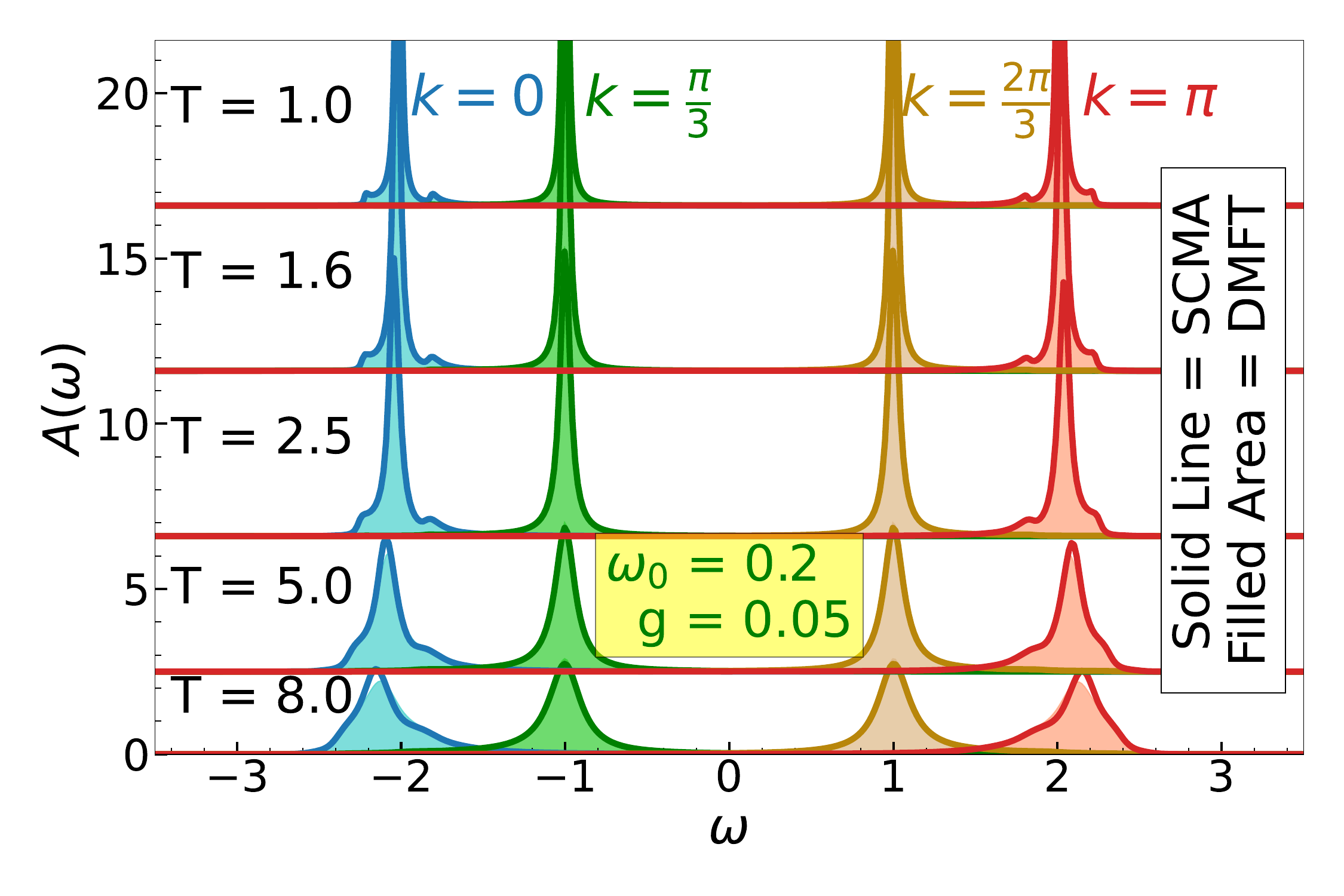} \\
   \includegraphics[width=3.05in,trim=0cm 0cm 0cm 0cm]{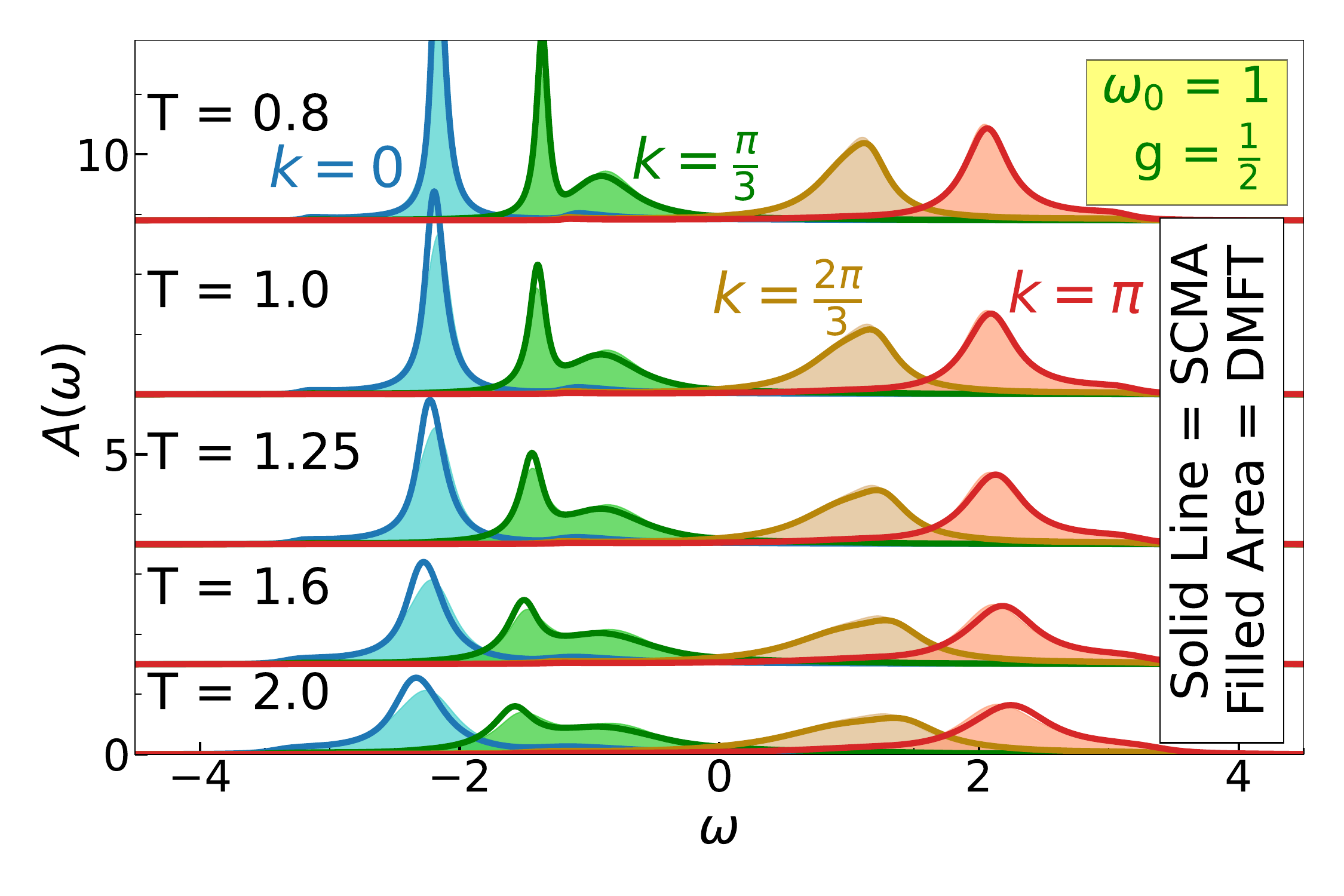} \\
   \includegraphics[width=3.05in,trim=0cm 0cm 0cm 0cm]{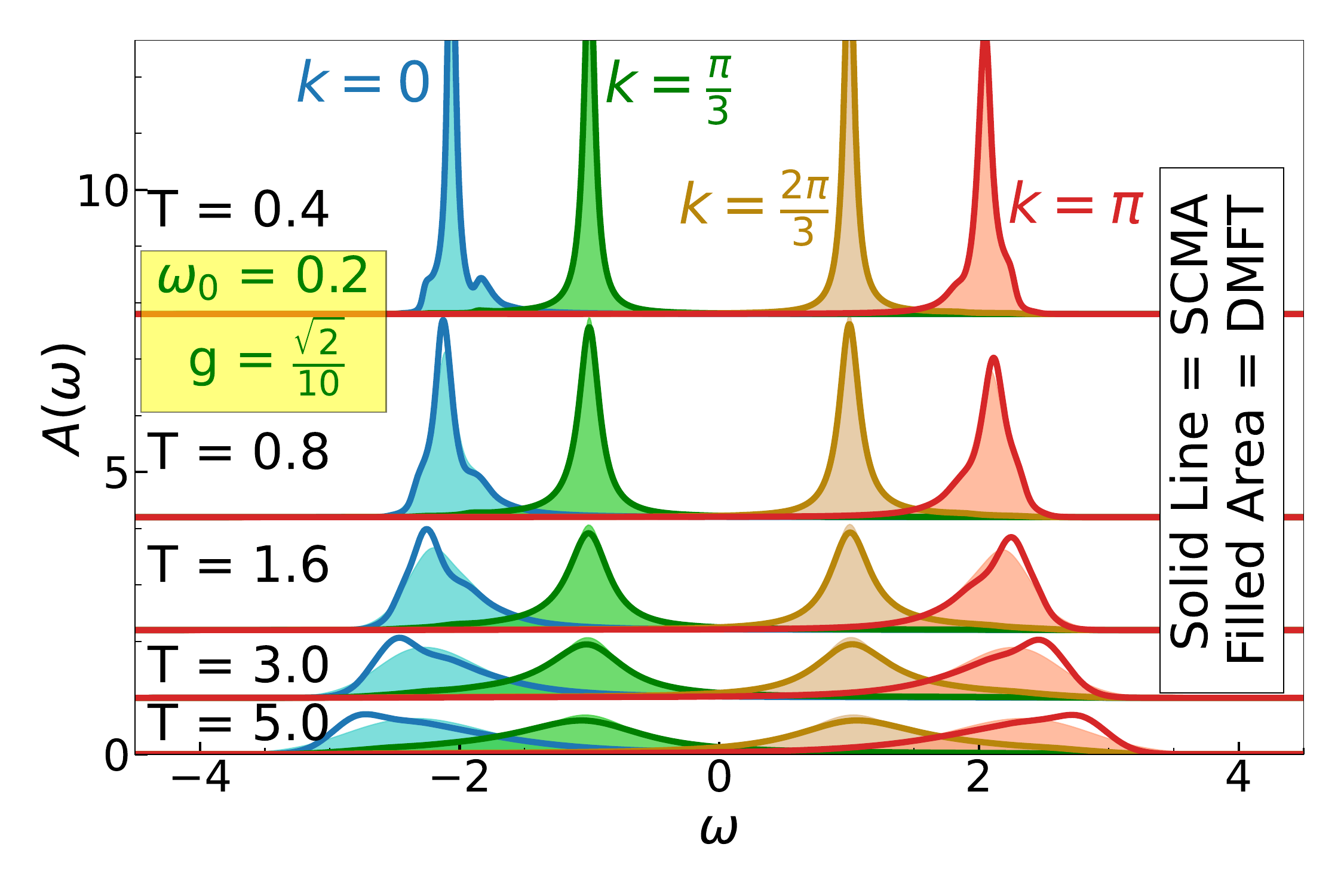}
  \end{array}$
 \caption{DMFT vs. SCMA spectral functions in the weak-coupling regime.}
\label{Supp:fig:SCMA}
 \end{figure}
In the SCMA, free fermionic propagator from Fig.~\ref{Supp:fig:Migdal_approx} is replaced with the interacting propagator, as shown in Fig.~\ref{Supp:fig:SCMA_approx}. The corresponding equation for the self-energy can be written as
\begin{equation} \label{Supp:eq:SCMA_Self-Consistent_Condition}
\Sigma_{ k}(\omega) = g^2  (b+1) G(\omega - \omega_0) 
 + g^2 b \,G(\omega + \omega_0),
\end{equation}  

where $G(\omega) = \frac{1}{N} \sum_k G_{ k}(\omega)$ is the local Green's function. Equation~\eqref{Supp:eq:SCMA_Self-Consistent_Condition} needs to be solved self-consistently, since the Green's function can be expressed in terms of the self-energy (via the Dyson equation). 

Using the expansion with respect to the free propagator, the formal solution for the self-energy can be written as an infinite series of non-crossing diagrams, as shown in Fig.~\ref{Supp:fig:SCMA_approx}. We see that  the first term represents the Feynman diagram in the Migdal approximation. It is thus not at all surprising that the SCMA range of validity is much larger than the one-shot Migdal approximation. 

We note that  the SCMA self-energy is momentum-independent, which follows from Eq.~\eqref{Supp:eq:SCMA_Self-Consistent_Condition},
making this method numerically cheap. 

\subsection{DMFT vs.~SCMA in the weak coupling limit}
\label{SCMA_vs_DMFT}

A comparison of the DMFT and SCMA spectral functions in the weak coupling limit  is shown in Fig.~\ref{Supp:fig:SCMA}. Results almost fully coincide. As the electron-phonon coupling increases, the SCMA spectral functions starts to deviate from the exact solution, as we see from the main text and from the remaining part of the Supplemental Material.

%

\newpage
\clearpage

\section{Strong coupling: Exact diagonalization}
\label{ED_method}

In the strong coupling regime we can approach the solution in the
thermodynamic limit by using a small number of lattice sites. In SM
Sec.~\ref{Sec:finitesize} we show that for $g=2$, $\omega_0=1$ we are
close to thermodynamic limit by considering a chain of just $N=4$ sites.
In this case we can reach a solution using the exact diagonalization
(ED). In the following we describe our implementation of the ED method.

We calculate the spectral function by diagonalizing the Holstein
Hamiltonian in the space spanned by the vectors
$U c_i^\dagger \ket{n_1n_2\ldots n_N}$, where $n_i$ is the number of
phonons at site $i\in\qty{1,\ldots,N}$, satisfying $\sum_i
n_i<n_{\mathrm{max}}$, while $U$ is the unitary operator of the
Lang-Firsov transformation \cite{SMLangFirsov} given as
\begin{equation}
U=e^{\frac{g}{\omega_0}\sum_i c_i^\dagger c_i\qty(a_i-a_i^\dagger)}.
\end{equation}
Both $N$ and $n_{\mathrm{max}}$ need to be increased until convergence is reached. The spectral function is then calculated as
\begin{equation} \label{eq:supp:exact_diag}
A_{\vb{k}}\qty(\omega)=\frac{1}{Z_p}\sum_p e^{-\beta E_p}
\sum_e \delta\qty(\omega+E_p-E_e)
\qty|\mel{p}{c_{\vb{k}}}{e}|^2,
\end{equation}
where $\ket{p}$ denotes purely phononic states, the energy of which is
$E_p$, $\ket{e}$ denotes the states with one electron and arbitrary
number of phonons, the energy of which is $E_e$ and $Z_p=\sum_p e^{-\beta
E_p}$ is the phononic partition function.

We found that convergent results for the spectral function when $g=2$,
$\omega_0=1$, $N=4$ could be obtained for $n_{\mathrm{max}}=16$.
The results are shown in Figs.~\ref{Supp:fig:Fig2_arg_1}-~\ref{Supp:fig:SpecF1_more_momenta}, as well as in Figs.~\ref{fig:zeroT}(b) and~\ref{fig:finiteT}(e)-(f) of the main text.
The spectral functions at $\vb{k}$ points different than
$k=\frac{2\pi}{N}i$, $i\in\qty{0,\ldots,N-1}$ were obtained by employing
so-called twisted boundary conditions, that is by changing the terms in
the Hamiltonian $t_0c_i^\dagger c_{i+1} \to
t_0e^{\mathrm{i}\phi}c_i^\dagger c_{i+1}$ and $t_0c_{i+1}^\dagger c_{i}
\to t_0e^{-\mathrm{i}\phi}c_{i+1}^\dagger c_{i}$. The spectral function
obtained from such a modified Hamiltonian corresponds then to the
spectral function at $k+\phi$.

\newpage
\clearpage

\section{Finite-size effects and HEOM depth}
\label{Sec:finitesize}

\begin{figure}[b]
 \includegraphics[width=3.4in,trim=0cm 0cm 0cm 0cm]{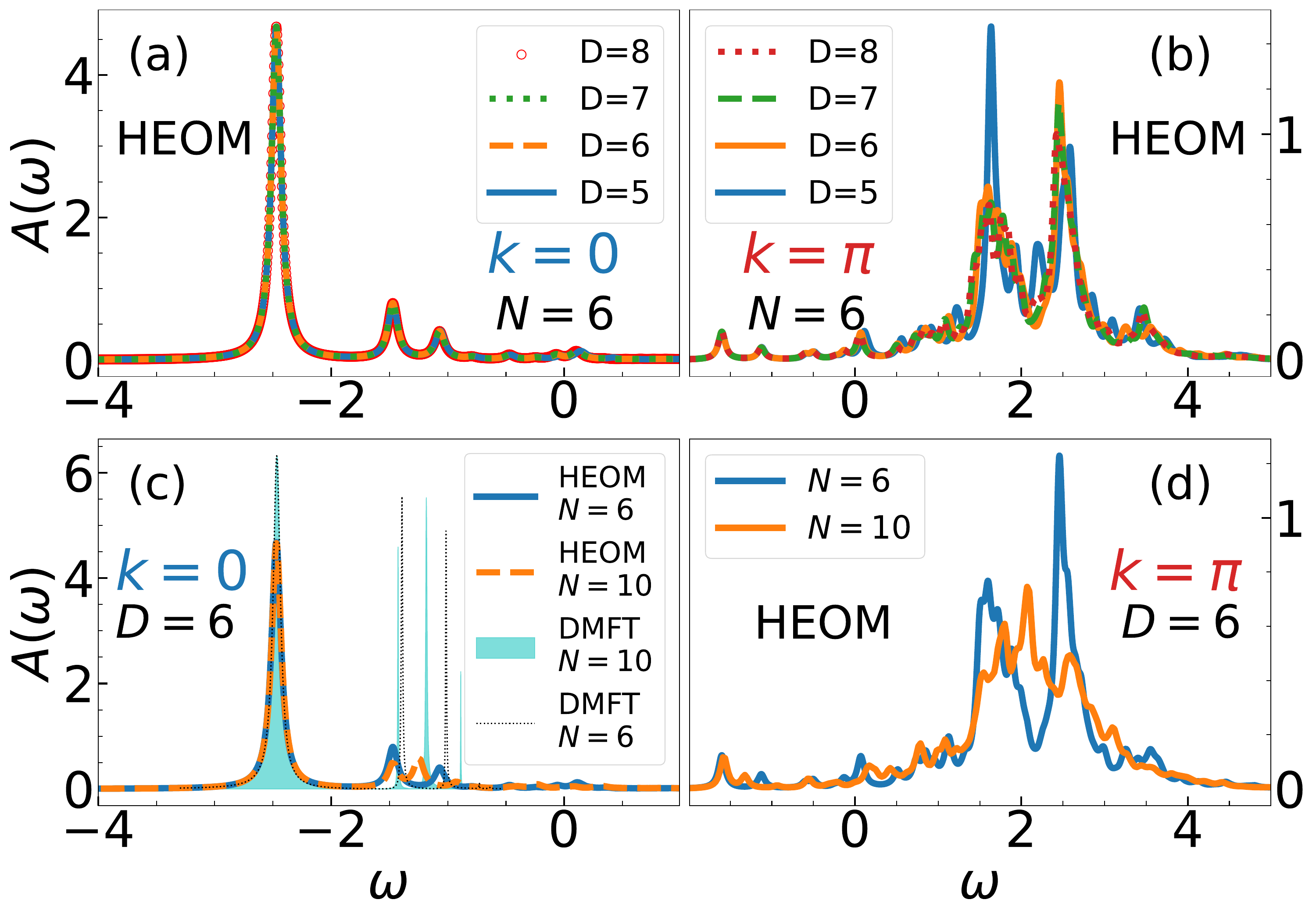}
 \caption{ Finite-$N$ and finite-$D$ effects in the HEOM method at intermediate coupling
$\omega_0=1$, $g=1$, $T=0$, which is the same regime as in Fig.~\ref{fig:zeroT}(a) of the main text. Here we use Lorentzian broadening  with $\eta=0.05$.
}
\label{Supp:fig:FiniteSize_w=1_g=1} 
\end{figure}
\begin{figure}[t]
 \includegraphics[width=3.0in,trim=0cm 0cm 0cm 0cm]{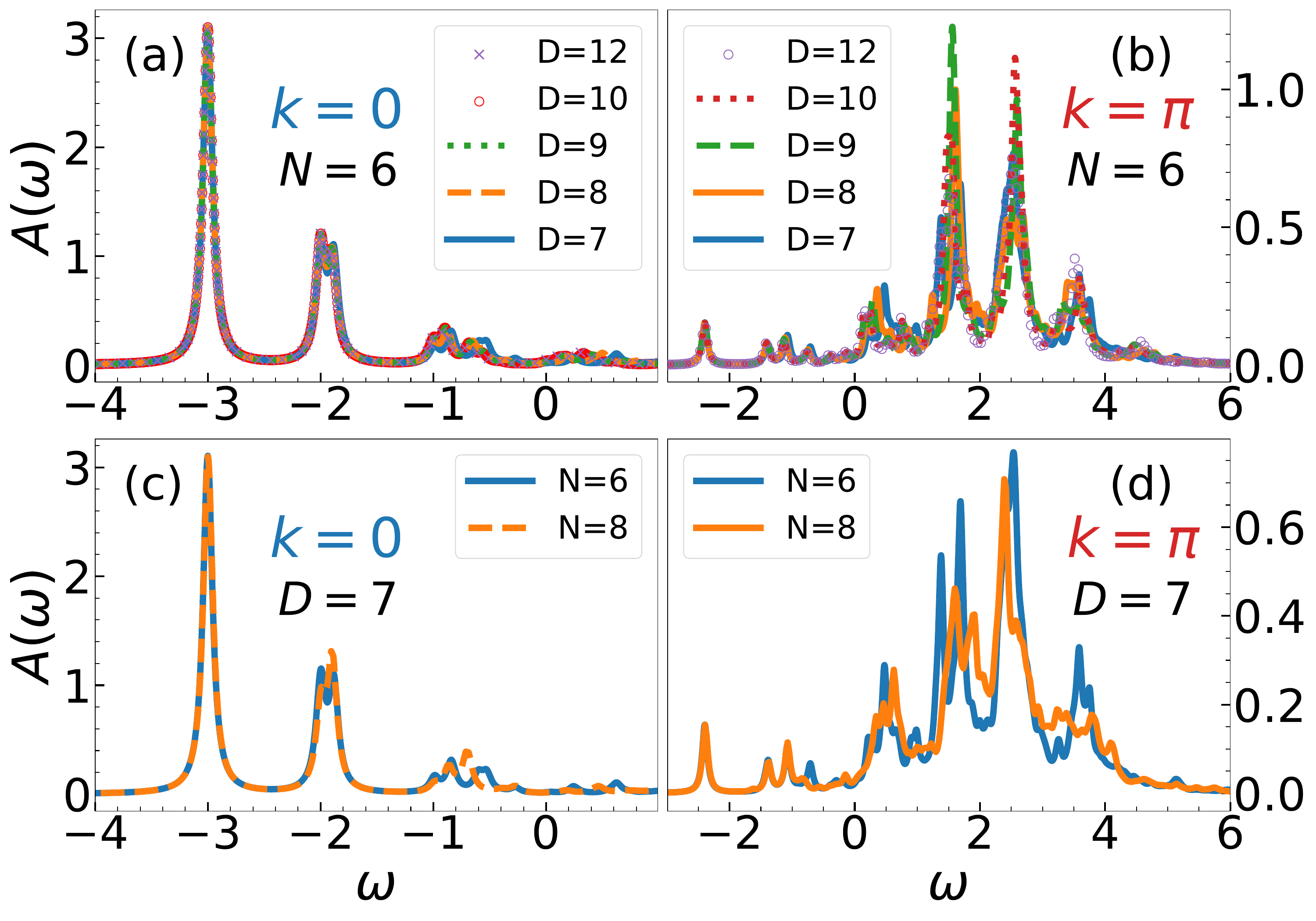}
 \caption{ Finite-$N$ and finite-$D$ effects in the HEOM at intermediate coupling 
$\omega_0=1$, $g=\sqrt{2}$, $T=0$, which is the same regime as in Fig.~\ref{fig:zeroT}(c) of the main text. Here we use Lorentzian broadening  with $\eta=0.05$.
}
\label{Supp:fig:FiniteSize_w=1_g=1.41} 
\end{figure}
\begin{figure}[h!]
 \includegraphics[width=3.2in,trim=0cm 0cm 0cm 0cm]{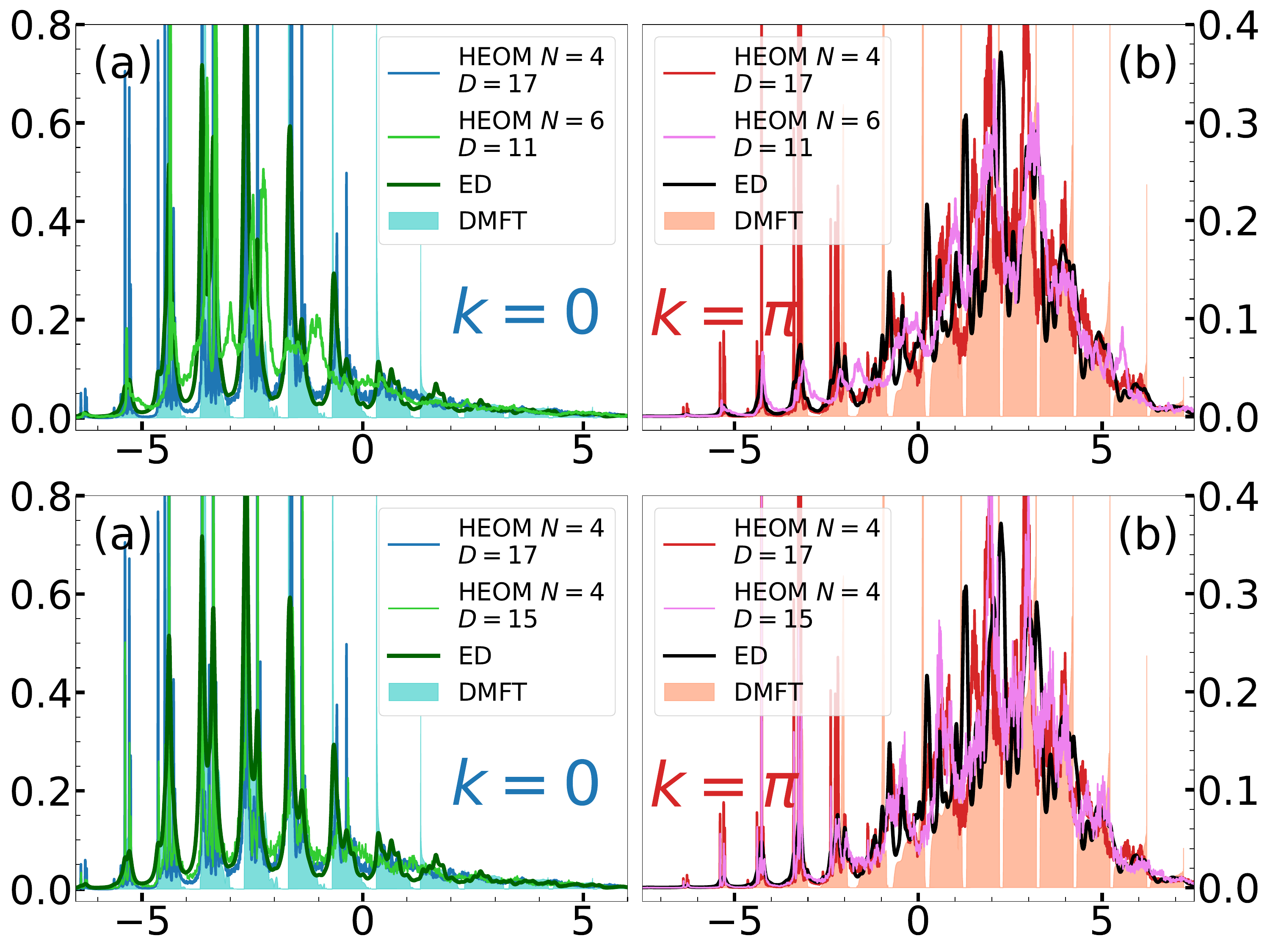}
 \caption{ Finite-$N$ and finite-$D$ effects in the strong coupling regime 
$\omega_0=1$, $g=2$, $T=0.4$, which is the same regime as in Figs.~\ref{fig:finiteT}(e)-(f) of the main text. ED spectral functions ($N=4$) are shown using Lorentzian broadening with $\eta = 0.05$, while other methods are shown without broadening. DMFT results are in thermodynamic limit.
}
\label{Supp:fig:FiniteSize_w=1_g=2} 
\end{figure}

The numerically exact HEOM, QMC and ED methods are implemented on a 1d lattice of length $N$. Results which are representative of the thermodynamic limit can be obtained by taking large enough $N$. Furthermore, the hierarchy of HEOM needs to be truncated using sufficient depth $D$. In the ED method the number of phonons in the Hilbert space need to be specified. 
All of these parameters should be as large as possible, but the practical numerical implementation is restricted by the available computer memory. 
Finite-$N$ and finite-$D$ analysis was performed in all parameter regimes where we have HEOM results. In Figs.~\ref{Supp:fig:FiniteSize_w=1_g=1},~\ref{Supp:fig:FiniteSize_w=1_g=1.41}, and~\ref{Supp:fig:FiniteSize_w=1_g=2} we briefly illustrate such analysis in the intermediate and strong coupling regime. 
%


The optimal value of $D$ strongly depends on the interaction strength and temperature. For large interaction we need large $D$ since many phonon states are populated even at $T=0$. Similarly, larger temperature also requires larger HEOM depth. As illustrated in Fig.~\ref{Supp:fig:FiniteSize_w=1_g=1}(a)-(b), for $\omega_0=1$, $g=1$ the convergence is nearly reached already for $D=6$. For  $g=\sqrt{2}$ (Fig.~\ref{Supp:fig:FiniteSize_w=1_g=1.41}(a)-(b)), we need slightly larger $D$. However, in the strong-coupling regime for $g=2$ we need much larger $D$, and from a comparison with the ED results for $N=4$ in Fig.~\ref{Supp:fig:FiniteSize_w=1_g=2}  we can conclude that the HEOM result has rather well converged only for $D=17$. We can also observe that the results at $k=0$ typically converge faster with respect to $D$ than the results at $k=\pi$. 


The value $N$ for which the spectral functions correspond to those in the thermodynamic limit also depends on the parameter regime: for larger interaction $g$ and for higher $T$ the chain length $N$ can be smaller, while for smaller $g$ and lower $T$ we need larger $N$.
In panels (c) and (d) of Figs.~\ref{Supp:fig:FiniteSize_w=1_g=1} and~\ref{Supp:fig:FiniteSize_w=1_g=1.41} we  see that for intermediate coupling there is some difference in spectral functions for $N=6$ and $N=10$ ($N=8$). At $k=0$ it is particularly visible in the first satellite structure for $g=1$.
Remarkably, the DMFT on a finite lattice $N=6$ ($N=10$) predicts very similar satellite structure as HEOM for the same $N$. This indicates that the correct satellite peak in Fig.~\ref{fig:zeroT}(a) of the main text should be closer to DMFT, while HEOM results have some artefacts because of the finite lattice size.
On the other hand, for $g=2$ it is enough to set $N=4$, as we now demonstrate.

\begin{figure}[h!]
 \includegraphics[width=2.98in,trim=0cm 0cm 0cm 0cm]{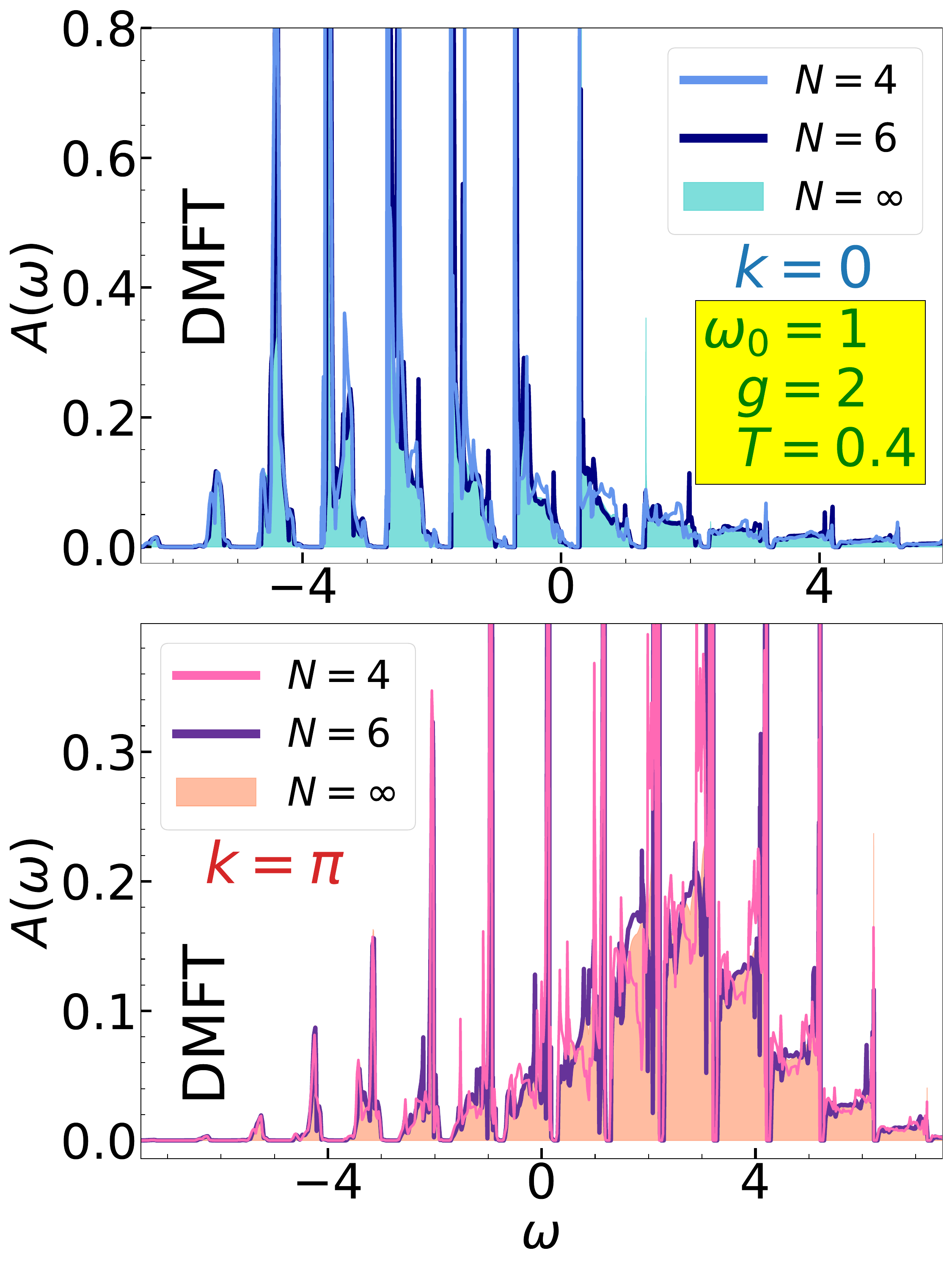}
 \caption{ DMFT spectral functions for different $N$.}
\label{Supp:fig:DMFT_finite_size} 
\end{figure}

It is very efficient to analyze the finite-size effects using the DMFT applied on a finite system with $N$ sites. This is very simple to implement in the DMFT loop. The only difference is in the self-consistency equation: instead of the integral over the density of states, the local Green function is obtained as an average over the $k$ vectors
\begin{equation}
 G(\omega) = \frac{1}{N} \sum_{i=1}^N G_{k_i}(\omega) .
\end{equation}
We can see from Fig.~\ref{Supp:fig:DMFT_finite_size} that there is very little difference between $N=4$, $N=6$ and thermodynamic limit for $g=2$, $\omega_0=1$. We showed only the results for $T=0.4$, but we checked that the conclusions remain true even for $T=0$. Therefore, setting $N=4$ in HEOM and ED calculations is enough. This left enough computer memory to use large $D=17$ in HEOM calculations. Then all three methods give very similar spectral functions as seen in Fig.~\ref{Supp:fig:FiniteSize_w=1_g=2}.



\begin{figure}[h!]
 \includegraphics[width=3.7in,trim=0cm 0cm 0cm 0cm]{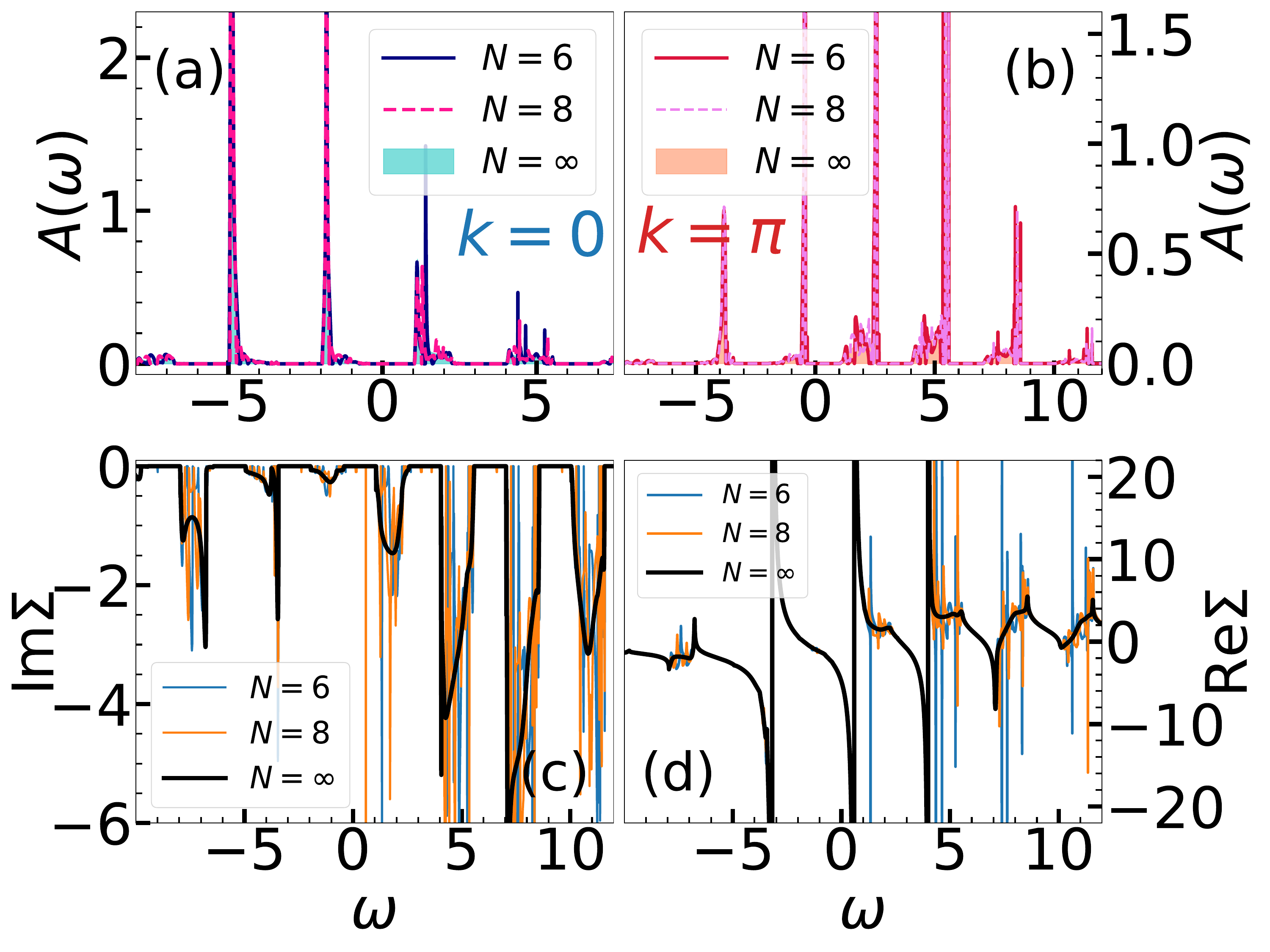}
 \caption{DMFT finite-size effects  close to the atomic limit 
$\omega_0=3$, $g=\sqrt{12}$, $T=1$
}
\label{Supp:fig:FiniteSize_w=3_g=sqrt12} 
\end{figure}

Fig.~\ref{Supp:fig:FiniteSize_w=3_g=sqrt12} shows the DMFT finite-size effects close to the atomic limit, both for the spectral function $A_k (\omega)$ and for the self-energy $\Sigma(\omega)$. The spectral functions are not strongly $N$-dependent. On the other hand, the details of the self-energy are much more sensitive to finite-size effects. Finite $N$ results show a kind of a stripe pattern, while $N=\infty$ results are smoother.

%

\newpage
\clearpage

\section{Atomic limit}
\label{Atomic_App}


Here we investigate the DMFT solution close to the atomic limit.
For decoupled sites ($t_0=0$), using the Lang-Firsov transformation \cite{SMLangFirsov, SMMahanbook}, the Green's function at $T=0$ is given by
\begin{subequations}\label{Supp:eq:atomic}
\begin{equation} \label{Supp:eq:atomic_T=0}
G(\omega) = \sum_{n=0}^\infty \frac{\alpha^{2n}e^{-\alpha^2}}{n!} \frac{1}{\omega - n\omega_0 - E_p + i0^+},
\end{equation} 
and at $T>0$
\begin{equation}\label{Supp:eq:atomic_T_finite}
G(\omega) = \sum_{n=-\infty}^\infty 
\frac{I_n\left( 2\alpha^2 \sqrt{b(b+1)} \right)}{\omega - n\omega_0 - E_p + i0^+}
e^{-(2b+1)\alpha^2 + n\omega_0/2T}.
\end{equation}
\end{subequations}
Here $E_p = -g^2/\omega_0$ is the ground-state energy, $I_n$ are the modified Bessel functions of the first kind and $b\equiv b(\omega_0)=(e^{\omega_0/T}-1)^{-1}$. We see that the atomic limit spectrum consists of a series of delta functions at a distance $\omega_0$ from each other. At $T=0$ the lowest energy peak is at $\omega = E_p$, which corresponds to the ground-state (polaron) energy. At finite temperatures  more delta peaks emerge even below the polaron peak.  

\begin{figure}[t]
 \includegraphics[width=3.4in,trim=0cm 0cm 0cm 0cm]{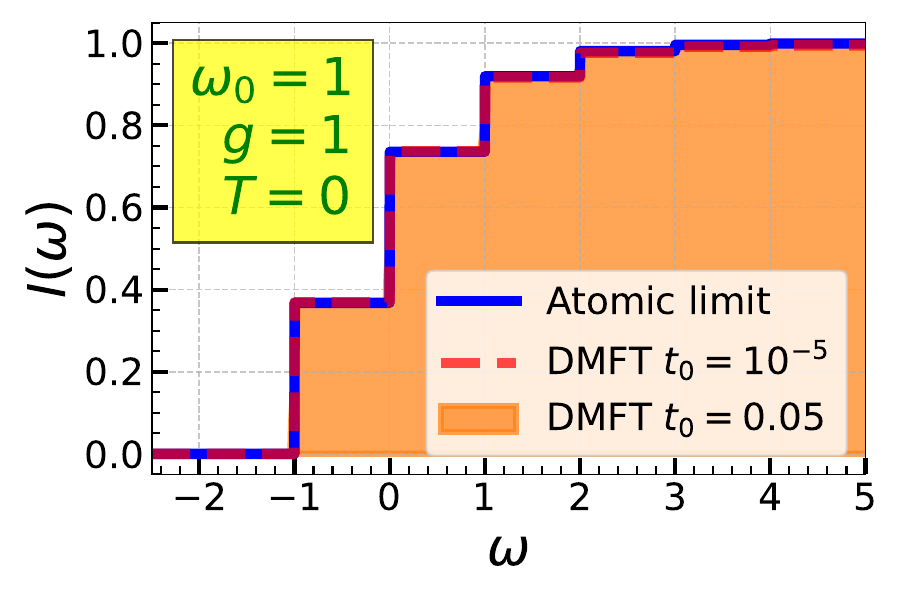}
 \caption{
  DMFT integrated spectral weight for $t_0=0.05$ (shaded) and $t_0 = 10^{-5}$ (red dashed line)  averaged over all momenta, $I(\omega) = \frac{1}{N}\sum_k \int_{-\infty}^\omega A_k(\nu) d\nu$, in comparison to the exact $t_0=0$ result (blue solid line).
}
\label{Supp:fig:atomic_limit_T=0} 
\end{figure}

The integrated DMFT spectral weight at $T=0$ is shown in Fig.~\ref{Supp:fig:atomic_limit_T=0} and compared to the exact atomic limit. It was calculated using the numerical procedure introduced in Sec.~\ref{Integral_A}. $I(\omega)$ features jumps at frequencies where $A(\omega)$ has peaks and the height of those jumps is equal to the weight of the peaks. Nonzero hopping in the DMFT solution introduces small momentum dependence of $I_k(\omega)$, which is why Fig.~\ref{Supp:fig:atomic_limit_T=0} shows the result averaged over all momenta. A more detailed comparison is presented in Table~\ref{Supp:Int_Spec_W_T=0}. It shows the numerical values of the DMFT $I(\omega)$ at the positions of delta peaks  (for a given $k$ and averaged over many $k$) in comparison with the analytical $t_0=0$ result from Eq.~\eqref{Supp:eq:atomic_T=0}. These delta peaks, positioned at $n\omega_0+E_p$, have the weights equal to $\alpha^{2n}e^{-\alpha^2} / n!$ for $n=0,1\dots$

\begin{figure}[h]
 \includegraphics[width=3.4in,trim=0cm 0cm 0cm 0cm]{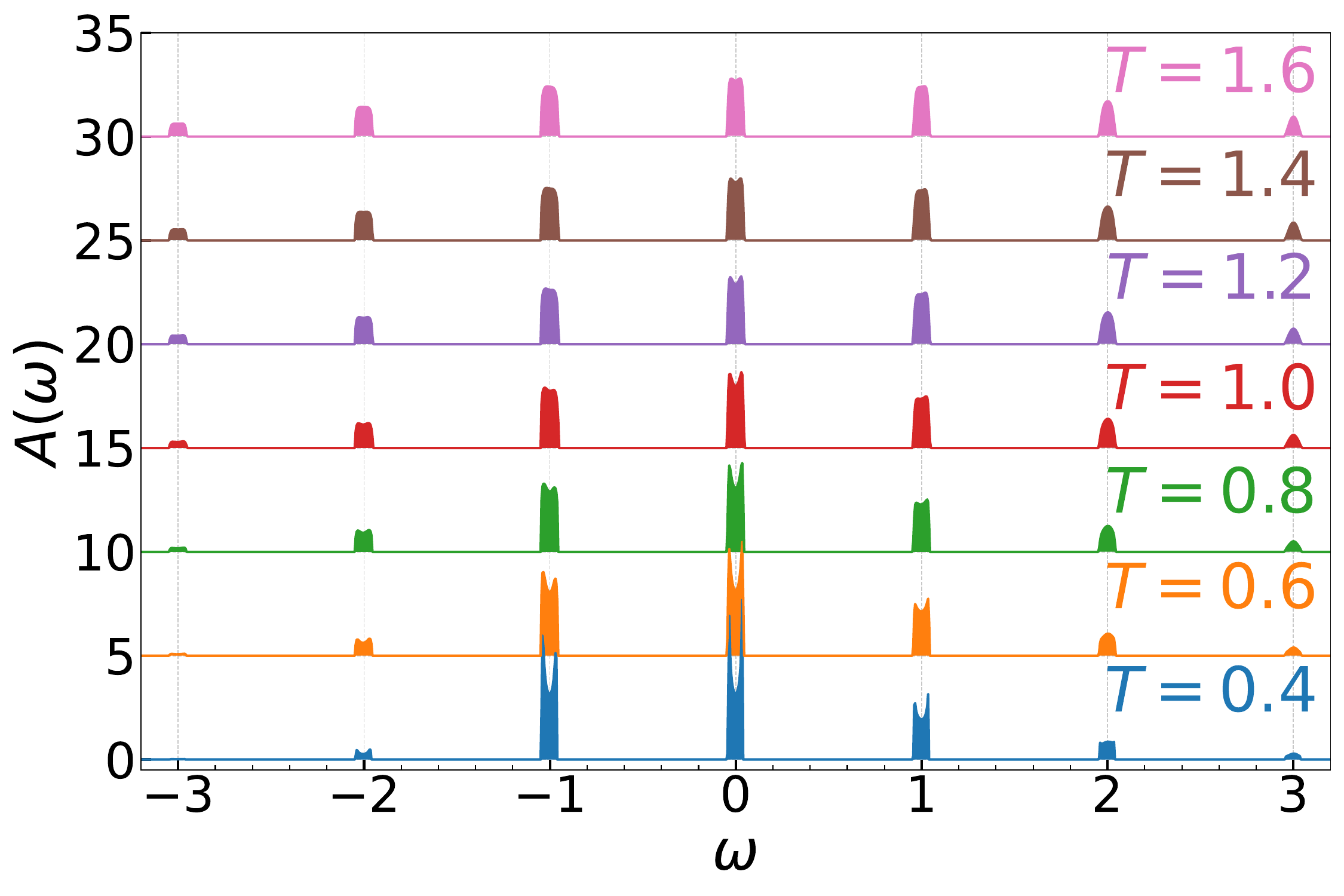}
 \caption{DMFT spectral functions $A(\omega) = \frac{1}{N} \sum_k  A_k(\omega)$ for $\omega_0=1$, $g=1$, $t_0=0.05$, at several temperatures.
}
\label{Supp:fig:atomic_limit_T_finite} 
\end{figure}

For $T>0$, the peaks are located both below and above $E_p$. The DMFT spectra averaged over $k$ are shown in Fig.~\ref{Supp:fig:atomic_limit_T_finite}.
They have a characteristic fork-shaped form at low $T$, which is the consequence of the 1d density of states. The weight of the peaks are very close to the { analytical result} $I_n( 2\alpha^2 \sqrt{b(b+1)} ) e^{-(2b+1)\alpha^2 + n\omega_0/2T}$. These spectral weights, averaged over momenta $k$, are given in Table~\ref{Supp:Int_Spec_W_T_finite}.

\newpage

\begin{center}

\large
\renewcommand{\arraystretch}{1.}%
\setlength{\tabcolsep}{4pt}
\captionof{table}{Integrated spectral weight $I(\omega)$ for  $\omega_0=1$, $g=1$ at $T=0$. The exact atomic limit corresponds to $t_0=0.00$. For $t_0=10^{-5}$ the DMFT solution has no $k$-dependence within the specified accuracy. We denote the $k$-values to be 'av.' if the answer is averaged over all momenta.    \label{Supp:Int_Spec_W_T=0}}

\begin{tabular}{l|c|cccccc}\hline
k & \backslashbox{$t_0$}{$\omega$} & $-2$ & $-1$ & $0$ & $1$ & $2$ & $3$\\\hline
 & $0.00$ &\color{red}{$0.00$}& \color{red}{$0.37$} & \color{red}{$0.74$} &\color{red}{$0.92$} &\color{red}{$0.98$} &\color{red}{$1.0$} \\\hline
all  & $10^{-5}$ & $0.00$ & $0.37$ & $0.74$ & $0.92$ & $0.98$ & $1.0$ \\\hline
av. & $0.05$ & $0.00$ & $0.37$ & $0.73$ & $0.92$ & $0.98$ & $1.0$ \\\hline
0 & $0.05$ & $0.00$ & $0.40$ & $0.76$ & $0.94$ & $0.99$ & $1.0$ \\\hline
$\pi/2$ & $0.05$ & $0.00$ & $0.37$ & $0.74$ & $0.92$ & $0.98$ & $1.0$ \\\hline
$\pi$ & $0.05$ & $0.00$ & $0.33$ & $0.71$ & $0.91$ & $0.98$ & $0.99$ \\\hline
\end{tabular}
\end{center}

%
%

\begin{center}

\large
\renewcommand{\arraystretch}{1.0}%
\setlength{\tabcolsep}{4pt}
\captionof{table}{Spectral weights of  the peaks { located at} $\omega = n \omega_0 + E_p$  for $n=-2,-1,0,1,2,3$. The DMFT spectra, obtained for $t_0=0.05$, are averaged over $k$. The atomic limit values ($t_0=0.00$) are obtained from the analytical formula. Here $\omega_0=1$, $g=1$. \label{Supp:Int_Spec_W_T_finite}}

\begin{tabular}{l|c|cccccc}\hline
T & \backslashbox{$t_0$}{$\omega$} & $-2$ & $-1$ & $0$ & $1$ & $2$ & $3$\\\hline
$0.4$ & $0.00$ &\color{red}{$0.03$}& \color{red}{$0.34$} & \color{red}{$0.35$} &\color{red}{$0.19$} &\color{red}{$0.07$} &\color{red}{$0.02$} \\\hline
$0.4$  & $0.05$ & $0.03$ & $0.34$ & $0.34$ & $0.18$ & $0.07$ & $0.02$ \\\hline
$0.6$ & $0.00$ &\color{red}{$0.06$}& \color{red}{$0.30$} & \color{red}{$0.33$} &\color{red}{$0.19$} &\color{red}{$0.08$} &\color{red}{$0.02$} \\\hline
$0.6$  & $0.05$ & $0.06$ & $0.30$ & $0.33$ & $0.19$ & $0.08$ & $0.02$ \\\hline
$0.8$ & $0.00$ &\color{red}{$0.09$}& \color{red}{$0.27$} & \color{red}{$0.30$} &\color{red}{$0.19$} &\color{red}{$0.09$} &\color{red}{$0.03$} \\\hline
$0.8$  & $0.05$ & $0.09$ & $0.27$ & $0.30$ & $0.19$ & $0.09$ & $0.03$ \\\hline
$1.0$ & $0.00$ &\color{red}{$0.10$}& \color{red}{$0.25$} & \color{red}{$0.28$} &\color{red}{$0.19$} &\color{red}{$0.09$} &\color{red}{$0.04$} \\\hline
$1.0$  & $0.05$ & $0.10$ & $0.25$ & $0.28$ & $0.19$ & $0.10$ & $0.04$ \\\hline
$1.2$ & $0.00$ &\color{red}{$0.11$}& \color{red}{$0.23$} & \color{red}{$0.26$} &\color{red}{$0.19$} &\color{red}{$0.10$} &\color{red}{$0.04$} \\\hline
$1.2$  & $0.05$ & $0.11$ & $0.23$ & $0.26$ & $0.19$ & $0.10$ & $0.04$ \\\hline
$1.4$ & $0.00$ &\color{red}{$0.12$}& \color{red}{$0.21$} & \color{red}{$0.24$} &\color{red}{$0.19$} &\color{red}{$0.11$} &\color{red}{$0.05$} \\\hline
$1.4$  & $0.05$ & $0.12$ & $0.21$ & $0.24$ & $0.19$ & $0.11$ & $0.05$ \\\hline
\end{tabular}
\end{center}

\newpage
\clearpage

\section{Spectral functions at $\mathbf{T=0}$: Additional results}
\label{T=0SUPP}

\begin{figure}[h]
 \includegraphics[width=3.2in,trim=0cm 0cm 0cm 0cm]{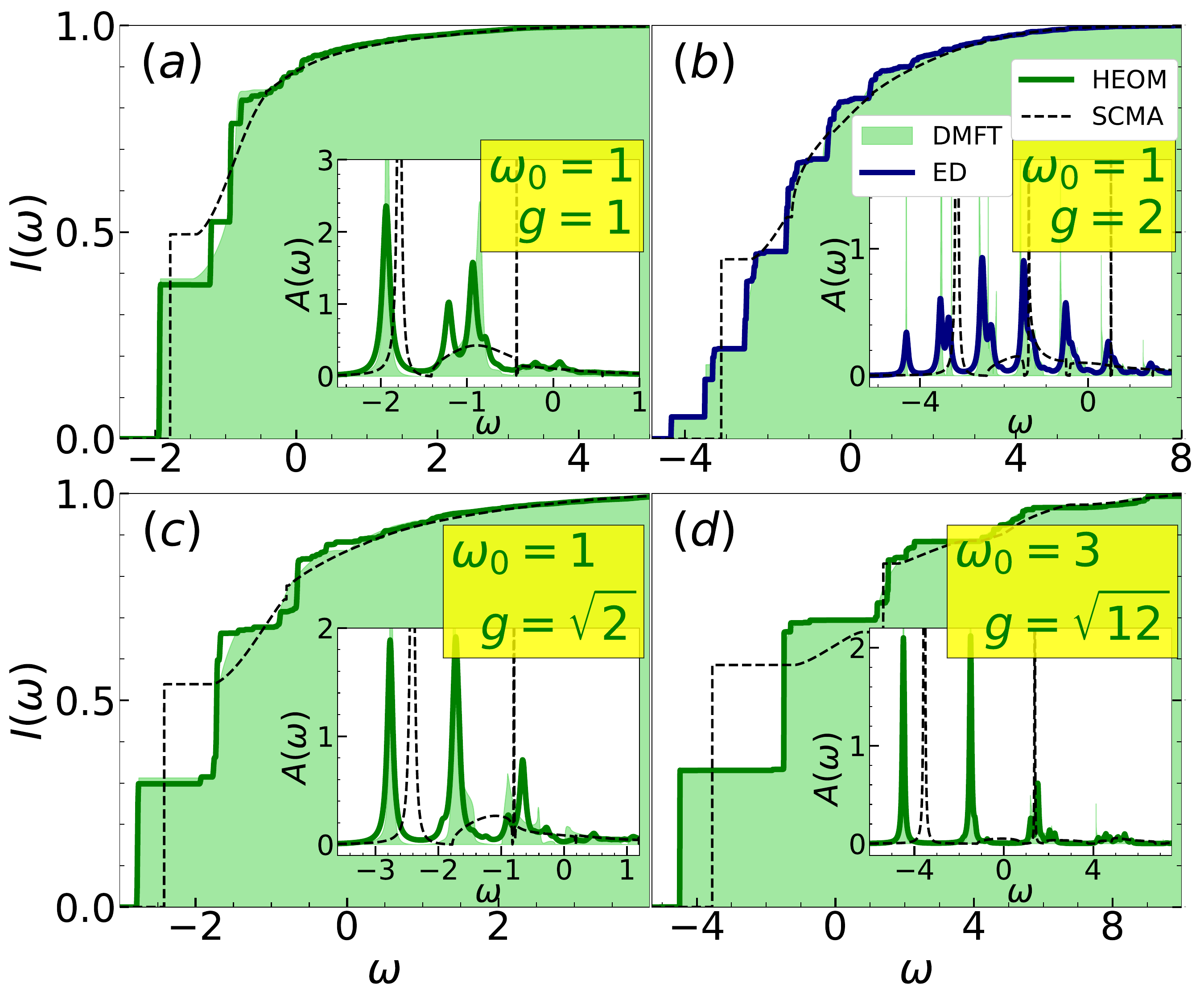}
 \caption{
Integrated spectral weight at $T=0$  with no broadening. The insets show spectral functions with $\eta=0.05$ Lorentzian broadening. Different panels have the following values of the momenta:  (a) $k=\frac{8\pi}{25}$, (b) $k=\frac{\pi}{4}$, (c) $k=\frac{\pi}{4}$, (d) $k=\frac{\pi}{3}$.
}
\label{Supp:fig:Fig2_arg_1} 
\end{figure}

\begin{figure}[!h]
 \includegraphics[width=3.2in,trim=0cm 0cm 0cm 0cm]{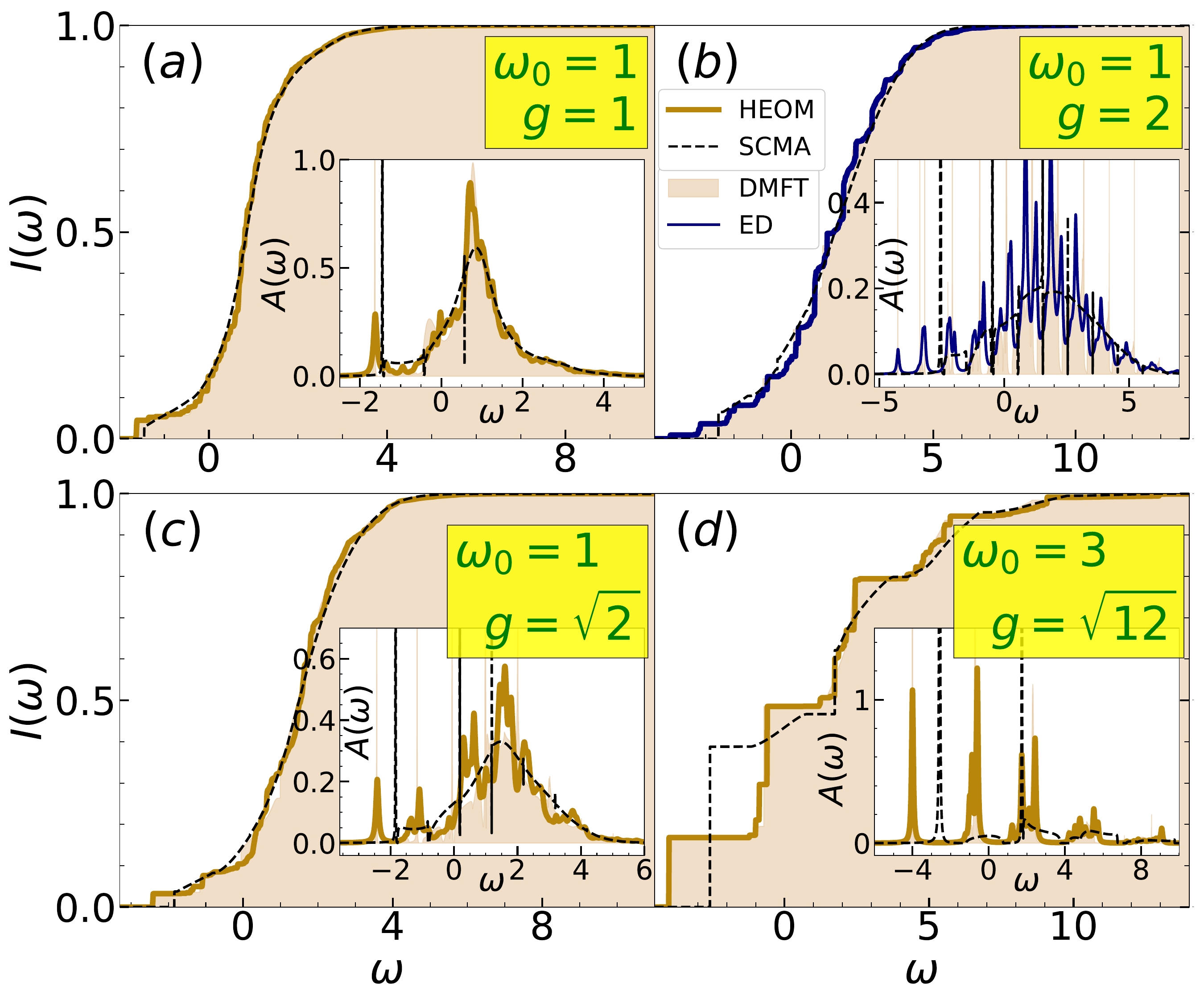}
 \caption{
Integrated spectral weight at $T=0$  with no broadening. The insets show spectral functions with $\eta=0.05$ Lorentzian broadening. Different panels have the following values of the momenta: (a) $k=\frac{16\pi}{25}$, (b) $k=\frac{3\pi}{4}$, (c) $k=\frac{3\pi}{4}$, (d) $k=\frac{2\pi}{3}$.
}
\label{Supp:fig:Fig2_arg_2} 
\end{figure}

\begin{figure}[!h]
 \includegraphics[width=3.2in,trim=0cm 0cm 0cm 0cm]{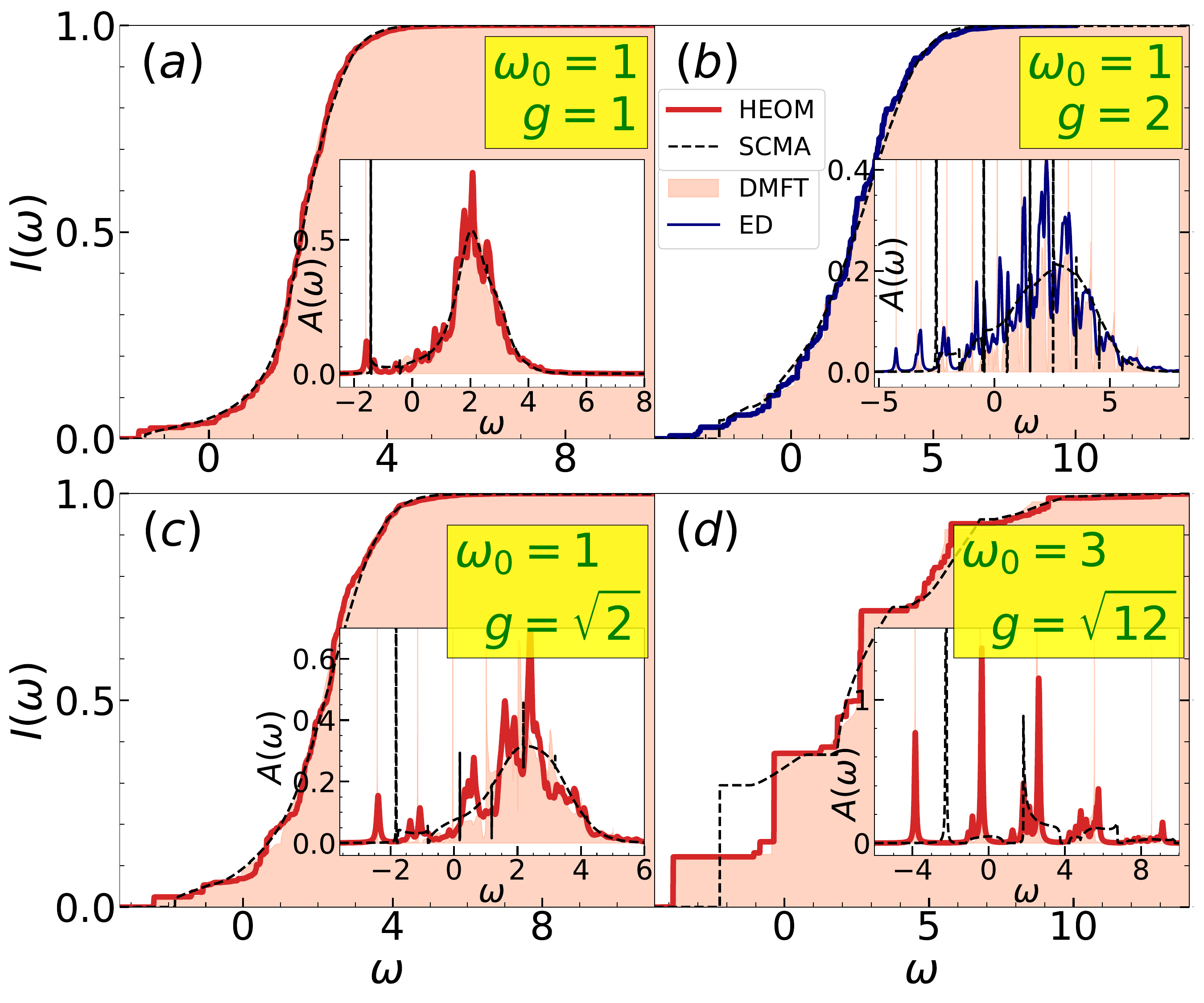}
 \caption{
Integrated spectral weight at $T=0$  with no broadening. The insets show spectral functions with ${\eta=0.05}$ Lorentzian broadening. Every panel is calculated for $k=\pi$.
}
\label{Supp:fig:Fig2_arg_pi} 
\end{figure}

Spectral functions and integrated spectral weights at ${T=0}$ for $k=0$ are shown in Fig.~\ref{fig:zeroT} of the main text. In Figs.~\ref{Supp:fig:Fig2_arg_1}~-~\ref{Supp:fig:Fig2_arg_pi}, we show the results for additional momenta. We note that the integrated spectral weight was calculated without broadening, using the numerical scheme described in Sec.~\ref{Integral_A}. The spectral functions are shown with a small Lorentzian broadening $\eta$,
\begin{equation}
A_{\eta}(\omega) = \frac{1}{\pi}\int_{-\infty}^{\infty} d\nu
\frac{\eta A(\nu)}{\eta^2 + (\omega - \nu)^2} ,
\end{equation}

We see that there is a very good agreement between DMFT and HEOM/ED results. 
In every regime where HEOM was implemented, we checked that the results were well converged with respect to the lattice size $N$ and the maximum hierarchy depth $D$. These values are shown in Table~\ref{Supp:tab:HEOM_par_0}.

We note that the HEOM/ED method imposes the periodic boundary conditions on a finite lattice. This means that the HEOM/ED spectral functions are available only for a discrete values of momenta, unlike the DMFT which is calculated in the thermodynamical limit.
Results for additional $k$-values are obtained using twisted boundary conditions.


\vspace*{1.5cm}

\begin{center}

\large
\renewcommand{\arraystretch}{1.0}%
\setlength{\tabcolsep}{20pt}
\captionof{table}{Lattice size $N$ and the maximum hierarchy depth $D$ used in the HEOM calculations which correspond to Figs.~\ref{Supp:fig:Fig2_arg_1}-\ref{Supp:fig:Fig2_arg_pi} and Fig.~\ref{fig:zeroT} from the main text.\label{Supp:tab:HEOM_par_0}}
\begin{tabular}{ l|c|c } 
 \hline
$\quad$Parameters                  & N  & D     \\
 \hline
 $\omega_0=1\;\;\;\;g=1$         & 10 & 6 \\ 
 $\omega_0=1\;\;\;\;g=\sqrt{2}$  & 8  & 7 \\ 
 $\omega_0=3\;\;\;\;g=\sqrt{12}$ & 6  & 9 \\ 
 \hline
\end{tabular}
\end{center}


\newpage
\clearpage

\section{Spectral functions at $\mathbf{T>0}$: Additional results}
\label{TfiniteSUPP}

\begin{figure}[t]
 \includegraphics[width=3.2in,trim=0cm 0cm 0cm 0cm]{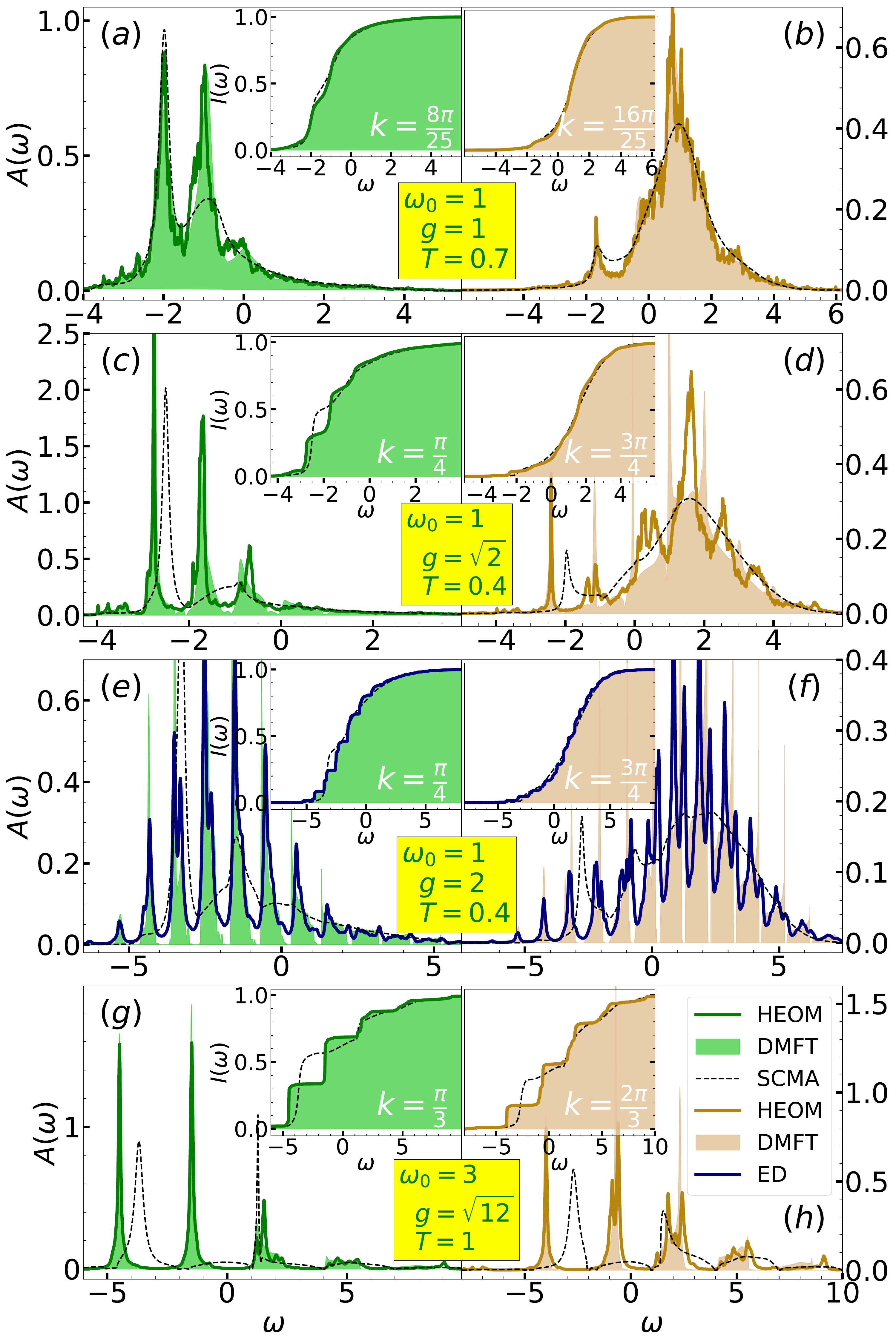}
 \caption{HEOM, DMFT, SCMA and ED spectral functions for different parameters. On the left panels $\pi/4 \leq k \leq \pi/3$, whereas $\pi/2 \leq k \leq 3\pi/4$ on the right. The integrated spectral weight is presented  in the insets without broadening. In panels (g) and (h) Lorentzian broadening with $\eta = 0.05$ is used for all spectral functions, while only ED is broadened in (e) and (f) using the same $\eta$. 
}
\label{Supp:fig:Fig3_addition} 
\end{figure}

\begin{figure}[t]
 \includegraphics[width=3.2in,trim=0cm 0cm 0cm 0cm]{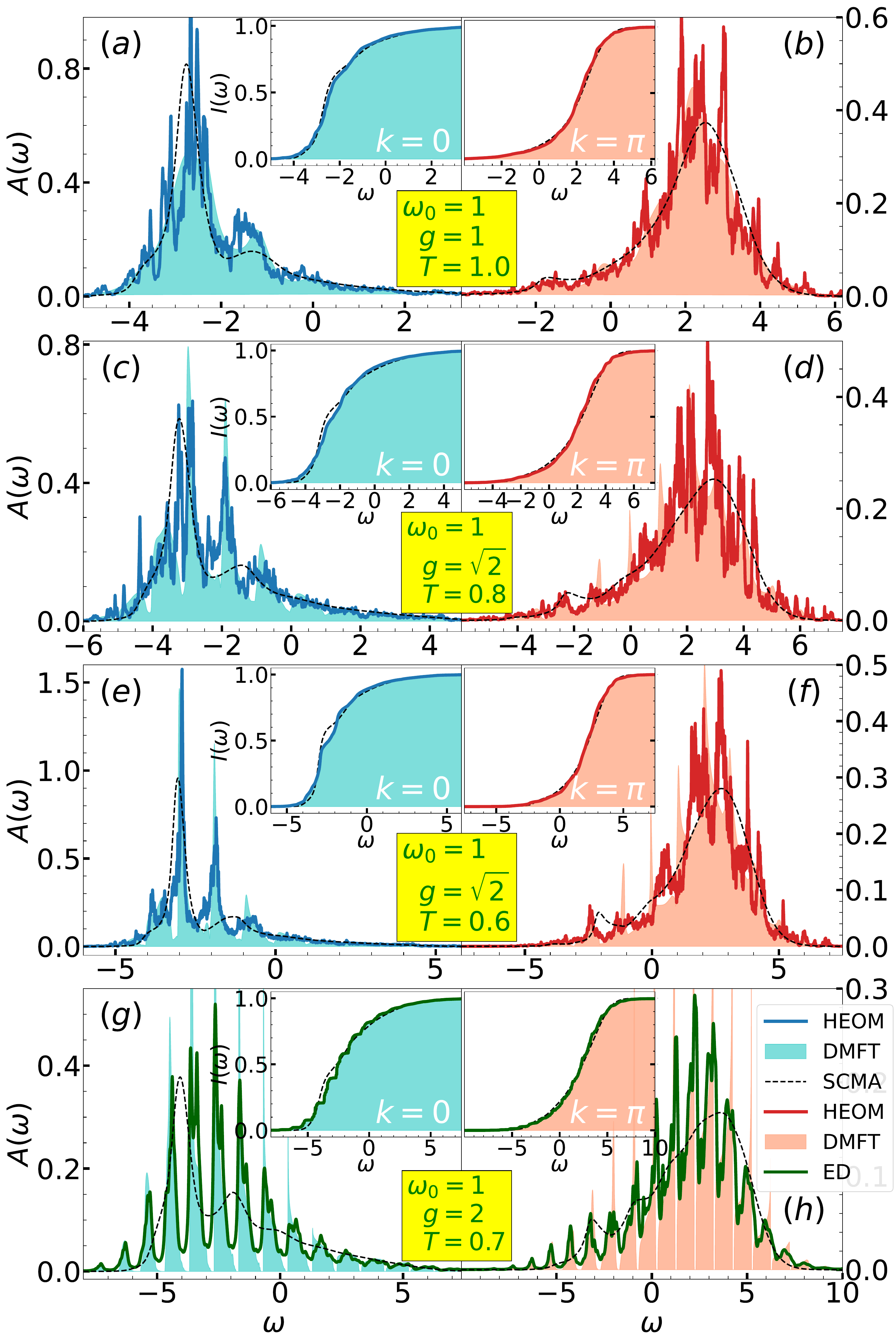}
 \caption{HEOM, 
DMFT, SCMA and ED spectral functions for different parameters. On the left panels $k=0$, whereas ${k=\pi}$ on the right. The integrated spectral weight is presented in the insets without broadening. The Lorentzian broadening with $\eta = 0.05$ is used only for ED spectral functions.
}
\label{Supp:fig:SpecF_1} 
\end{figure}

\begin{figure}[t]
 \includegraphics[width=3.2in,trim=0cm 0cm 0cm 0cm]{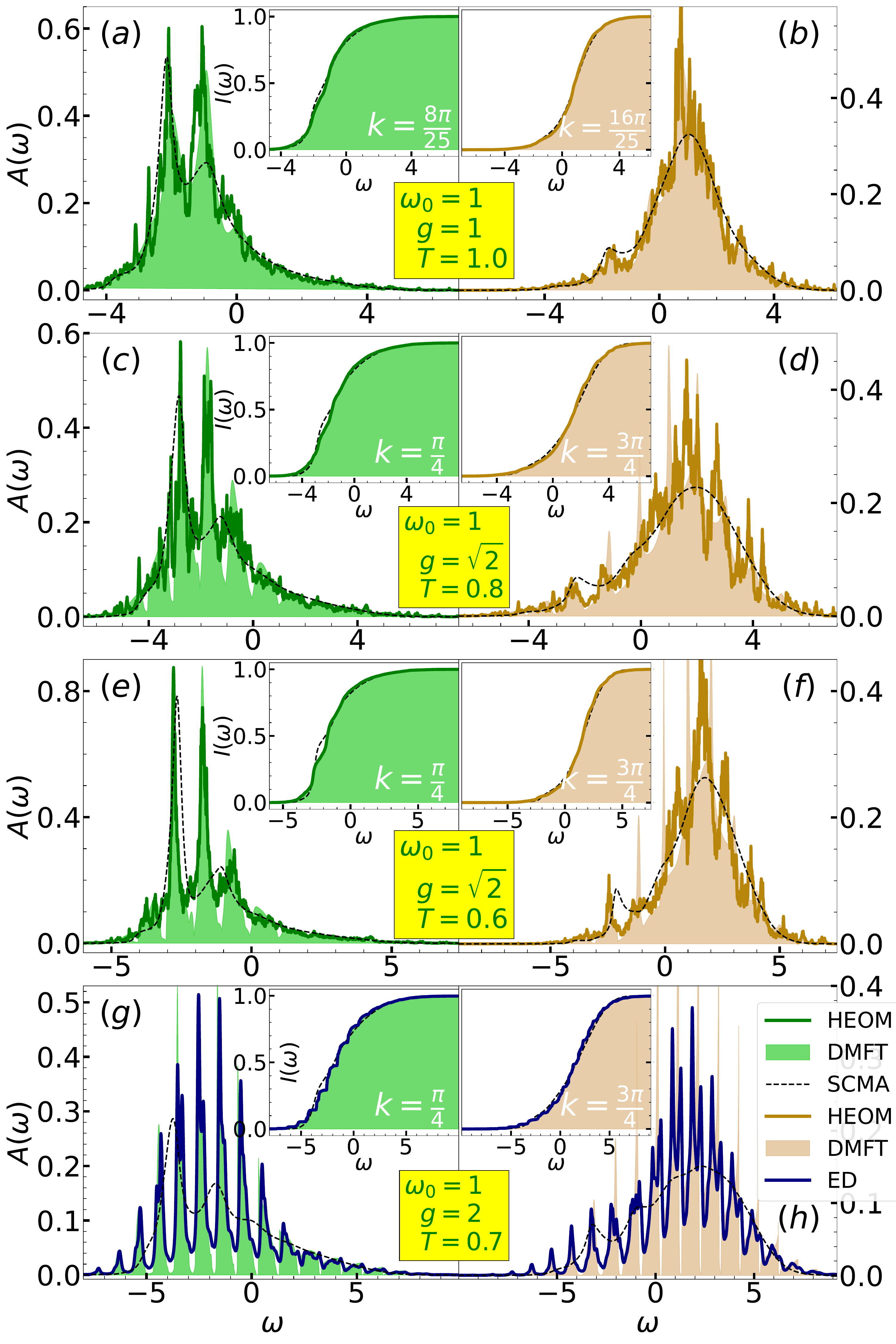}
 \caption{HEOM, DMFT, SCMA and ED spectral functions for different parameters. On the left panels $\pi/4 \leq k \leq \pi/3$, whereas $\pi/2 \leq k \leq 3\pi/4$ on the right. The integrated spectral weight is presented in the insets without broadening. The Lorentzian broadening with $\eta = 0.05$ is used only for ED spectral functions.
}
\label{Supp:fig:SpecF1_more_momenta} 
\end{figure}

Spectral functions for $k=0$ and $k=\pi$, shown in Fig.~\ref{fig:finiteT} of the main text, are supplemented with the results for different $k$ in Fig.~\ref{Supp:fig:Fig3_addition}. Overall, the agreement of DMFT and HEOM/ED spectra is very good which confirms that the nonlocal correlations are not pronounced. 
Results for different temperatures are shown in Figs.~\ref{Supp:fig:SpecF_1} and~\ref{Supp:fig:SpecF1_more_momenta}. We checked that the HEOM results are well converged with respect to lattice size $N$ and maximum hierarchy depth $D$. The values of $N$ and $D$, used in the calculations, are shown in Table~\ref{Supp:tab:HEOM_par_fin}.

\begin{center}
\large
\renewcommand{\arraystretch}{1.0}%
\setlength{\tabcolsep}{15pt}
\captionof{table}{Lattice size $N$ and the maximum hierarchy depth $D$ used in the HEOM calculations which correspond to Figs.~\ref{Supp:fig:Fig3_addition}~-~\ref{Supp:fig:SpecF1_more_momenta} and Fig.~\ref{fig:finiteT} from the main text.\label{Supp:tab:HEOM_par_fin}}

\begin{tabular}{ l|c|c } 
 \hline
$\quad$Parameters                         & N  & D     \\
 \hline
 $\omega_0=1\;\;g=1\;\;\;\;\;\;\;\;T=0.7$   & 10 & 6     \\ 
 $\omega_0=1\;\;g=1\;\;\;\;\;\;\;\;T=1$     & 10 & 6      \\ 
 $\omega_0=1\;\;g=\sqrt{2}\;\;\;\;\;T=0.4$  & 8  & 8     \\
 $\omega_0=1\;\;g=\sqrt{2}\;\;\;\;\;T=0.6$  & 8  & 7     \\
 $\omega_0=1\;\;g=\sqrt{2}\;\;\;\;\;T=0.8$  & 8  & 7     \\
 $\omega_0=1\;\;g=2\;\;\;\;\;\;\;\;T=0.4$   & 4  & 17    \\  
 $\omega_0=3\;\;g=\sqrt{12}\;\;\;T=1$       & 6  & 9     \\ 
 \hline
\end{tabular}
\end{center}

%

%
%
%

\begin{figure}[!t]
 \includegraphics[width=3.2in,trim=0cm 0cm 0cm 0cm]{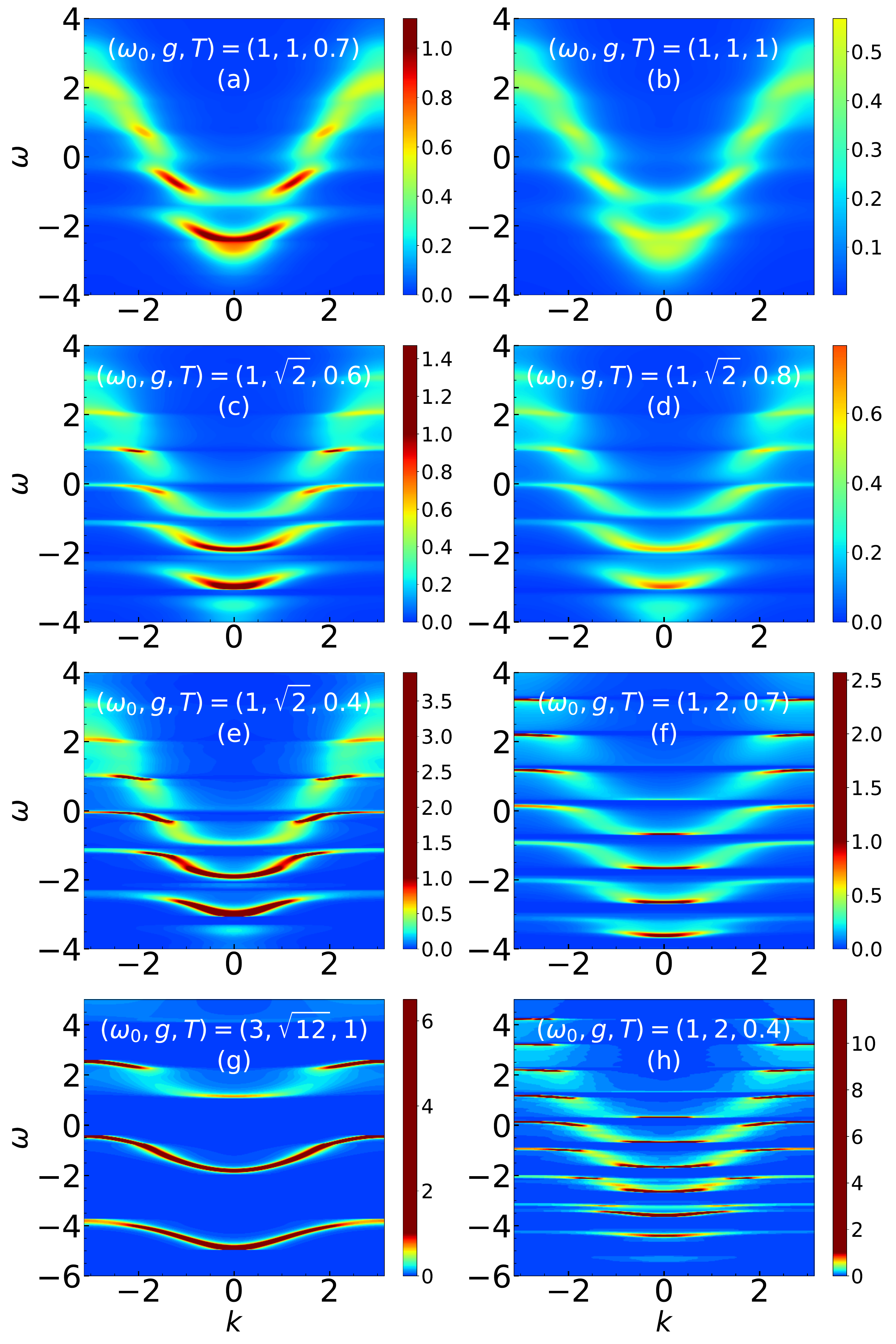}
 \caption{
The DMFT spectral functions $A_k(\omega)$ for parameters as in Figs.~\ref{Supp:fig:Fig3_addition}~-~\ref{Supp:fig:SpecF1_more_momenta}. The same color coding
is used in all plots.
}
\label{Supp:fig:DMFT_Heom_parameters} 
\end{figure}

\begin{figure}[!t]
 \includegraphics[width=3.2in,trim=0cm 0cm 0cm 0cm]{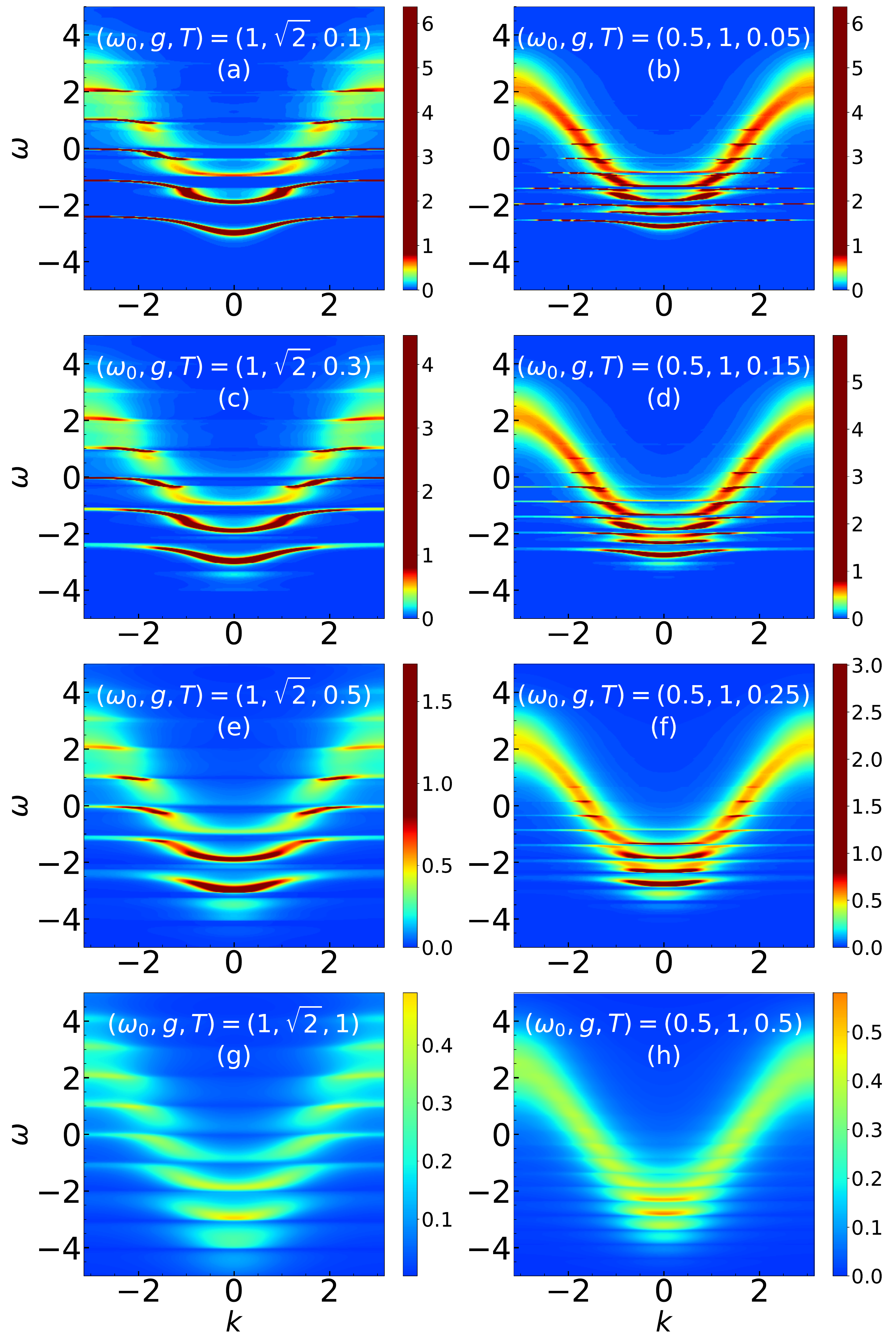}
 \caption{
The DMFT spectral functions $A_k(\omega)$ for parameters as in Fig.~2 of Ref.~\cite{SMBonca_Lanczos_2019}. The same color coding
is used in all plots.
}
\label{Supp:fig:DMFT_Bonca_parameters} 
\end{figure}

It is common to present the spectral functions as color plots in the $k-\omega$ plane. In Fig.~\ref{Supp:fig:DMFT_Heom_parameters} we show the DMFT color plot for parameters as in Figs.~\ref{Supp:fig:Fig3_addition}~-~\ref{Supp:fig:SpecF1_more_momenta}. 
For comparison purposes, in Fig.~\ref{Supp:fig:DMFT_Bonca_parameters} we also show the DMFT color plot for the same parameters as in the finite-$T$ Lanczos results from Fig.~2 of Ref.~\cite{SMBonca_Lanczos_2019}. Small difference in DMFT vs. Lanczos method color plots is due to the more pronounced peaks in the DMFT spectra. 


\newpage
\clearpage

\section{HEOM self-energies}
\label{self_energies}

The results for the spectral functions, as well as for the effective mass and ground state energy, have shown that the DMFT gives an excellent approximate solution of 1d Holstein model in the whole parameter space. This indicates that the self-energy is approximately local which we explicitly demonstrate in this Section. Since $\Sigma_k(\omega) = \Sigma_{-k}(\omega)$ we will show only the results for $k \ge 0$.

\begin{figure}[b]
 \includegraphics[width=3.4in,trim=0cm 0cm 0cm 0cm]{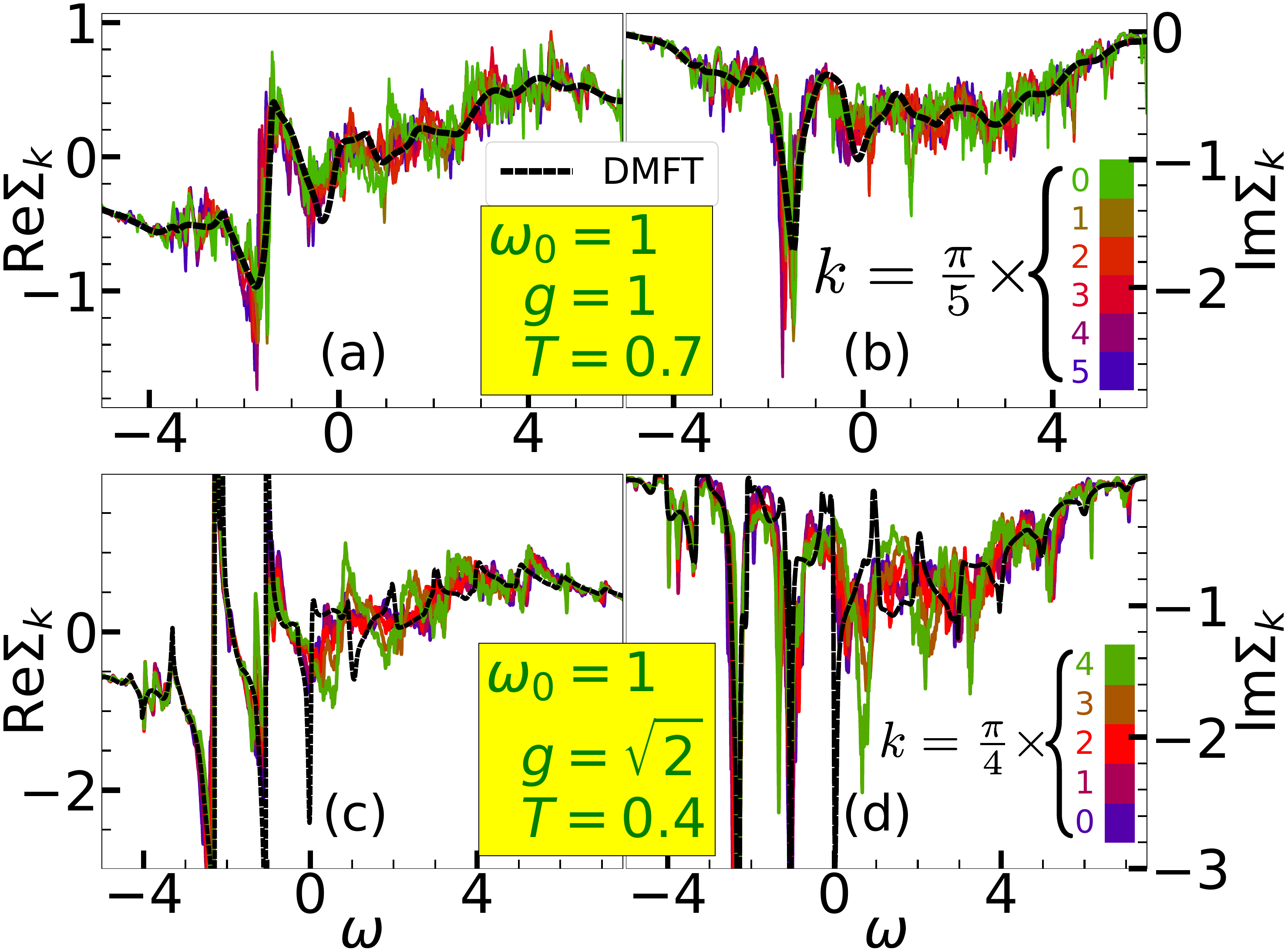}
 \caption{HEOM and DMFT self-energies for intermediate coupling.
}
\label{Supp:fig:selfen_intermediate} 
\end{figure}

In Fig.~\ref{Supp:fig:selfen_intermediate} we present the HEOM and DMFT self-energies in the intermediate coupling regime. Panels (a) and (b) of Fig.~\ref{Supp:fig:selfen_intermediate} show that the self-energies are nearly local, whereas the DMFT solution interpolates in between. The self-energy is approximately local also for $g=\sqrt{2}$, Fig.~\ref{Supp:fig:selfen_intermediate}(c)-(d). There is a visible discrepancy only at higher momenta, which reflects in a shift of the spectral functions with respect to the DMFT solution in Fig.~\ref{fig:finiteT}(d) of the main text.


\begin{figure}[t]
 \includegraphics[width=3.4in,trim=0cm 0cm 0cm 0cm]{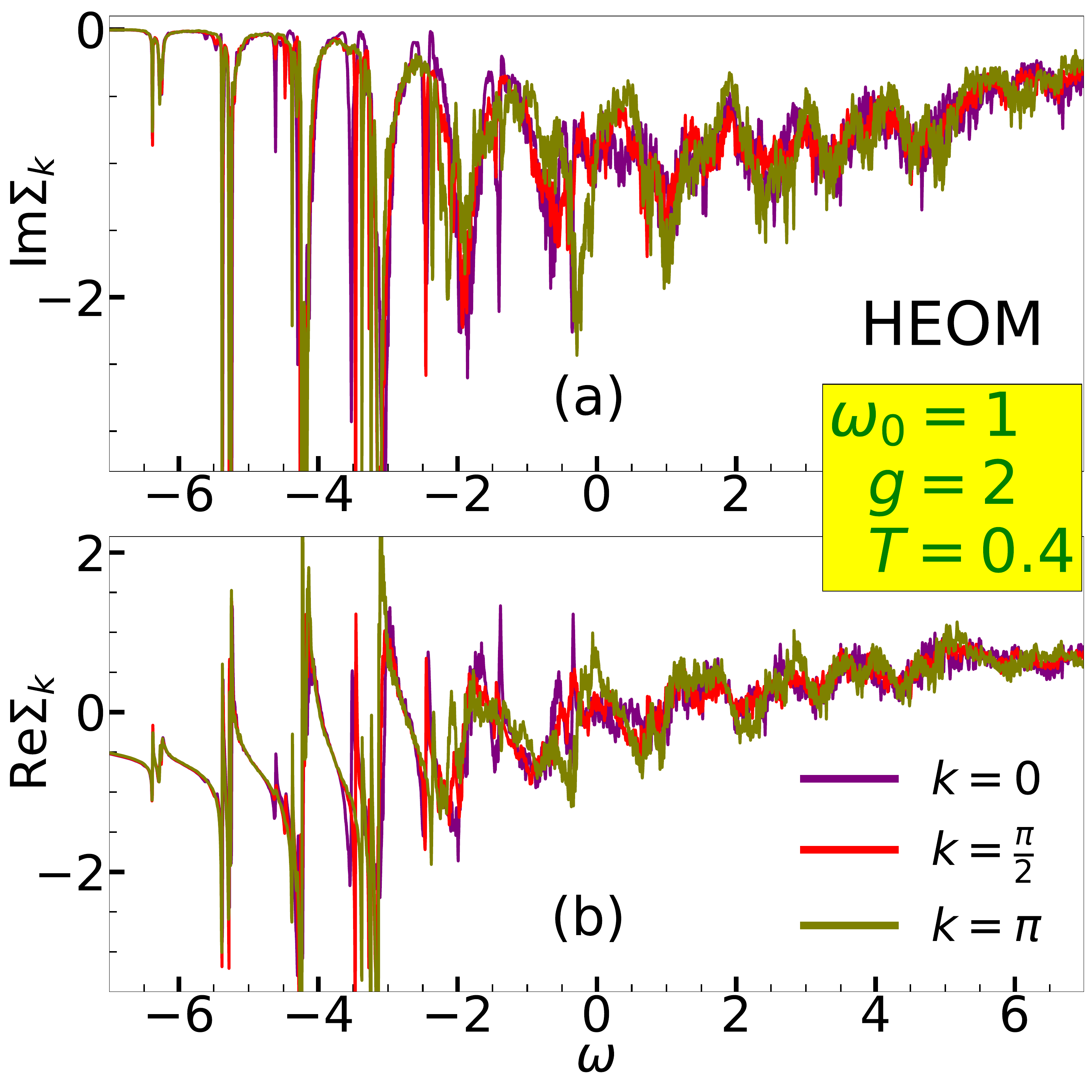}
 \caption{HEOM self-energies for strong coupling. Here $N=4$ and $D=17$. 
}
\label{Supp:fig:selfen_strong} 
\end{figure}

The results for the strong coupling are presented in Fig.~\ref{Supp:fig:selfen_strong}. The DMFT solution for $\mathrm{Im}\Sigma$ falls to zero between the peaks, as opposed to the HEOM solution where such behavior is observed only for the first few peaks. This is why, for the sake of clarity, the DMFT self-energy is omitted. This is consistent with Fig.~\ref{Supp:fig:FiniteSize_w=1_g=2} where the HEOM results feature the dips, while DMFT solution has gaps.  Nevertheless, the presented HEOM results are enough to conclude that the self-energy is nearly local. This is particularly important conclusion since these parameters correspond to strongly renormalized effective mass, $m^*/m \approx 10$.

\begin{figure}[h!]
 \includegraphics[width=3.4in,trim=0cm 0cm 0cm 0cm]{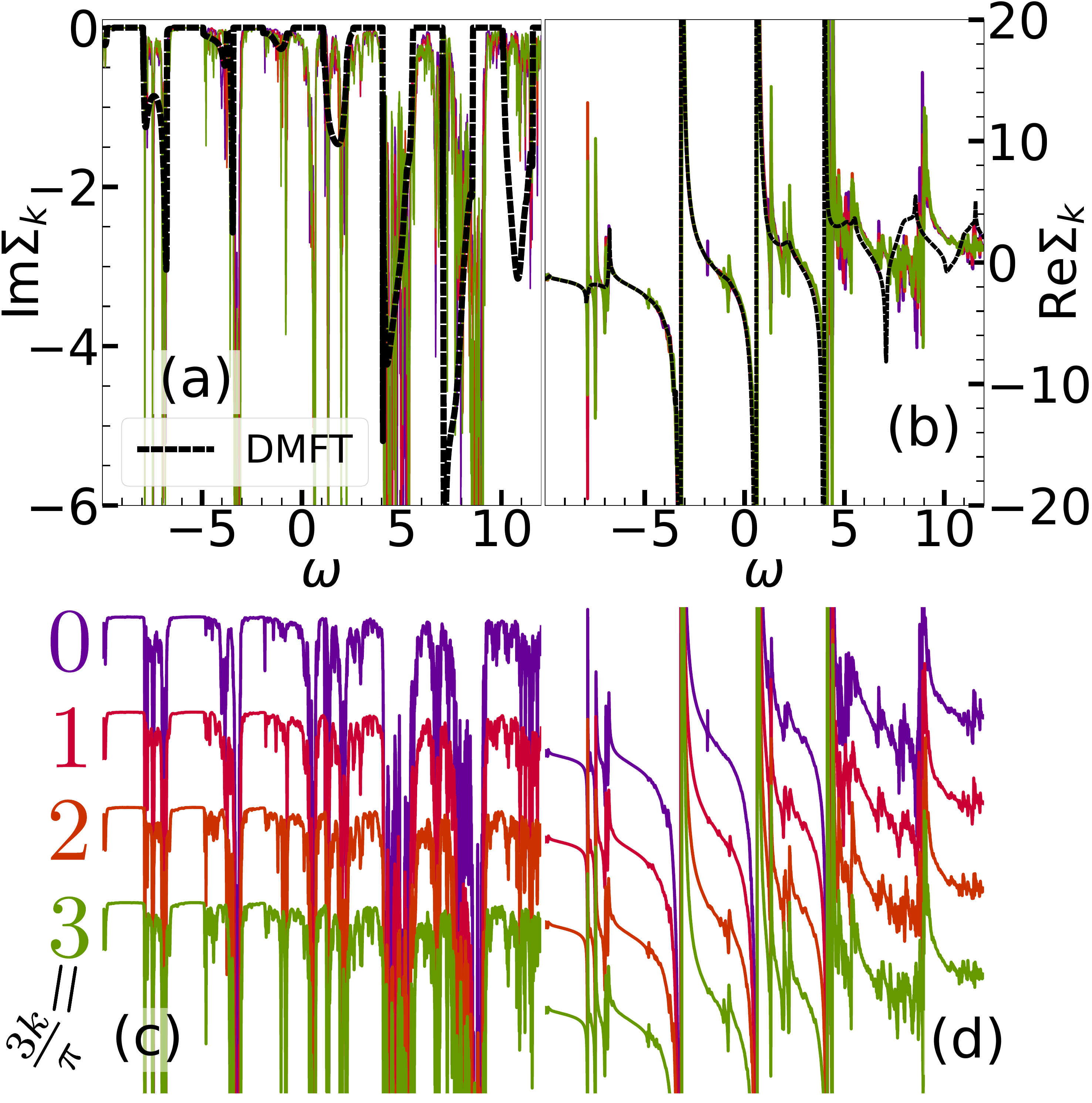}
 \caption{Panels (a) and (b) show HEOM and DMFT self-energies  close to the atomic limit 
$\omega_0=3$, $g=\sqrt{12}$, $T=1$. Panels (c)-(d) show the same HEOM results as in (a)-(b) but shifted for different values of momenta $k$.
}
\label{Supp:fig:selfen_atomic} 
\end{figure}

The regime close to the atomic limit is investigated in Fig.~\ref{Supp:fig:selfen_atomic}. Panels (c) and (d) show that the results are nearly local, but have a kind of stripe pattern, unlike the DMFT solution which is in thermodynamic limit. This is here just a consequence of the finite-size effects, as shown in Fig.~\ref{Supp:fig:FiniteSize_w=3_g=sqrt12}. As discussed in Sec.~\ref{Sec:finitesize}, even though the finite-size effects are visible as stripes in the self-energies, they will not significantly affect the spectral functions. This is why we see a very good agreement between the DMFT and $N=6$ HEOM spectral functions in panels (g) and (h) of Fig.~\ref{fig:finiteT} in the main text.

\newpage
\clearpage

\section{Correlation functions}
\label{CorrF}

Here we present a detailed comparison between QMC, HEOM and DMFT correlation functions.  The QMC correlation function is defined by
\begin{equation}
C_{k}(\tau) =  \langle c_{k}(\tau) c_{k}^\dagger \rangle_{T,0},
\end{equation}
where $c_{k}(\tau) = e^{\tau H}c_{k}e^{-\tau H}$ and $0\leq \tau \leq 1/T$. In Sec.~\ref{Supp:rel_spec_corr} we proved the following relation
\begin{equation}
C_{k}(\tau) = 
\int_{-\infty}^\infty d\omega \,
e^{-\omega\tau} A_{ k}(\omega). \label{Supp:eq:Corr_From_Spec}
\end{equation}
Eq.~(\ref{Supp:eq:Corr_From_Spec}) can now be used to check whether the spectral functions that we calculated using other methods are consistent with the QMC results. A calculation of the spectral functions from the QMC data would assume an analytical continuation which is an ill-defined procedure, particularly problematic when the spectrum has several pronounced peaks. Therefore, we have to settle for a comparison on the imaginary axis.

\begin{figure}[t]
 \includegraphics[width=3.7in,trim=0cm 0cm 0cm 0cm]{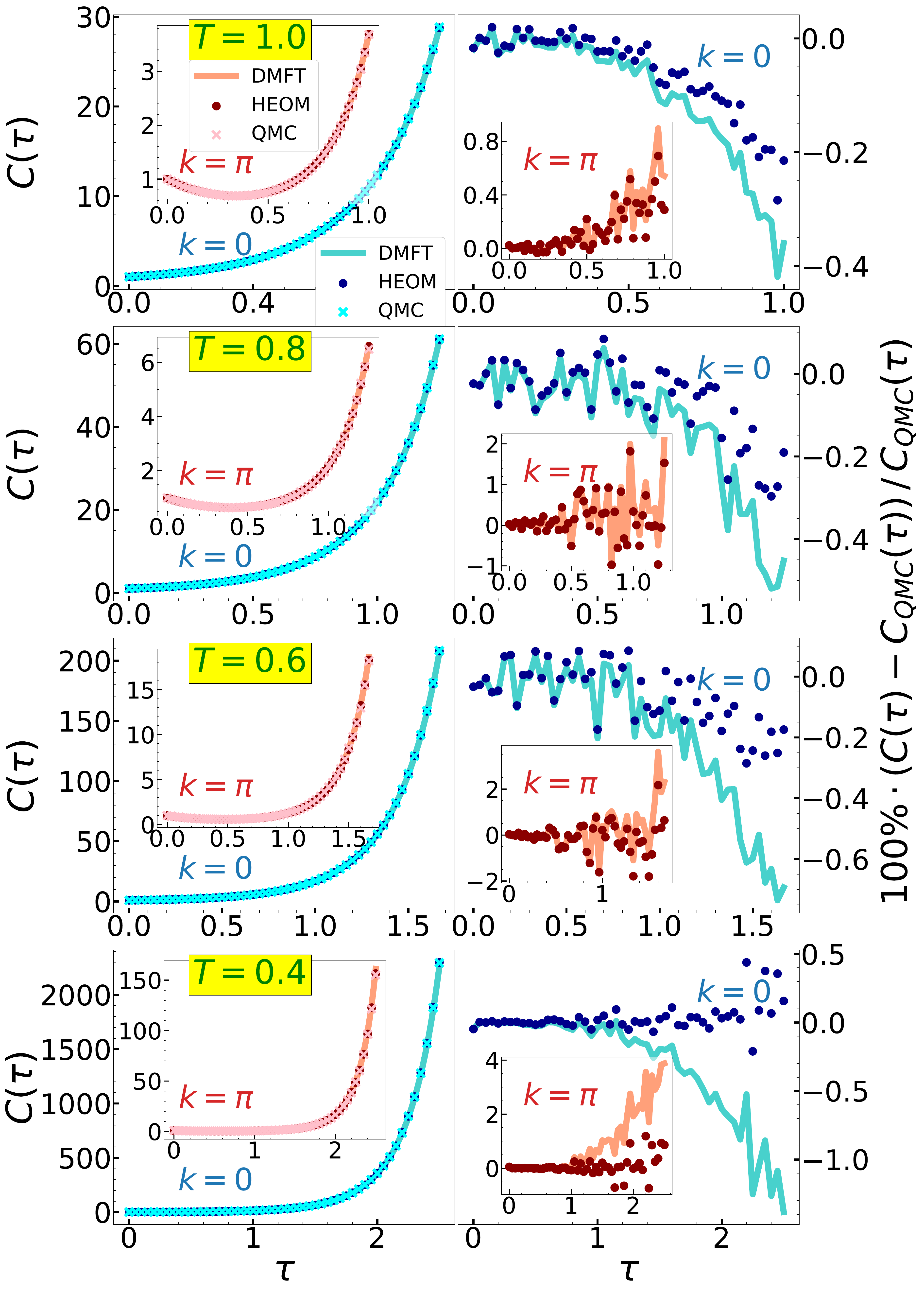}
 \caption{
 DMFT, HEOM and QMC correlation functions for $\omega_0=1, \;g=\sqrt{2}$ at $k=0$ and $k=\pi$ at several temperatures. The right panels show the relative discrepancy between DMFT and HEOM results with respect to QMC.
}
\label{Supp:fig:CorrF_with_err} 
\end{figure}

Fig.~\ref{Supp:fig:CorrF_with_err} shows the imaginary time QMC, DMFT and HEOM correlation functions and their deviation from the QMC result, for parameters as in Fig.~\ref{fig:corF} of the main text. We see that the deviation is very small, the relative discrepancy being just a fraction of a percent at $T=1$. The discrepancy between the DMFT and QMC increases at lower temperatures when the nonlocal correlations are expected to be more important, but it remains quite small even at $T=0.4$. As we can see, the DMFT gives better results at $k=0$ than at $k=\pi$. 

\begin{figure*}[!t]
 \includegraphics[width=7.2in,trim=0cm 0cm 0cm 0cm]{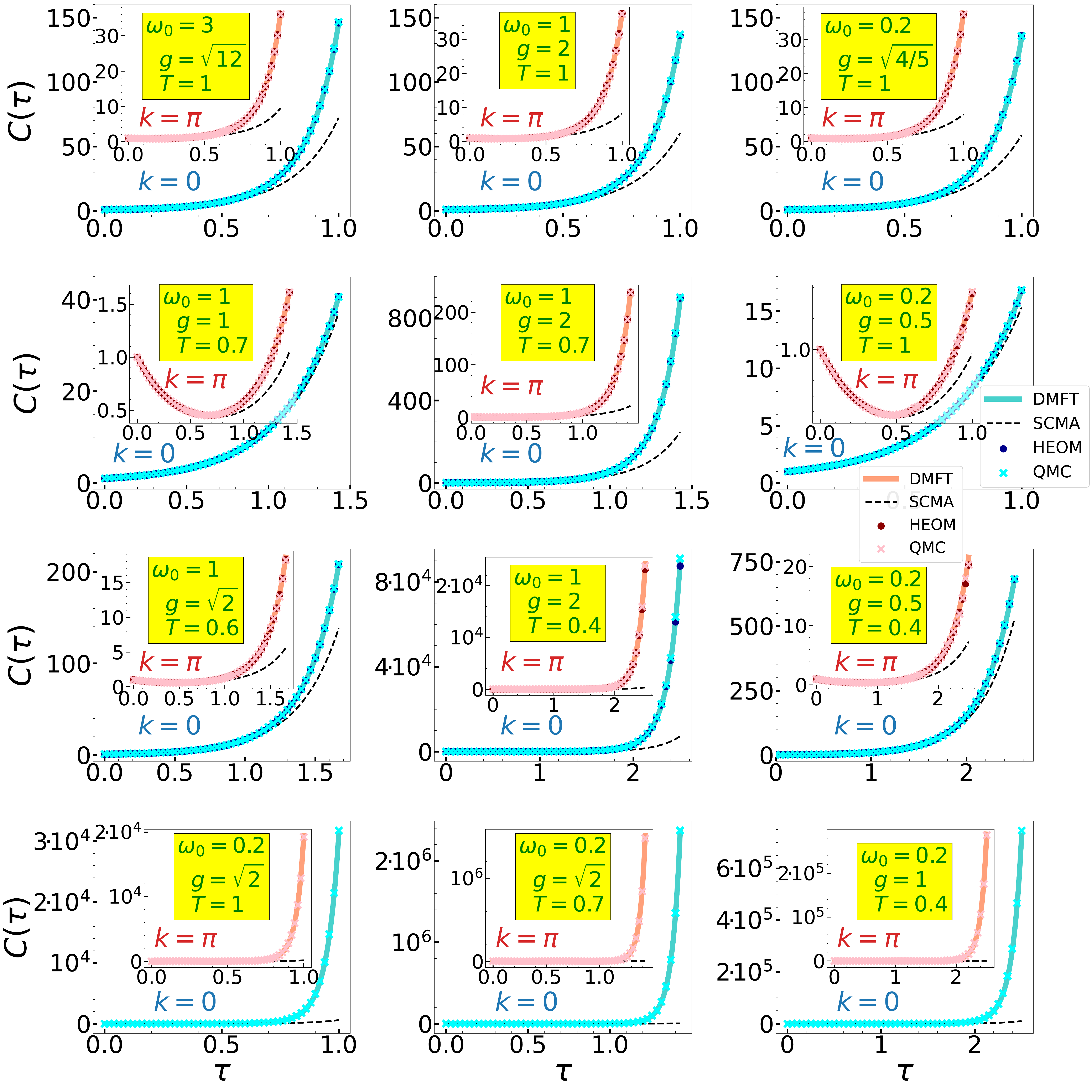}
 \caption{
 Comparison of DMFT, HEOM, QMC and SCMA correlation functions over a wide range of parameters. The HEOM results are not available for the parameters in the last row.
}
\label{Supp:fig:CorrF} 
\end{figure*}

In Fig.~\ref{Supp:fig:CorrF} we present the correlation function comparison over a broad set of parameters. The DMFT, HEOM and QMC are in excellent agreement, with the relative discrepancy of the order of one percent for $\tau \sim 1/T$. The SCMA results are also included for comparison.

From Eq.~\eqref{Supp:eq:Corr_From_Spec} we see that the correlation function unevenly treats different frequencies from the spectral function. Because of the exponential term, it takes into account low-frequency contributions with much larger weight. Thus, the correct DMFT and HEOM predictions about correlation function reveal that the low-frequency parts of the corresponding spectral functions behave appropriately and fall off fast enough. This is very important property for calculating quantities where the low-frequency part gives large contribution to the result, which would be the case for optical conductivity.

Let us now estimate how much a Gaussian centered at frequency $a$, 
\begin{equation}
A^{G}_{k}(\omega) = \frac{W}{\sigma\sqrt{2\pi}} e^{-\frac{(\omega-a)^2}{2\sigma^2}},
\end{equation}
would  contribute to the correlation function. Here $W$ is the spectral weight and $\sigma$ is the standard deviation of the Gaussian. This could model a tiny peak present due to the noise, or a real physical contribution. 
The corresponding part of the correlation function $C^{G}_{k}$ can be singled out since Eq.~\eqref{Supp:eq:Corr_From_Spec} is linear in $A_{k}$. It can be evaluated analytically, giving
\begin{equation} \label{Supp:eq:gaussian_contr}
C^{G}_{k}(\tau) = W e^{\frac{\sigma^2 \tau^2}{2}-a\tau}.
\end{equation}
We see that the spectral weight contributes linearly, while the position of the delta peak contributes exponentially (note that $a$ can be negative). The width of the Gaussian $\sigma$, as well as the imaginary time $\tau$, are quadratic inside the exponential. Hence, Eq.~\eqref{Supp:eq:gaussian_contr} explicitly shows that precise calculation of the correlation function requires very accurate spectral functions at low frequencies. Even a small error or noise can produce a completely wrong result. Reliable comparison of $C_k(\tau)$ was made possible only due to the high precision of both DMFT and HEOM calculations.

\newpage
\clearpage


\section{technical note: Numerical calculation of the integrated spectral weight}
\label{Integral_A}

We describe a numerical scheme for calculating the integrated spectral weight. Integrated spectral weight is defined as
\begin{equation} \label{Supp:eq:cumul_def}
I_k(\omega) = \int_{-\infty}^\omega A_k(\nu) d \nu,
\end{equation}
where $A_k(\nu)$ is the spectral function. Straightforward numerical integration of Eq.~\eqref{Supp:eq:cumul_def} can sometimes lead to the conclusion that the spectral sum rule $I_k(\infty) = 1$ is violated. This happens because the numerical representation of $A_k(\nu)$ on a finite grid does not detect the possible presence of delta function peaks without introducing artificial broadening. This is why our numerical scheme calculates $I_k(\omega)$ directly from the self-energy $\Sigma(\omega)$.

Let us suppose that the self-energy data $\{\Sigma_0, \Sigma_1 ... \Sigma_{N-1}\}$  are known on a grid $\{\omega_0, \omega_1 ... \omega_{N-1}\}$. The integrated spectral weight can then be rewritten as
\begin{align}
I_k(\omega_l) &= -\frac{1}{\pi} \mathrm{Im} \int_{-\infty}^{\omega_l} \frac{d\nu}{\nu - \Sigma(\nu) - \varepsilon_k} \nonumber \\ 
          &\approx -\frac{1}{\pi} \mathrm{Im} \sum_{q=0}^{l-1} \int_{\omega_q}^{\omega_{q+1}}
\frac{d\nu}{\nu - \Sigma(\nu) - \varepsilon_k}. \label{Supp:eq:Cumul_Spec_Fun_Def}
\end{align}
The delta peaks in Eq.~\eqref{Supp:eq:Cumul_Spec_Fun_Def} occur whenever our subintegral function is (infinitely) close to the singularity, i.e. when ${\mathrm{Im} \Sigma(\nu) \to 0^-}$ and ${\nu - \mathrm{Re}\Sigma(\nu) - \varepsilon_k \approx 0}$. These are most easily taken into account by using the linear interpolation of the denominator in Eq.~\eqref{Supp:eq:Cumul_Spec_Fun_Def} and evaluating the integral analytically
\begin{align}
I_k(\omega_l) &\approx -\frac{1}{\pi} \mathrm{Im} \sum_{q=0}^{l-1} \int_{\omega_q}^{\omega_{q+1}}
\frac{d\nu}{\nu - \varepsilon_k - \left[\Sigma_q + \Sigma^\prime_q(\nu - \omega_q) \right]} \nonumber \\
&= -\frac{1}{\pi} \mathrm{Im} \sum_{q=0}^{l-1} \frac{1}{1- \Sigma^\prime_q} \ln \left[ \frac{\omega_{q+1} - \varepsilon_k - \Sigma_{q+1}}{\omega_q - \varepsilon_k - \Sigma_q} \right], \label{Supp:eq:Cumul_Spec_Fun}
\end{align}

where $\Sigma^\prime_q =(\Sigma_{q+1}-\Sigma_q) /(\omega_{q+1}-\omega_q)$. In the last line of Eq.~\eqref{Supp:eq:Cumul_Spec_Fun} we used that $\ln x - \ln y = \ln (x/y)$, which holds since ${\mathrm{Im}\Sigma_q < 0}$ (for every $q$).

In the limit when $\Delta \omega_q = \omega_{q+1} - \omega_q$ is small, Eq.~\eqref{Supp:eq:Cumul_Spec_Fun} predicts that the contribution which corresponds to the interval $(\omega_q, \omega_{q+1})$ is equal to
\begin{equation} \label{Supp:eq:Cumul_with_delta}
\frac{1}{1-\frac{\Sigma_{q+1}-\Sigma_q}{\omega_{q+1}-\omega_q}}
\approx
\frac{1}{1-\partial_\omega \Sigma},
\end{equation}
if the interval contains a delta peak, whereas it is 
\begin{equation} \label{Supp:eq:Cumul_no_delta}
-\frac{1}{\pi}\mathrm{Im}\left[  \frac{\Delta \omega_q}{\omega_q - \varepsilon_k - \Sigma_q} \right]
\end{equation}
otherwise. The analytical result for the contribution of the delta peak coincides with Eq.~\eqref{Supp:eq:Cumul_with_delta}, while Eq.~\eqref{Supp:eq:Cumul_no_delta} is exactly the term we would get using the standard Riemann sum. Having in mind that the Riemann sum approach is completely justified in the absence of delta peaks, we conclude that the integration scheme presented in Eq.~\eqref{Supp:eq:Cumul_Spec_Fun} is perfectly well-suited for the calculation of the integrated spectral weight.

\newpage
\clearpage

\section{technical note: equivalence of spectral functions from different definitions}
\label{def_spectral}
Throughout this paper we compared spectral and correlation functions obtained with various methods. Each method uses different definition of the spectral function. The purpose of this Section is to show that all of them are equivalent in the case we are considering, which is a single electron in a system. We also present the relation which connects the spectral function with the imaginary-time correlation function obtained from QMC calculation.

\subsection{Spectral function from greater Green's function}
\label{Sec:spec_from_greater}
In the HEOM method, the most natural starting point is the greater Green's function \cite{SM2021_Jankovic_Vukmirovic}
\begin{equation}
G^{>}_{\mathbf{k}}\left(t\right)=-i
\left\langle c_{\mathbf{k}}\left(t\right) c_{\mathbf{k}}^\dagger\right\rangle_{T,0} .
\end{equation}
Here $c_{\mathbf{k}}$ and $c_{\mathbf{k}}^\dagger$ are the electron annihilation and creation operators, while
$$c_{\mathbf{k}}\left(t\right)=e^{iHt}c_{\mathbf{k}}\left(0\right)e^{-iHt}.$$ 
The notation $\left\langle  \ldots \right\rangle_{T,0}$ denotes the thermal overage over the space of states containing zero electrons
\begin{equation}
\left\langle  x \right\rangle_{T,0}
=\frac{\sum_p \mel{p}{e^{- H_{\mathrm{ph}}/T}x}{p}}{\sum_p \mel{p}{e^{-H_{\mathrm{ph}}/T}}{p}}= \frac{1}{Z_p} \sum_p \mel{p}{e^{- H_{\mathrm{ph}}/T}x}{p}.
\end{equation}
Here $\ket{p}$ denotes the states containing no electrons and arbitrary number of phonons, $H_{\mathrm{ph}}$ is purely phononic part of the Hamiltonian and $Z_p$ is the phononic partition function. The spectral function is now defined as
\begin{equation} \label{supp:eq:spec_def_111}
A_{\mathbf{k}}\left(\omega\right)=-\frac{1}{2\pi}
\mathrm{Im}G^{>}_{\mathbf{k}}\left(\omega\right),
\end{equation}
where 
\begin{equation}
G^{>}_{\mathbf{k}}\left(\omega\right)=
\int_{-\infty}^{\infty}\mathrm{d}t \: e^{i\omega t} \:
G^{>}_{\mathbf{k}}\left(t\right).
\end{equation}
These expressions can be cast into explicit form using the Lehmann spectral representation (using the basis of energy eigenstates $H |n \rangle = E_n |n \rangle$)
\begin{equation}
G_{\mathbf{k}}^{>}\left(t\right)=\frac{-i}{Z_p}
\sum_{p,e} e^{- E_p/T} 
e^{iE_p t}
\mel{p}{c_{\mathbf{k}}}{e}
e^{-iE_e t}
\mel{e}{c_{\mathbf{k}}^\dagger}{p},
\end{equation}
where $\ket{e}$ denotes the states containing one electron and an arbitrary number of phonons. The spectral function can now be obtained by taking the Fourier transform of previous expression and using Eq.~\eqref{supp:eq:spec_def_111}
\begin{equation}\label{eq:spec1}
A_{\mathbf{k}}\left(\omega\right)=
\frac{1}{Z_p}
\sum_p e^{-E_p/T} 
\sum_e 
\delta\left(\omega+E_p-E_e\right)
\left|\mel{p}{c_{\mathbf{k}}}{e}\right|^2.
\end{equation}

\subsection{Spectral function from retarded and time-ordered Green's function}
\label{Sec:spec_from_retarded}
In the DMFT/SCMA, we can start from the time-ordered Green's function \citep{SMCiuchi_1997} with just a single electron inserted into the system
\begin{equation} \label{Supp:eq:time_ordered}
G_{\mathbf{k}}(t) = -i \langle T c_{\mathbf{k}}(t) c_{\mathbf{k}}^\dagger \rangle_{T,0} .
\end{equation}
As in the case of the greater Green's function, here we average only over the phonon degrees of freedom. This means that~\eqref{Supp:eq:time_ordered} gives nonvanishing contribution only for ${t>0}$
\begin{equation}
G_{\mathbf k}(t) = -i \theta(t) \langle c_{\mathbf{k}}(t) c_{\mathbf{k}}^\dagger \rangle_{T,0} .
\end{equation}
In our case of a single electron in the system, this coincides with the retarded Green's function. Ref \cite{SMCiuchi_1997} explains in detail how is this connected to the polaron impurity problem. Now, the spectral function can be obtained as
\begin{equation} \label{supp:eq:spec_dmft}
A_{\mathbf k}(\omega) = -\frac{1}{\pi} \mathrm{Im} G_{\mathbf k}(\omega),
\end{equation}
where
\begin{equation} \label{supp:eq:gw111}
G_{\mathbf{k}}\left(\omega\right)=\lim_{\varepsilon\to 0^+}
\int_{-\infty}^{\infty}\mathrm{d}t \: e^{i\left(\omega+i\varepsilon \right) t} \:
G_{\mathbf{k}}\left(t\right).
\end{equation}
%
Let us now check whether the definitions of spectral functions from Secs.~\ref{Sec:spec_from_greater} and~\ref{Sec:spec_from_retarded} are in agreement with one another. This can be easily checked by utilizing the Lehmann spectral representation 
\begin{equation}
G_{\mathbf{k}}\left(t\right)=
\frac{-i\theta\left(t\right)}{Z_p}\sum_{p,e}
e^{- E_{p}/T} 
e^{i\left(E_{p}-E_{e}\right)t}
\left|\mel{p}{c_{\mathbf{k}}}{e}\right|^2.
\end{equation}
The spectral function is now obtained by performing the Fourier transform, using Eq.~\eqref{supp:eq:spec_dmft} and the Plemelj-Sokhotski theorem $\mathrm{Im}\lim_{\varepsilon\to 0^+}\frac{1}{x+i\varepsilon}=-\pi\delta\left(x\right)$. We obtain the result which coincides with~\eqref{eq:spec1}. Furthermore, these results also coincide with Eq.~\eqref{eq:supp:exact_diag}. This confirms that all of these approaches are consistent with one another. 


\subsection{Spectral function from grand canonical ensemble}

It is also quite common to work within the grand canonical ensemble, not restricting ourselves explicitly to a single electron in a system. Here we use the usual definition of the retarded Green's function 
\begin{equation} \label{definicija_grin}
G_{\mathbf{k}}\left(t\right)=-i\theta\left(t\right)
\left\langle \left\{ c_{\mathbf{k}}\left(t\right) , c_{\mathbf{k}}^\dagger \right\} \right\rangle_{T},
\end{equation}
where
\begin{equation}
c_{\mathbf k}(t) = e^{i K t} c_{\mathbf k} 
e^{-i K t},
\end{equation}
$K=H-\mu N$ and $N$ being the electron number operator.  The notation $\langle \dots \rangle_{T}$ denotes the average value in the grand canonical ensemble and $\{,\}$ is the anticommutator. The spectral function is obtained by substituting $G_{\mathbf{k}}\left(t\right)$ from~\eqref{definicija_grin} into Eqs.~\eqref{supp:eq:gw111} and~\eqref{supp:eq:spec_dmft}. A more explicit form can be obtained using the Lehmann spectral representation (using the basis of energy eigenstates $K|n\rangle = K_n |n \rangle$)
\begin{equation}\label{eq:spec2}
\begin{split}
A_{\mathbf{k}}\left(\omega\right)=&
\frac{1}{Z}\sum_{n_1n_2}
e^{-\beta K_{n_1}} 
\left[
\left|\mel{n_1}{c_{\mathbf{k}}}{n_2}\right|^2
\delta\left(K_{n_1}-K_{n_2}+\omega\right)
\right.
 \\
&\left.
+\left|\mel{n_1}{c_{\mathbf{k}}^\dagger}{n_2}\right|^2
\delta\left(K_{n_2}-K_{n_1}+\omega\right)
\right],
\end{split}
\end{equation}
where $Z=\mathrm{Tr} \qty(e^{-\beta K})$ is the partition function. Let us now consider what happens in the case we are interested in, which is the zero density limit. This corresponds to $\mu\to -\infty$. 

We note first that the dominant terms in the partition function $Z$ in this limit are from the states with zero electrons
\begin{equation}
Z=\sum_n e^{-\beta K_n} = \sum_p e^{-\beta K_p}=Z_p.
\end{equation}
The states containing a larger number of electrons introduce an additional term $e^{\beta \mu N}$ which is exponentially small when $\mu\to -\infty$. Consequently, we have shown that $Z$ from Eq.  \eqref{eq:spec2} is the same as $Z_p$ from Eq.  \eqref{eq:spec1} in the limit  $\mu\to -\infty$.

Next, we consider the sum in Eq. \eqref{eq:spec2}. Due to the $e^{-\beta K_{n_1}}$ factor, the dominant contribution to the sum over $n_1$ comes from the states $\ket{n_1}$ containing zero electrons. The states containing a larger number of electrons introduce an additional term $e^{\beta \mu N}$ which is exponentially small when $\mu\to -\infty$. Therefore, the sum over $n_1$ in Eq. \eqref{eq:spec2} can be replaced by a sum over $p$, where $\ket{p}$ denote the states containing no electrons. The second term containing $\mel{n_1}{c_{\mathbf{k}}^\dagger}{n_2}$ in Eq. \eqref{eq:spec2} is then zero, while the first term containing $\mel{n_1}{c_{\mathbf{k}}}{n_2}$ is different from zero only when $\ket{n_2}$ is the state containing one electron. The sum in Eq. \eqref{eq:spec2} then reads as
\begin{equation}
A_{\mathbf{k}}\left(\omega\right)=
\frac{1}{Z_p}\sum_{p,e}
e^{-\beta K_{p}} 
\left|\mel{p}{c_{\mathbf{k}}}{e}\right|^2
\delta\left(K_{p}-K_{e}+\omega\right),
\end{equation}
We further note that the last equation can be also expressed in the form
\begin{equation}\label{eq:akom4510}
A_{\mathbf{k}}\left(\omega-\mu\right)=
\frac{1}{Z_p}\sum_{p,e}
e^{-\beta E_{p}} 
\left|\mel{p}{c_{\mathbf{k}}}{e}\right|^2
\delta\left(E_{p}-E_{e}+\omega\right).
\end{equation}
The right hand side in previous equation coincides with Eq.~\eqref{eq:spec1}. This proves that the spectral function within the grand canonical formalism needs to be considered in the limit  $\mu \to -\infty$ and also the result needs to be shifted $A_{\mathbf k} (\omega) \to A_{\mathbf k}(\omega-\mu)$ if we want our result to coincide with Eq.~\eqref{eq:spec1}. 

All of these results give us to flexibility to work within different formalisms knowing that all of them give the same result. Hence, we proved that the definitions of spectral functions within HEOM, DMFT, SCMA and ED are all in agreement.

\subsection{Relation between the spectral function and imaginary-time correlation function}
\label{Supp:rel_spec_corr}
In QMC we calculate the quantity
\begin{equation}\label{supp:eq:cktau}
C_{\vb{k}}(\tau) =  \langle c_{\vb{k}}(\tau) c_{\vb{k}}^\dagger \rangle_{T,0},
\end{equation}
where
\begin{equation}
c_{\mathbf k}(\tau) = e^{ \tau H } c_{\mathbf k} e^{-  \tau H }.
\end{equation}
Again, using the Lehmann spectral representation in Eq.~\eqref{supp:eq:cktau} we get
\begin{equation}\label{eq:cktau725227}
 C_{\vb{k}}(\tau)=
 \frac{1}{Z_p}\sum_{p,e} e^{-\beta E_p} 
 \abs{\mel{p}{c_{\vb{k}}}{e}}^2
 e^{\tau\qty(E_{p}-E_{e})}.
\end{equation}
By performing straightforward integration, one then finds from Eqs.~\eqref{eq:spec1} and \eqref{eq:cktau725227}
\begin{equation}
C_{\vb{k}}(\tau) = 
\int_{-\infty}^\infty d\omega \,
e^{-\omega\tau} A_{\vb{k}}(\omega).
\end{equation}
This proves Eq.~\eqref{Supp:eq:Corr_From_Spec}, which connects the correlation functions from QMC with spectral functions, obtained from other methods.

\bibliographystyle{apsrev4-1}

\end{document}